\documentclass[useAMS,usenatbib,fleqn]{mnras}
\usepackage{graphicx}
\usepackage{times}
\usepackage{amssymb, amsmath}
\usepackage[section]{placeins}
\usepackage[english]{babel}
\usepackage{lineno}
\PassOptionsToPackage{hyphens}{url}
\usepackage{ulem}
\usepackage{hyperref}
\usepackage{url}
\newcommand{\hinvMpc}{h^{-1} {\rm Mpc}}      
\newcommand{\hinvMsun}{h^{-1} {\rm M_\odot}} 
\newcommand{\Mr}{M_{\rm r}}                  
\newcommand{\Mh}{M_{\rm h}}                  
\newcommand{\Mnl}{M_{\rm nl}}
\newcommand{\wpp}{w_{\rm p}}
\newcommand{\rp}{r_{\rm p}}
\newcommand{\fred}{f_{\rm p-red}}
\newcommand{\fblue}{f_{\rm p-blue}}

\def\ni{\bar{n}_i}
\def\nj{\bar{n}_j}
\def\ng{\bar{n}_{\rm g}}

\def\nijg{\frac{\ni\nj}{\ng^2}}
\def\meanNsati{\langle N_{\rm sat} (M_i)\rangle}
\def\meanNsatj{\langle N_{\rm sat} (M_j)\rangle}
\def\meanNceni{\langle N_{\rm cen} (M_i)\rangle}
\def\meanNcenj{\langle N_{\rm cen} (M_j)\rangle}
\def\Mi{M_i}
\def\Mj{M_j}
\def\whhcc{w_{\rm hh,cc}}
\def\whhcs{w_{\rm hh,cs}}
\def\whhss{w_{\rm hh,ss}}

\def\aap{A\&A}

\def\apj{ApJ}
\def\apjs{ApJS}
\def\apjl{ApJL}

\def\mnras{MNRAS}
\def\aj{AJ}
\def\nat{Nature}

\title[CCMD: Colour-Magnitude-Halo Mass]{The Conditional 
Colour-Magnitude Distribution: I. A Comprehensive Model of the 
Colour-Magnitude-Halo Mass Distribution of Present-Day Galaxies}
\author[H. Xu et al.]{
Haojie Xu$^{1,2}$\thanks{E-mail: haojie.xu@sjtu.edu.cn},
Zheng Zheng$^{1,3,2}$\thanks{E-mail: zhengzheng@astro.utah.edu},
Hong Guo$^{4}$,
Ying Zu$^{2,5,6}$,
Idit Zehavi$^{7}$
\newauthor
and
David H. Weinberg$^{5,6}$\\
\\
$^1$
Department of Physics and Astronomy, University of Utah, 115 South 1400 East,
Salt Lake City, UT 84112, USA\\
$2$
Department of Astronomy, Shanghai Jiao Tong University, Shanghai 200240, China\\ 
$^3$
Tsung-Dao Lee Institute, Shanghai 200240, China\\
$^4$
Key Laboratory for Research in Galaxies and Cosmology, Shanghai Astronomical 
Observatory, Shanghai 200030, China\\
$^5$
Department of Astronomy, Ohio State University, Columbus, OH 43210, USA\\
$^6$
Center for Cosmology and Astro-Particle Physics, Ohio State University, 
Columbus, OH 43210, USA\\
$^7$
Department of Astronomy and Department of Physics, Case Western Reserve 
University, 10900 Euclid Avenue, Cleveland, OH 44106, USA
}
\begin{document}
\maketitle

\begin{abstract}
We formulate a model of the conditional colour-magnitude distribution (CCMD) to
describe the distribution of galaxy luminosity and colour as a function of halo
mass. It consists of two populations of different colour distributions, dubbed
pseudo-blue and pseudo-red, respectively, with each further separated into
central and satellite galaxies. We define a global parameterization of these
four colour-magnitude distributions and their dependence on halo mass, and we
infer parameter values by simultaneously fitting the space densities and
auto-correlation functions of 79 galaxy samples from the Sloan Digital Sky
Survey defined by fine bins in the colour-magnitude diagram (CMD). The model
deprojects the overall galaxy CMD, revealing its tomography along the halo
mass direction. The bimodality of the colour distribution is driven by central
galaxies at most luminosities, though at low luminosities it is driven by the
difference between blue centrals and red satellites. For central galaxies, the
two pseudo-colour components are distinct and orthogonal to each other in the
CCMD: at fixed halo mass, pseudo-blue galaxies have a narrow luminosity range
and broad colour range, while pseudo-red galaxies have a narrow colour range
and broad luminosity range. For pseudo-blue centrals, luminosity correlates
tightly with halo mass, while for pseudo-red galaxies colour correlates more
tightly (redder galaxies in more massive haloes). The satellite fraction is
higher for redder and for fainter galaxies, with colour a stronger indicator
than luminosity. We discuss the implications of the results and
further applications of the CCMD model.
\end{abstract}


\section{Introduction}
\label{sec:intro}

Among all the galaxy physical properties directly measurable from observation with 
galaxy surveys, such as luminosity, colour, surface brightness, morphology, and 
environment, luminosity and colour jointly are the two most predictive ones 
for the underlying action and cessation of star formation inside galaxies. 
At fixed luminosity and colour, it has been shown that there is little residual 
correlation between environment and surface brightness or morphology 
\citep{Blanton05c}. In principle, galaxy luminosity can be 
regarded as a zero-th order proxy to
galaxy stellar mass, while in detail it is also linked to the galaxy star 
formation history and the age of the stellar population. Galaxy colour is an 
approximate indicator of star formation activity and status, 
and it is also related to galaxy metallicity and dust content. 

In this paper, we develop a formalism to study the luminosity and colour 
distribution of galaxies through galaxy clustering and apply it to galaxy survey
data. The model substantially extends previous methods to connect galaxies 
with dark matter haloes, presenting
a quantitative global description of the joint colour-magnitude distribution as 
a function of halo mass for galaxies in the Sloan Digital Sky Survey 
(SDSS; \citealt{York00}) main galaxy redshift survey \citep{Strauss02}. 

In the local universe, the distribution of galaxy luminosity and colour
(or the colour-magnitude diagram; CMD) shows ordered patterns 
\citep[e.g.][]{Strateva01, Blanton03a, Baldry04}. 
Towards the low luminosity part in the CMD, galaxies form a diffuse `blue cloud',
which mainly consists of late-type galaxies with star formation.
There is also a tight `red sequence' of galaxies stretching from 
the faint end all the way to the luminous end, which is mainly made of 
early-type galaxies lacking star formation. 
A `green valley' lies in between the blue cloud and red sequence.
The bimodal colour distribution of galaxies has already emerged early in the
history of the universe (e.g. $z \sim$1--2.5; \citealt{Faber07,Brammer09,Coil17}).
The formation of the bimodality involves mechanisms responsible for transforming 
the young active blue galaxies to old passive red ones, through quenching the 
star formation. The quenching mechanisms can be broadly divided into `internal'
and `external' processes. Star-formation and active galactic nuclei (AGN)
feedback that heats or blows away the gas and secular process that uses up 
the gas slowly belong to the former. Processes that quench star formation 
inside high density environment belong to the latter, such as ram-pressure 
stripping to remove the cold gas \citep[e.g.][]{Gunn72} and strangulation to cut 
off the cold gas supply \citep[e.g.][]{Peng15}. \citet{Faber07} investigate 
the evolution of luminosity function (LF) of blue and red galaxies since 
$z \sim 1$ and find 
that there is little evolution in the number density and stellar mass of 
blue galaxies, while those of red galaxies have dramatically increased. 
A `mixed' scenario is thus proposed to explain the new red galaxies in which 
blue star-forming galaxies are quenched by gas-rich major mergers and then move 
along the red sequence by a series of gas-poor mergers. 

In the standard paradigm of galaxy formation and evolution 
\citep[e.g.][]{White78}, galaxies form and evolve inside dark matter haloes. 
To some degree, all the physical processes affecting a galaxy are 
unavoidably connected to its parent halo. Linking galaxy properties (e.g.
luminosity and colour) to haloes forms an important step towards learning about
the galaxy formation processes. Without resorting to numerically expensive 
cosmological hydrodynamic simulation or semi-analytic galaxy formation models 
(SAM), there are empirical approaches within the halo model to establish the 
relation between galaxy luminosity and colour and dark matter haloes.
One approach is to construct galaxy group catalogues
\citep[e.g.][]{Yang05b,Berlind06, Yang07}, with each group representing
a dark matter halo. Using a galaxy group catalogue, \citet{Yang08} study the
conditional luminosity functions (CLFs) of red and blue galaxies and 
\citet{vandenBosch08} investigate the satellite properties. When interpreting
the results from group catalogues, systematic errors such as group membership 
determination, central/satellite designation, and halo mass assignment need to be
accounted for \citep[e.g.][]{Campbell15, Lin16}.
With haloes and sub-haloes identified in high-resolution $N$-body simulations, 
a connection between galaxies and haloes/sub-haloes can be established through
the sub-halo abundance matching (SHAM) approach \citep{Kravtsov04, Conroy06}, 
which matches the cumulative LF with the cumulative abundance of haloes/sub-haloes
(e.g. in terms of mass or circular velocity). The SHAM approach has been extended
to include galaxy colour through `age matching' \citep{Hearin13}, which assumes 
a one-to-one mapping of halo/sub-halo formation redshift onto galaxy colour in each
fixed luminosity bin.

The other powerful method of establishing galaxy-halo relation, in particular 
in terms of galaxy luminosity and colour, is through
galaxy clustering. As galaxy clustering depends on galaxy luminosity
(e.g. \citealt{Norberg01, Norberg02, Zehavi02, Zehavi05, Zehavi11})
and halo clustering depend on halo mass \citep[e.g.][]{Mo96}, comparing 
galaxy clustering and halo clustering would establish the link between
galaxy luminosity and halo mass. More sophisticated models have been
developed to interpret galaxy clustering, which include the  
halo occupation distribution framework (HOD; e.g. 
\citealt{Jing98,Seljak00,Scoccimarro01,Berlind02,Zheng05}) and the 
CLF method \citep{Yang03,Yang05}. Such models transform galaxy clustering
measurements to the informative, physical relation between galaxies and dark
matter haloes, which encodes the complex physics of galaxy formation and 
evolution and helps test galaxy formation theory.

With well-motivated parameterizations, HOD and CLF have been successfully applied
to model the luminosity dependent galaxy clustering 
\citep[e.g.][]{vandenBosch03,Yang05a,Zehavi05,Tinker05,Zheng07b,Zheng09,Zehavi11,
Guo13,vandenBosch13}. In general, a tight correlation between galaxy luminosity
and halo mass is inferred, and the increase in the amplitude of the projected 
two-point correlation function (2PCF) with luminosity reflects an overall shift 
in the halo mass scale. 

Modelling the dependence of galaxy clustering on galaxy colour or the joint
dependence on luminosity and colour turns out to be less well formulated, and
it is usually carried out on a case-by-case basis. 
\citet{Zehavi05a} model the joint 
luminosity-colour dependence of galaxy clustering by further parameterizing the 
blue and red fraction of galaxies as a function of halo mass, separated into
central and satellite components. To model the colour dependent galaxy 
clustering in fine colour bins at fixed luminosity (rather than a simple red/blue division as in \citealt{Zehavi05a}), \citet{Zehavi11} employ a simplified HOD 
model by explicitly assuming the relative normalisation of central 
and satellite mean occupation functions to be responsible for the 
colour-dependent clustering. It is not straightforward to generalise the
models in \citet{Zehavi05a} and \citet{Zehavi11} to model galaxy clustering
for galaxy samples in fine bins of luminosity and colour, covering the whole
colour-magnitude plane.
\citet{Skibba09b} provide a prescription of halo mass dependent colour-magnitude
relation for central and satellite galaxies, which assumes that the colour 
distribution depends on luminosity while the luminosity is related to halo mass.
The colour-magnitude relations used in the model are derived from the CMD through
fitting two Gaussian components at each fixed luminosity, with modifications for
satellites. The model can reasonably explain the colour dependent galaxy clustering.

In this paper, we aim at a more complete description of the luminosity and
colour distribution of galaxies inside dark matter haloes, to enable
the modelling of galaxy clustering for galaxy samples with any luminosity
and colour cuts. We first develop the \textit{global} parameterization of the
galaxy colour-magnitude distribution as a function of halo mass, with galaxies
separated into centrals and satellites and model colour populations. The model
is named the conditional colour-magnitude distribution (CCMD). Given a set of CCMD
parameters, the HOD of any galaxy sample defined by cuts in luminosity and
colour is readily computed. Combining the derived HOD with the haloes identified 
in $N$-body simulations, the clustering statistics can be calculated for the
sample. The CCMD formalism can be used to \textit{simultaneously} model the
abundances and clustering measurements of galaxy samples in fine bins of 
luminosity and colour across the whole CMD. The subsequent constraints on 
CCMD parameters allow a de-projection of galaxy CMD along the halo mass
direction and a decomposition into contributions of central/satellite and
different model colour components. This will help our understanding of the
bimodality and the role halo mass plays in the transition from blue to 
red galaxy populations for central and satellite galaxies. Throughout the
paper, we model the $g-r$ colour distributions of galaxies as a sum of
two Gaussians, which we refer to as `pseudo-blue' and `pseudo-red'.
These distributions overlap, however, so both components may contribute
to the population of galaxies at a given luminosity and colour 
(see \S~\ref{sec:para} and Fig.~\ref{fig:ngdm}).

The paper is organised as follows. In section~\ref{sec:data}, we describe
the construction of galaxy samples defined in fine luminosity and colour bins
in the CMD and the measurements of galaxy clustering statistics (the projected
2PCFs). In section~\ref{sec:para}, we formulate and parameterize the CCMD 
and present the method to calculate 2PCFs of galaxies based on the CCMD and 
an $N$-body simulation. In section~\ref{sec:res}, we apply the CCMD formalism 
to simultaneously model the clustering measurements in fine bins of galaxy 
luminosity and colour and present the modelling results and the constraints 
on the CCMD. In this section, we also present the derived quantities and relations
from the CCMD and compare them with those from previous work.
Finally, we summarize and discuss our results in section~\ref{sec:dis}.

Throughout the paper, we adopt a spatially flat $\Lambda$ cold dark matter 
cosmology with $\Omega_{\rm m}=0.307$, $\Omega_{\rm b}=0.048$, $h=0.678$,
$n_{\rm s}=0.96$, and $\sigma_8=0.823$, 
following the constraints from {\rm Planck} 
\citep{PlanckCollaboration14,PlanckCollaboration16} 
and consistent with those in the simulation we use in our model 
(see details in section~\ref{sec:para}). We use $\log$ for base-10 logarithm.

\section{Galaxy Samples and 
Clustering Measurements}
\label{sec:data}

We investigate the colour and luminosity dependence of galaxy clustering 
with the Sloan Digital Sky Survey Data Release 7 Main Galaxy Sample
(SDSS DR7; \citealt{York00, Strauss02, Stoughton02, Abazajian09}).
All the galaxy positions and properties are extracted from {\it bright1},
the large-scale structure sample of the NYU Value-Added Galaxy 
Catalog\footnote{\url{http://sdss.physics.nyu.edu/lss.html}} 
(NYU-VAGC; \citealt{Blanton05b,Adelman-McCarthy08,Padmanabhan08}).
For the purpose of this work,
we further limit our analysis to the galaxies with 
$-22 < M_{\rm r} < -18$ and $0 < g-r < 1.2$.

All the magnitudes and colour used in this work have been $K$-
and evolution-corrected to $z \sim 0.1$, the median
redshift of galaxies in the SDSS DR7 Main Sample. The magnitude 
is calculated by setting $h=1$, where $h$ is the Hubble constant in units
of 100${\rm km\, s^{-1}\, Mpc^{-1}}$. We use magnitude, absolute magnitude, 
and luminosity interchangeably.

\subsection{Galaxy Sample Construction with 
Luminosity and Colour Bins}
\label{subsec:binning}

To measure and model the luminosity and colour dependence of
galaxy clustering, we construct galaxy samples in fine bins 
of luminosity and colour. There are two competing requirements. 
On the one hand, the bins should be 
narrow enough to capture the main features in the 
colour-magnitude distribution of galaxies.
On the other hand, the bins should be
broad enough to include a large number of galaxies to reach 
reasonable signal-to-noise ratios for clustering measurements.

We first divide galaxies ($-22<M_{\rm r}<-18$) into $16$ magnitude bins
of bin width $\Delta M_{\rm r} = 0.25$ mag. For each magnitude bin,
a volume-limited galaxy sample is constructed based on the luminosity
bounds. The volume-limited luminosity bin sample is then further divided
into sub-samples of different colours. 

In many previous studies, tilted luminosity-dependent dividing lines are 
usually adopted to construct 
galaxy samples of different colours \citep[e.g.][]{Baldry04, Zehavi05a,
Zehavi11, Zu15, Zu16}, which roughly follow the valley and ridges in the CMD. 
As we will show, the model we develop 
has a global parameterization over the whole colour and luminosity distribution 
of galaxies (\S~\ref{sec:para}). Therefore  
the details of the boundaries of colour sub-samples do not matter, 
as long as the main features (e.g. colour bimodality) are captured 
in defining the samples. We apply simple colour cuts to construct 
the colour sub-samples for each luminosity-bin sample. 

\begin{figure}
\includegraphics[width =\columnwidth]{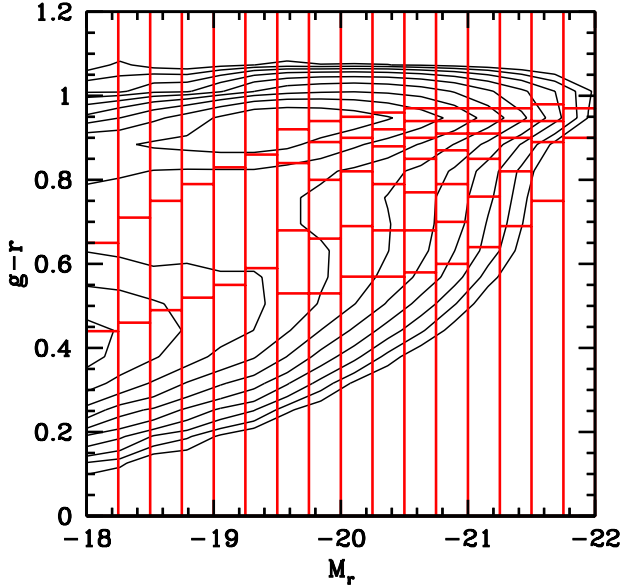}	
\caption{
Construction of galaxy samples in fine bins of colour and luminosity.
The contours show the number density of galaxies in the CMD, illustrating 
the colour-magnitude distribution of galaxies.
The vertical red lines (plus the two vertical axes) represent the 
magnitude boundaries for volume-limited luminosity-bin samples, while 
the horizontal red lines (plus the two horizontal axes) mark the 
colour cuts within each luminosity-bin sample. In total, $79$
galaxy samples are constructed for clustering measurements and modelling.
See the text (\S~\ref{subsec:binning}) for detail.
}
\label{fig:binning}
\end{figure}

With the above two requirements taken into consideration, the number 
of colour sub-samples ranges from $3$ to $8$ for different luminosity-bin 
samples. As a whole, we end up with 
$79$ galaxy sub-samples defined by fine bins in colour and luminosity and 
covering the galaxy CMD, which are shown in Fig.~\ref{fig:binning}. 
The contours show the number density distribution of galaxies in the CMD, 
computed with the $1/V_{\rm max}$ method.
For low-luminosity samples (e.g. $-18.25<M_{\rm r}<-18$), the 
corresponding small survey volumes limit
the number of colour sub-samples to 3 for each sub-sample to have
reasonable clustering measurements. For each of the two samples just 
above $L^\ast$ ($M_{\rm r} \sim -20.44$; \citealt{Blanton03b}), we can 
afford to form 8 colour sub-samples and still have at least 5,000 galaxies 
per sub-sample. For
high-luminosity samples, while the survey volumes are large, the
steep drop of galaxy LF (number density) limits the
number of colour sub-samples (e.g. 3 for the most luminous sample we
consider in this work, $-22<M_{\rm r}<-21.75$). 
Appendix~\ref{subsec:galsampleinfo} lists the details of the $79$ galaxy samples .

\subsection{Galaxy Clustering Measurements}
\label{subsec:clustering}

We measure the projected two-point correlation functions 
(projected 2PCFs) for the $79$ galaxy samples defined by 
fine bins in luminosity and colour.

We first evaluate the redshift-space 2PCFs $\xi(r_{\rm p},\pi)$ using the
Landy-Szalay estimator \citep{Landy93},
\begin{equation}
\label{eqn:LS}
\xi(r_{\rm p},\pi)=\frac{{\rm DD}-2{\rm DR}+{\rm RR}}{\rm RR},
\end{equation}
where $r_{\rm p}$ ($\pi$) is the galaxy pair separation perpendicular 
(parallel) to the line of sight, and DD, RR, and RR are the normalised 
numbers of data-data, data-random, and random-random galaxy pairs in 
the corresponding bins around $r_p$ and $\pi$. We adopt uniform $r_{\rm p}$ bins 
in logarithmic space with a bin width $\Delta \log r_{\rm p} = 0.2$, with 
$r_p$ ranging from $0.13 \hinvMpc$ to $20.48 \hinvMpc$.  The $\pi$ bins
are uniform in linear space from $0$ to $40 \hinvMpc$ with a bin width 
$\Delta \pi = 2 \hinvMpc$. To account for the survey geometry and angular
selection function, we use the random catalogues from the NYU-VAGC. 
The corresponding random catalogue for each galaxy sample contains $\sim$50 
times as many galaxies. 

The projected 2PCF $w_{\rm p}$ is obtained by integrating 
the $\xi(r_{\rm p},\pi)$ along the $\pi$ direction,
\begin{equation}
w_{\rm p}(r_{\rm p})=2\int_0^{\pi_{\rm max}} d\pi\xi(r_{\rm p},\pi),
\end{equation}
with $\pi_{\rm max} = 40 \hinvMpc$.
The projected 2PCF has the redshift-space distortion effects 
largely removed and represents the galaxy clustering in real space.
As shown later, our model accounts for any residual redshift-space
distortion effects caused by a finite $\pi_{\rm max}$.

We adopt the sample mean from jackknife re-sampling method as our estimation 
of $w_{\rm p}$, which has no noticeable difference from that estimated with the full sample. The covariance matrix is also estimated with the jackknife 
method. The footprint of the galaxy sample is divided into $N=144$ 
spatially contiguous and equal area sub-regions. We measure $w_{\rm p}$ 
for $144$ times leaving out one different sub-region at each time, and
the covariance is calculated as $143$ times the variance of the $144$ measurements
\citep[e.g.][]{Zehavi05a, Zehavi11, Guo15, Xu16, Zu16, Guo17}.

At a fixed magnitude bin, different colour sub-samples share the same 
survey volume (section~\ref{subsec:binning}), and hence the measurements of
their projected 2PCFs are strongly correlated. In this work,
we take into account the covariance of the $w_{\rm p}$ measurements 
among all the colour sub-samples at a fixed magnitude bin. 
In practice, galaxy samples in the adjacent magnitude bins partially overlap 
in volume so that their $w_{\rm p}$ measurements should also be
correlated. Ideally, one would like to have a big covariance
matrix to account for the correlations among $w_{\rm p}$ measurements 
from all galaxy samples, which would have $948\times 948$ elements (12 
data points per sample times 79 galaxy samples). 
However, to have a precise estimation of such a covariance matrix 
is not practical, especially with only 
$144$ jackknife sub-samples. We therefore neglect the covariance for 
$w_{\rm p}$ measurements of samples across different magnitude bins.

In this paper, as the first step of CCMD modelling, we limit ourselves to constrain 
the model with auto-correlation functions. The constraints can be tested and further
tightened with other clustering statistics, such as cross-correlation functions and
galaxy lensing measurements. See \S~\ref{sec:dis} for more discussion.

\section{Conditional Colour-Magnitude
Distribution (CCMD) Parametrization \&
Global Modelling of Colour-Luminosity 
Dependent Galaxy Clustering}
\label{sec:para}

Within the HOD/CLF framework, modelling the luminosity-dependent galaxy 
clustering has become a routine procedure. The halo occupation function
for a luminosity-bin or luminosity-threshold galaxy sample can be obtained
through an integral over the CLF \citep[e.g.][]{Yang03}. With the HOD
approach, the halo occupation function for a luminosity-threshold sample
is well parametrized \citep[e.g.][]{Kravtsov04,Zheng05,Zheng07b}, and that
for a luminosity-bin sample can be obtained by differencing those of two
threshold samples \citep[e.g.][]{Zehavi05,Zehavi11}. For modelling the 
colour-dependent galaxy clustering, most previous work adopts simple 
parametrizations on the red/blue fraction \citep[e.g.][]{Zehavi05}, on
relating colours to satellite fraction \citep[e.g.][]{Zehavi11}, or on
colour distribution \citep[e.g.][]{Skibba09b}. 

To model the luminosity-colour dependent clustering of galaxies in fine bins
of luminosity and colour over the whole galaxy CMD, parametrizing the HOD of 
each individual sample of galaxies would be ineffective. First, the total number
of parameters in the model would be large. For example, even if we apply 5
parameters for each sample, we would end up with a total of nearly 400 parameters
for the $\sim 79$ samples. Second, it is difficult to make the parametrizations and
the modelling results of individual samples fully compatible with each other. 
For example, at a fixed halo mass, over a sufficiently large luminosity range, 
the mean occupation numbers of central galaxies from different samples
should have the constraint of adding up to unity, which may not be the case
if samples are modelled individually.

A better strategy to model the luminosity-colour dependent clustering is to 
develop a global parametrization of the galaxy-halo relation. Here `global' 
means that we describe the overall galaxy distribution as a function of galaxy
luminosity/colour and halo mass, not those of individual samples. Then the 
halo occupation for each individual sample can be derived based on the luminosity
and colour cuts of the sample. The global paramerization can avoid the two above
problems (see more details in the following subsections). For such a purpose, 
we introduce the formalism of the conditional colour-magnitude distribution
(CCMD), which parametrizes the colour and luminosity distribution of galaxies
as a function of halo mass. It can be regarded as a substantial extension to the CLF 
formalism, which parameterizes the galaxy luminosity distribution as a function of 
halo mass. The CCMD adds the dimension of galaxy colour and describes 
its distribution. As with 
previous work \citep[e.g.][]{Kravtsov04,Zheng05} the CCMD separates 
contributions from central and satellite galaxies. Motivated by the bimodal 
colour distribution of galaxies, the CCMD also divides galaxies into red-like and
blue-like populations (dubbed as pseudo-red and pseudo-blue, respectively; see
below). We describe the CCMD parameterization of the four populations, 
pseudo-red and pseudo-blue central galaxies (\S~\ref{subsec:cenpara}) and
pseudo-red and pseudo-blue satellite galaxies (\S~\ref{subsec:satpara}).

With the CCMD parameterization, the halo occupation function can be obtained by 
integration for any galaxy sample given its colour and luminosity cuts. We 
then employ an accurate and fast method based on a $N$-body simulation to compute
the model 2PCFs (\S~\ref{subsec:simubased}) and use the Markov Chain Monte Carlo
(MCMC) to explore the CCMD parameter space from jointly modelling the clustering
measurements of the 79 SDSS galaxy samples. 

\subsection{CCMD of Central Galaxies}
\label{subsec:cenpara}

In general, the CCMD of galaxies is made of two parts, the description of
colour and magnitude distribution of galaxies at a fixed halo mass and 
then the parameterization of the key features of that distribution 
as a function of halo mass. 

For central galaxies, at fixed halo mass, the CLF is usually modelled as a 
log-normal function in luminosity or a Gaussian function in magnitude
\citep[e.g.][]{Yang03}, which is supported by the results from the SDSS 
group catalogue \citep{Yang08}. At fixed luminosity, the bimodal colour 
distribution of galaxies can be well described by a superposition of 
two Gaussian components \citep[e.g.][]{Baldry04}. 
The two components can be broadly referred to as `red' and
`blue' galaxies. However, the two components can substantially overlap with
each other. To avoid any possible confusion, hereafter we will call them 
pseudo-red and pseudo-blue components, and the usual red and blue samples can 
be defined by choosing a cut in colour. 

The above descriptions of luminosity and colour distribution motivate us to 
model the colour-magnitude distribution of central galaxies of each 
pseudo-colour component at a fixed halo mass as a 2D Gaussian distribution. 
We need 5 numbers to describe a two-dimensional (2D) normalised Gaussian 
function: the centre 
(2 numbers), the width (2 numbers, each for one direction), and 
the correlation (i.e. to describe the orientation of the contours 
corresponding to the 2D Gaussian distribution). There are then 10 numbers
from the two 2D Gaussian functions for the pseudo-red and pseudo-blue 
central galaxies. As we require the sum of the number of central galaxies to be unity 
at fixed halo mass, we also need a number to describe the the relative 
fraction of pseudo-red and pseudo-blue central galaxies. We parameterize it
as the fraction $\fred$ of pseudo-red central galaxies, and the fraction
of pseudo-blue central galaxies is simply $\fblue=1-\fred$. In total, we have
11 numbers to describe the colour-magnitude distribution of central galaxies
at a fixed halo mass.

For the convenience of presentation, we refer to the magnitude direction as 
the $x$ direction and the colour direction as the $y$ direction.
The differential colour-magnitude distribution of each pseudo-colour central
galaxy component can be written as
\begin{equation}
\frac{{\rm d}^2 \langle N_{\textrm{cen, comp}} \rangle}{\rm dx\, dy} =
f_{\rm comp}\frac{1}{2\pi\sigma_{\rm x}\sigma_{\rm y}
\sqrt{1-\rho^2}}\exp\left[-\frac{Z^2}{2(1-\rho^2)}\right],
\end{equation}
with
\begin{equation}
Z^2 =  \frac{(x-\mu_{\rm x})^2}{\sigma^2_{\rm x}}+\frac{(y-\mu_{\rm y})^2}
 {\sigma^2_{\rm y}} - \frac{2\rho(x-\mu_{\rm x})(y-\mu_{\rm y})}
 {\sigma_{\rm x}\sigma_{\rm y}}
\end{equation}
and
\begin{equation}
\rho=\frac{Cov(x,y)}{\sigma_{\rm x}\sigma_{\rm y}}.
\end{equation}
Here `comp' is `p-red' (`p-blue') for the pseudo-red (pseudo-blue)
population, $\mu_{\rm x}$ is the mean magnitude,
$\mu_{\rm y}$ is the mean colour, $\sigma_{\rm x}$ and $\sigma_{\rm y}$ are
the standard deviations in magnitude and colour, and $\rho$ ($Cov(x,y)$) is 
the coefficient of the correlation (covariance) between magnitude and colour. 

In the CCMD framework, the above $11$ numbers for the colour-magnitude
distribution of central galaxies are all functions of halo mass. We 
parameterize the halo mass dependence for each of them with motivations based
on previous work. For each pseudo-colour central galaxy population, we first
parameterize the halo-mass dependent mean and scatter in magnitude. From 
modelling the luminosity-dependent galaxy clustering with SDSS DR7, \citet{Zehavi11} 
obtains an empirical relation between median luminosity
$L_{\textrm{cen}}$ of central galaxies and halo mass $\Mh$ (see Eq.11 in their 
paper):
\begin{equation}
L_{\textrm{cen}}  = L_{\rm t} \left( \frac{\Mh}{M_{\rm t}} \right)
^{\alpha_{\rm M}} \exp\left(-\frac{M_{\rm t}}{\Mh}+1\right),
\end{equation}
which describes the relation as a power law with an exponential cutoff towards 
the low halo mass end and is characterised by the high-mass end power-law
index $\alpha_{\rm M}$, the transition (cutoff) mass scale $M_{\rm t}$, and 
the median luminosity $L_{\rm t}$ at the cutoff mass scale.
We adopt this functional form to parameterize the relation between 
the mean magnitude of central galaxies and halo mass for each pseudo-colour central
galaxy population. Rewriting the equation in magnitude, we have for each central
population
\begin{equation}
\label{eqn:mux}
\mu_{\rm x} = x_{\rm t}-2.5\alpha_{\rm M}\log
\left( \frac{\Mh}{M_{\rm t}} \right) - 1.086
\left(-\frac{M_{\rm t}}{\Mh}+1\right),
\end{equation}
with $\mu_{\rm x}$ the mean absolute magnitude of the central galaxy population 
in haloes of mass $M_{\rm h}$ and $x_{\rm t}$ that in haloes of transition mass 
$M_{\rm t}$.

For the scatter in the magnitude of central galaxies of each population, 
we parameterize it to follow a linear relation with halo mass, motivated
by \citet{Yang08} (see their fig.4),
\begin{equation}
\label{eqn:sigx}
\sigma_{\rm x} = a_{\rm 0,\sigma_x} 
+ a_{\rm 1,\sigma_x} (\log \Mh - \log M_{\rm nl}),
\end{equation}
with two free parameters, $a_{\rm 0,\sigma_x}$ and $a_{\rm 1,\sigma_x}$. 
The quantity $M_{\rm nl}$ is the $z=0$ nonlinear mass for collapse, and for
our adopted cosmology $\log [M_{\rm nl}/(\hinvMsun)]=12.35$. We note that the
coefficient parameters use subscripts to denote the quantity to be described,
and such a notation will be adopted in the following cases.

For the mean colour $\mu_{\rm y}$ and colour scatter $\sigma_{\rm y}$ of 
central galaxies, \citet{vandenBosch08} show that both of them vary with 
galaxy magnitude monotonically. \citet{Baldry04} introduce a hyperbolic 
tangent function to accurately describe the colour distribution as a function 
of magnitude in the galaxy CMD. We therefore parameterize both $\mu_{\rm y}$ 
and $\sigma_{\rm y}$ as a function of the mean magnitude $\mu_{\rm x}$, which 
is then linked to halo mass. With the insights from \citet{Baldry04} and 
\cite{vandenBosch08}, we express each as a combination of a linear function
and a hyperbolic tangent function,
\begin{equation}
\label{eqn:muy}
\mu_{\rm y} = a_{\rm 0,y} + a_{\rm 1,y}(\mu_{\rm x}-\mu_{\ast})
+q_{\rm 0,y}\textrm{tanh}\left(\frac{\mu_{\rm x}-q_{\rm 1,y}}
{q_{\rm 2,y}}\right)
\end{equation}
and
\begin{equation}
\label{eqn:sigy}
\sigma_{\rm y} = 
a_{\rm 0,\sigma_y} + a_{\rm 1,\sigma_y}(\mu_{\rm x}-\mu_\ast) 
+q_{\rm 0,\sigma_y} \rm{tanh}\left(\frac{\mu_x-q_{\rm 1,\sigma_y}}{q_{\rm 2,\sigma_y}}\right),
\end{equation}
where $a$'s and $q$'s are free parameters and $\mu_*=-20.44$ corresponds to 
the $r$-band characteristic luminosity in the local galaxy LF
\citep{Blanton03b}.

The correlation $\rho$ between central galaxy luminosity and colour at
fixed halo mass is not well constrained in the literature. We propose a 
linear relation to describe its halo mass dependence,
\begin{equation}
\label{eqn:rho}
\rho = a_{\textrm{0,$\rho$}} + a_\textrm{1,$\rho$}(\log M_h - \log M_{\rm nl}), 
\end{equation}
with $a_\textrm{1,$\rho$}$ and $a_{\textrm{0,$\rho$}}$ as the two 
free parameters.
 
For the fraction of pseudo-red central galaxies, the galaxy CMD suggests that 
it increases with increasing galaxy luminosity and hence halo mass. We 
parameterize it with a ramp-like function,
\begin{equation}
\label{eqn:fred}
f_{\textrm{p-red}} = \frac{1}{2}a_{\textrm{0,f}}\left[ 1 +
\textrm{erf}\left(\frac{\log 
M_h-a_{\textrm{2,f}}}{a_\textrm{1,f}}\right)\right],
\end{equation}
where ${\rm erf}$ is the error function.

In summary, the colour-magnitude distribution of central galaxies is described
as a combination of contributions from two galaxy populations, the pseudo-red 
and pseudo-blue central galaxies. At fixed halo mass, the distribution of 
each population is modelled as a 2D Gaussian. The halo-mass
dependences of the quantities in each 2D Gaussian function are described
with 17 parameters: 3 for the mean magnitude ($x_{\rm t}$, $M_{\rm t}$, and
$\alpha_{\rm M}$; Eq.~\ref{eqn:mux}), 2 for the scatter in magnitude 
($a_{0,\sigma_{\rm x}}$ and $a_{1,\sigma_{\rm x}}$; Eq.~\ref{eqn:sigx}),
5 for the mean colour ($a_{\rm 0,y}$, $a_{\rm 1,y}$, $q_{\rm 0,y}$, 
$q_{\rm 1,y}$, and $q_{\rm 2,y}$; Eq.~\ref{eqn:muy}), 5 for the scatter in 
colour ($a_{\rm 0,\sigma_y}$, $a_{\rm 1,\sigma_y}$, $q_{\rm 0,\sigma_y}$, 
$q_{\rm 1,\sigma_y}$, and $q_{\rm 2,\sigma_y}$; Eq.~\ref{eqn:sigy}), and 2 
for the colour-magnitude correlation ($a_{0,\rho}$ and $a_{1,\rho}$;
Eq.~\ref{eqn:rho}). Therefore, the two 2D Gaussian functions for the two 
central galaxy populations have 34 parameters. There are also 3
parameters ($a_{\rm 0,f}$, $a_{\rm 1,f}$, and $a_{\rm 2,f}$; Eq.~\ref{eqn:fred})
in the fraction of the pseudo-red central galaxies, which specifies the 
relative normalisation of the two Gaussian functions. In total, we use 37
parameters to describe the CCMD of central galaxies. The halo-mass dependences
of the parameters describing central galaxies in the best-fit model are
illustrated in Fig.~\ref{fig:msmscen}.

\subsection{CCMD of Satellite Galaxies}
\label{subsec:satpara}

For satellite galaxies, we also divide their CCMD into contributions from two 
populations, namely the pseudo-blue and pseudo-red satellite galaxies. 

Using a group catalogue, \citet{Yang08} study the CLFs of blue and red 
satellite galaxies and find that each can be well described by a modified 
Schechter function
\begin{equation}
\frac{d \langle N_{\textrm{sat}} \rangle}{d \log L} =
\phi_{\rm s} \left( \frac{L}{L_{\rm s}^\ast} 
\right)^{\alpha_{\rm s} + 1} \exp \left[ -\left( \frac{L}{L_{\rm s}^\ast} \right)^2 \right], 
\end{equation}
where $\phi_{\rm s}$ is the normalisation, $\alpha_{\rm s}$ is the faint-end 
slope, and $L_{\rm s}^\ast$ is the characteristic 
luminosity where the CLF transits from a power-law form to an exponential form.

We adopt the above CLF form and add the colour dimension to construct 
the luminosity and colour distribution of satellite galaxies for each
pseudo-colour population. Motivated by the double-Gaussian fit to the
galaxy colour distribution at fixed luminosity, we assume that for each
population satellite colour follows a Gaussian distribution at fixed 
luminosity. In terms of absolute magnitude and colour, the satellite CCMD at 
fixed halo mass for a pseudo-colour population reads
\begin{multline}
\frac{{\rm d}^2 \langle N_{\textrm{sat,comp}} \rangle}{\rm dx\, dy} 
 =
 \frac{1 }{\sqrt{2\pi}\sigma_{\textrm{y,sat}}} \exp\left[- \frac{
 (y-\mu_{\textrm{y,sat}})^2}{2\sigma^2_{\textrm{y,sat}}} \right]  \\
 \times 0.4\phi_{\rm s}
10^{-0.4(\alpha_{\rm s}+1)(x-x_{\rm s}^\ast)}  
\exp \left[ -10^{-0.8(x-x_{\rm s}^\ast)}  \right], 
\end{multline}
where $x_{\rm s}$ is the absolute magnitude corresponding to $L_{\rm s}^\ast$,
the Gaussian function represents the colour distribution (see below), and
similar to the case with central galaxies `comp' here can be either `p-blue'
or `p-red'.

Each quantity on the right-hand side of the above equation varies with halo
mass, which needs to be parameterized. As suggested in \citet{Yang08} and
\citet{vandenBosch13}, a quadratic function of halo mass is a good description
for the halo-mass dependence of either the normalisation $\phi_{\rm s}$
and the faint-end slope $\alpha_{\rm s}$. Therefore we adopt the following
parameterizations,
\begin{multline}
\label{eqn:phis}
\log \phi_{\rm s} = a_{\textrm{0,$\phi$}} + a_{\textrm{1,$\phi$}} (\log \Mh-\log \Mnl) \\ 
+ a_{\textrm{2,$\phi$}} (\log \Mh -\log \Mnl)^2
\end{multline}
and
\begin{multline}
\label{eqn:alphas}
\alpha_{\rm s} = a_{\textrm{0,$\alpha$}} + a_{\textrm{1,$\alpha$}}(\log  \Mh - \log \Mnl)\\
 + a_{\textrm{2,$\alpha$}} (\log  \Mh - \log \Mnl)^2,
\end{multline}
where the $a$'s are free parameters.

\cite{Yang08} suggest that there is a luminosity gap between the median 
luminosity of central galaxy $L_\textrm{cen}$ and the characteristic 
luminosity $L_{\rm s}^\ast$ of satellite galaxies independent of halo mass. 
In terms of absolute magnitude, we have
\begin{equation}
\label{eqn:xs}
x_{\rm s}^\ast = \mu_x + \Delta_{\rm sc},
\end{equation}
where $\Delta_{\rm sc}$ (independent of halo mass) is the luminosity gap in 
absolute magnitude. This $\Delta_{\rm sc}$ parameter and the halo-mass dependent
$\mu_{\rm x}$ (Eq.~\ref{eqn:mux}) give our parameterization of the 
dependence of $x_{\rm s}^\ast$ on halo mass.

The mean and standard deviation of the colour of satellite galaxies is found to
follow a linear relation with stellar mass \citep{vandenBosch08}. Motivated by 
this result, we parameterize the mean and standard deviation of the satellite 
colour through luminosity,
\begin{equation}
\label{eqn:muys}
\mu_{\textrm{y,sat}} = a_{\textrm{0,y,sat}} + a_{\textrm{1,y,sat}}(x-\mu_{\ast})
\end{equation}
and
\begin{equation}
\label{eqn:sigys}
\sigma_{\textrm{y,sat}} = a_{\textrm{0, $\sigma_{\textrm{y,sat}}$}} + a_{\textrm{1, $\sigma_{\textrm{y,sat}}$}}(x-\mu_{\ast}),
\end{equation}
where the $a$'s are parameters to link the satellite colour distribution with
the magnitude and $\mu_\ast=-20.44$ corresponds to the characteristic luminosity
in the $r$-band LF of local galaxies \citep{Blanton03b}.

In summary, in the parameterization of the satellite CCMD, we introduce 
parameters related to the CLF and satellite colour distribution. For each
pseudo-colour satellite population, there are 7 parameters to describe the
CLF as a function of halo mass: 3 for $\phi_{\rm s}$ ($a_{0,\phi}$,
$a_{1,\phi}$, and $a_{2,\phi}$; Eq.~\ref{eqn:phis}), 3 for $\alpha_{\rm s}$
($a_{0,\alpha}$, $a_{1,\alpha}$, and $a_{2,\alpha}$; Eq.~\ref{eqn:alphas},
and 1 for $x_{\rm s}$ ($\Delta_{\rm sc}$; Eq.~\ref{eqn:xs}). There are 4 
parameters to characterise the satellite colour distribution: 2 for the
mean $\mu_{\rm y,sat}$ ($a_{\rm 0,y,sat}$ and $a_{\rm 1,y,sat}$;
Eq.~\ref{eqn:muys}) and 2 for the scatter $\sigma_{\rm y,sat}$ 
($a_{\rm 0,\sigma_{y,sat}}$ and $a_{\rm 1,\sigma_{y,sat}}$; Eq.~\ref{eqn:sigys}).
Overall, we end up with 22 free parameters to describe the CCMD of 
the satellite population (pseudo-blue plus pseudo-red). The halo-mass dependences
of the parameters describing satellite galaxies in the best-fit model are
illustrated in Fig.~\ref{fig:msmssat}. 

Together with the 37 parameters introduced in the CCMD of central galaxies,
we have a total of 59 parameters to fully connect the colour-magnitude 
distribution of galaxies with halo mass. At first glimpse, the total number 
of parameters is large. However, we note that the parameterization is intended 
to cover the whole range of the galaxy CMD in Fig.~\ref{fig:binning}, not for 
a single galaxy sample. With this global parameterization, the halo occupation
function for each of the 79 galaxy samples we construct can be obtained by 
integrating the CCMD over the range of the colour and magnitude cuts of the
sample. Instead of the global parameterization, if the model were built by
parameterizing each individual sample with, say, 5 parameters, we would 
end up with $\sim 400$ parameters in total. Therefore the number of parameters
in our global parameterization in fact is small, about 59/79$\sim$0.7 parameter 
per sample. Furthermore, with the global parameterization, no inconsistency 
would arise among the halo occupations of different samples, which avoids 
a problem seen in the results from individual parameterizations of different
samples \citep[e.g.][]{Zheng07b}. We will use a total of
948 data points to constrain these 59 parameters.

\subsection{Calculation of Projected 2PCFs and 
CCMD Modelling of Galaxy Clustering}
\label{subsec:simubased}

With the global parameterization of the CCMD of galaxies, for a given galaxy
sample, the mean occupation function (separated into central and satellite
contributions) can be obtained by integrating the CCMD within the boundary 
set by the sample's colour and magnitude cuts. We then adopt the accurate and
efficient method laid out in \citet{Zheng16} to compute the projected 2PCF
for each sample.

The basic idea of the method is to tabulate all the necessary information 
for computing the various 2PCF components, using dark matter haloes identified 
in an $N$-body simulation and within fine bins of halo mass. The galaxy 2PCF 
is then obtained by a simple summation over halo mass bins with  
the halo occupation of galaxies properly accounted for. 
The method is equivalent to populating 
galaxies into haloes to form a mock galaxy catalogue and using the 2PCF 
measurements from the mock as the model predictions, which ensures the accuracy 
of the calculation. The calculation is also fast, enabling an efficient
exploration of the HOD or CCMD parameter space.

In detail, we use haloes identified in the MultiDark simulation 
with Planck cosmology. We assume that central galaxies reside at the 
potential minimum of haloes and satellite galaxies follow the spatial 
distribution of dark matter particles inside haloes. 
Rather than determining the positions of satellites in a halo 
based on a radial profile of analytic form, we assign the positions of randomly chosen dark matter particles to satellites.
Then with the simulation, the following 
quantities are measured in fine bins of halo mass: 
$\ni$, the number density of haloes in the $i$-th mass bin (mass $M_i$);
$f_{\rm cs}(\boldsymbol{r};M_i)$, the normalised central-satellite pair 
distribution profile in haloes of mass $M_i$, calculated using the positions 
of the potential 
minimum (position to put the central galaxy) and dark matter particles 
(positions for satellite galaxies) in each halo; 
$f_{\rm ss}(\boldsymbol{r};M_i)$, the normalised satellite-satellite pair 
distribution profile in haloes of mass $M_i$, calculated using the 
positions of dark matter particles in each halo;
$\xi_{\rm hh,cc}(\boldsymbol{r};M_i,M_j)$, the two-point cross-correlation 
function between haloes of masses $M_i$ and $M_j$, where halo positions 
come from the potential minimum (positions to put central galaxies);
$\xi_{\rm hh,cs}(\boldsymbol{r};M_i,M_j)$, the two-point cross-correlation 
function between haloes of masses $M_i$ and $M_j$, where for each halo pair 
the position of the potential minimum (position to put a central galaxy) of 
one halo and the position of a random dark matter particle 
(position to put a satellite galaxy) in the other halo are used;
$\xi_{\rm hh,ss}(\boldsymbol{r};M_i,M_j)$, the two-point cross-correlation function
between haloes of masses $M_i$ and $M_j$, where the positions of dark matter 
particles (positions to put satellite galaxies) are used.

The above quantities are all calculated in redshift space (i.e. with 
$\boldsymbol{r}$ the redshift-space pair separation vector) with the same
binning scheme as used in our measurements of the SDSS galaxy clustering. Then 
we integrate each component along the line of sight to obtain the projected
counterpart for the projected 2PCF. That is,
\begin{equation}
w_{\rm f,cs}(\rp; M_i)
=2\int_0^{\pi_{\rm max}} {\rm d}\pi\, f_{\rm cs}(\rp, \pi; M_i),
\end{equation}
\begin{equation}
w_{\rm f,ss}(\rp; M_i)
=2\int_0^{\pi_{\rm max}} {\rm d}\pi\, f_{\rm ss}(\rp, \pi; M_i),
\end{equation}
\begin{equation}
w_{\rm hh,cc}(\rp; M_i, M_j)
=2\int_0^{\pi_{\rm max}} {\rm d}\pi\, \xi_{\rm hh, cc}(\rp, \pi; M_i, M_j),
\end{equation}
and similarly for $w_{\rm hh,cs}(\rp; M_i, M_j)$ and 
$w_{\rm hh,ss}(\rp; M_i, M_j)$. Those $w$'s and $\ni$ are tabulated.
The projected 2PCF $\wpp(\rp)$ for a given 
galaxy sample is the sum of the one-halo and two-halo terms (contributed
by intra-halo and inter-halo galaxy pairs, respectively), 
$\wpp(\rp) = \wpp^{\rm 1h}(\rp) + \wpp^{\rm 2h}(\rp)$. The two terms are
calculated as \citep{Zheng16}
\begin{multline}
\wpp^{\rm 1h}(\rp) = 
\sum_{i}2\frac{\ni}
{\ng^2}\langle N_{\rm cen}(M_i) N_{\rm sat}(M_i) \rangle 
w_{\rm f,cs}(\rp;M_i) \\ 
+ \sum_{i}\frac{\ni}
{\ng^2}\langle N_{\rm sat} (M_i) [N_{\rm sat}(M_i)-1] \rangle 
w_{\rm f,ss}(\rp;M_i)
\label{eqn:wp1h}
\end{multline}
and
\begin{multline}
\wpp^{\rm 2h}(\rp)  =
      \sum_{i, j}  \nijg \meanNceni\meanNcenj \whhcc(\rp ; \Mi,\Mj)  \\
 +  \sum_{i,j} 2\nijg \meanNceni\meanNsatj \whhcs(\rp ; \Mi,\Mj) \\
 +  \sum_{i,j}  \nijg \meanNsati\meanNsatj \whhss(\rp ; \Mi,\Mj),
 \label{eqn:wp2h}
\end{multline}
where 
$\ng=\sum_{i} \ni [\langle N_{\rm cen} (M_i) \rangle 
+ \langle N_{\rm sat} (M_i) \rangle]$ is the number density of the galaxy 
sample. The mean occupation functions of central and satellite galaxies, 
$\langle N_{\rm cen} (M) \rangle$ and $\langle N_{\rm sat} (M)\rangle$,
are computed through integrating the CCMD with the integral limits set by
the colour and magnitude cuts of the given sample. For the mean number
of one-halo central-satellite pairs in haloes of mass $M_i$ 
(Eq.~\ref{eqn:wp1h}), we assume that there is no correlation between central 
and satellite galaxies, which leads to 
$\langle N_{\rm cen} (M_i)N_{\rm sat}(M_i) \rangle 
= \langle N_{\rm cen} (M_i)\rangle \langle N_{\rm sat}(M_i) \rangle $.
For the mean number of one-halo satellite-satellite pairs (Eq.~\ref{eqn:wp1h}), 
we assume that the number of satellites follows a Poisson distribution, 
meaning that $\langle N_{\rm sat} (M_i) [N_{\rm sat}(M_i)-1] \rangle 
= \langle N_{\rm sat} (M_i) \rangle^2$.

As the model follows the same binning and integration procedure as in 
the observation, it is immune to any effect caused by binning and any 
residual redshift-space distortion is automatically accounted for. That
is, the computed projected 2PCF for a given galaxy sample is equivalent 
to that measured in a mock galaxy catalogue from populating haloes with 
the corresponding HOD \citep{Zheng16}.

The tables are computed with Rockstar haloes \citep{Behroozi13} from the 
MultiDark MDPL2 simulation\footnote{\url{https://www.cosmosim.org/cms/simulations/mdpl2/}}. 
The MDPL2 simulation \citep{Klypin16} assumes a spatially flat cosmology 
consistent with the constraints from 
{\it Planck} \citep{PlanckCollaboration14,PlanckCollaboration16}, 
with $\Omega_{\rm m}=0.307$, $\Omega_{\rm b}=0.048$, $h=0.678$, $n_{\rm s}=0.96$, 
and $\sigma_8=0.823$. The simulation has a box size of $1h^{-1}{\rm Gpc}$ (comoving) 
on a side with $3840^3$ particles. With the corresponding mass resolution 
of $1.51\times 10^9 \hinvMsun$, there are about 100 particles per halo at 
the low mass end relevant for our modelling and the simulation is well suited for 
modelling the clustering measurements of the galaxy samples we construct. 
We note that the halo mass we adopt in this work is the virial mass $M_{\rm vir}$. 
There is a small offset between $M_{\rm vir}$ and $M_{\rm 200b}$, 
where $M_{\rm 200b}$ is the halo mass with haloes defined as having mean 
density of 200 times that of the background universe. Based on the MDPL2
halo catalogue, we find that the mean offset can be well described as
$\log M_{\rm 200b} - \log M_{\rm vir} = 0.0052(\log M_{\rm vir}-13.5) + 0.0498$, 
where halo mass is in units of $h^{-1}M_\odot$. This offset will be
applied when we make comparisons with some results in previous work.

To constrain the CCMD, we model the clustering measurements and number 
densities of the 79 galaxy samples in fine bins of colour and magnitude. 
The $\chi^2$ is constructed as
\begin{multline}
\chi^2 = (\boldsymbol{w_{\rm p}} -
 \boldsymbol{w^\ast_{\rm p}})^{\rm T}\boldsymbol{\sf{C^{-1}_{w_{\rm p}}}}
 (\boldsymbol{w_{\rm p}} -
 \boldsymbol{w^\ast_{\rm p}}) \\
  + (\boldsymbol{n_{\rm g}} - 
\boldsymbol{n^\ast_{\rm g}})^{\rm T}\boldsymbol{\sf{C^{-1}_{n_{\rm g}}}}
(\boldsymbol{n_{\rm g}} - \boldsymbol{n^\ast_{\rm g}}),
\end{multline}
where $\boldsymbol{w_{\rm p}}$ ($\boldsymbol{w^\ast_{\rm p}}$) 
and $\boldsymbol{n_{\rm g}}$ ($\boldsymbol{n^\ast_{\rm g}}$)
are the 2PCF and number density model (data) vectors from all the samples,
and $\boldsymbol{\sf{C_{w_{\rm p}}}}$ and $\boldsymbol{\sf{C_{n_{\rm g}}}}$
are the corresponding covariance matrices. The covariances among samples 
within the same magnitude bin (hence the same survey volume) have been 
accounted for (see \S~\ref{subsec:clustering}).

There is noise in the covariance matrix for galaxy samples in each luminosity bin
as it is estimated with a limited 
number of jackknife sub-samples. To reduce the effect of noise, we apply the 
singular-value decomposition (SVD) technique to the normalised covariance 
matrix and remove eigenvalues smaller than $\sqrt{2/N}$ 
\citep[e.g.][]{Gaztanaga05, Sinha17}, where $N=144$ is the number of jackknife 
sub-samples. 
This SVD procedure effectively reduces the data points, and we
end up with 897 effective data points 
(818 $w_{\rm p}$ and 79 $n_{\rm g}$ points).
To account for the mean bias in inverting the matrix, we also scale
the inverse matrix by $(N-N_{\rm dom}-2)/(N-1)$, a correction proposed by 
\citet{Hartlap07}, where $N_{\rm dom}$ is the number of effective data points.

With the likelihood of the model given the data, which is proportional to
$\exp(-\chi^2/2)$, we employ the MCMC method to 
explore the CCMD parameter space. Flat priors in linear space are adopted 
for all the parameters. 
Given the large dimension of the parameter space
(59 parameters; see \S~\ref{sec:para}), we typically require the length of 
the chain to be about $10^8$ and the convergence is tested by comparing 
the results from different realisations of chains. We present additional
tests of the results in Appendix~\ref{sec:AppendixB}.

\section{CCMD Modelling Results}
\label{sec:res}

With the CCMD parameterization, we simultaneously model the clustering 
measurements and number densities of the 79 samples of galaxies in fine
bins of colour and magnitude. The minimum $\chi^2$ from the MCMC chain is
found to be $952$. Given the number of the degrees of freedom 
$n_{\rm d.o.f.}= 818+79-59=838$ (818 effective $\wpp$ points plus 79 $n_{\rm g}$
points and 59 CCMD parameters), the expected 1$\sigma$ range of the 
$\chi^2$ distribution is $838\pm 41$. The minimum $\chi^2$ we find is 
within the 2.8$\sigma$ range. See Appendix~\ref{sec:AppendixB} for 
more tests with parameter constraints.

In this section, we present the modelling results. We start with an 
assessment on how well the model works by comparing the observables inferred 
from the model and those from the observational data.
We then focus on the constraints on the CCMD parameters and inferred 
quantities and relations. Finally we compare different aspects of the 
galaxy-halo connection inferred from our model with previous work.
 
We note that by design the CCMD model implements no galaxy assembly bias
effect, as the distribution of galaxy luminosity and colour is assumed 
to only depend on halo mass, not other halo properties that are related to halo environment and assembly history. Therefore the modelling results and the subsequent galaxy-halo connection presented in this paper are valid 
under such an assumption. For future work, we can extend the framework 
to parameterize the galaxy-halo relation to depend on an environment or assembly sensitive halo property, which would incorporate a certain form of assembly bias. As we discuss in section~\ref{sec:dis}, our modelling results here, in combination with observational data, can help reveal and constrain galaxy assembly bias.

\subsection{Observation versus Model}
\label{subsec:obsvsmodel}

\begin{figure*}
\includegraphics[width = \textwidth]{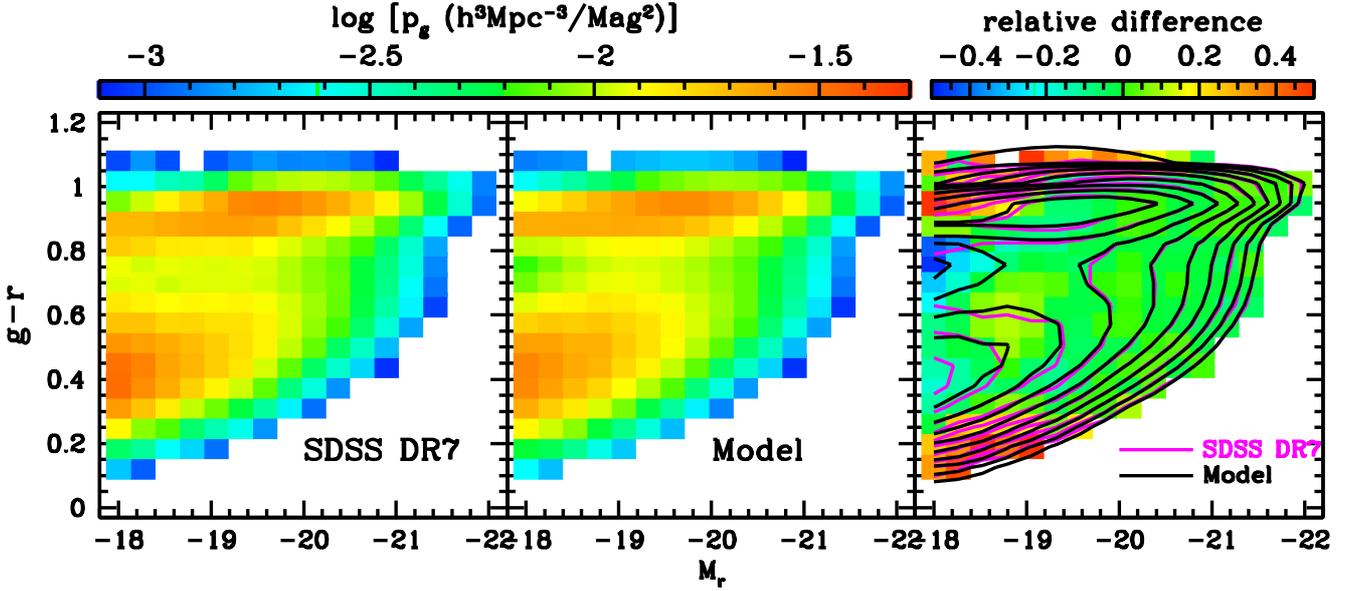}
\caption{
Comparison of galaxy abundances in the CMD between observation and best-fitting 
model prediction. The left panel shows the measured galaxy number density 
distribution calculated in fine bins of galaxy colour and magnitude, for bins 
with $p_{\rm g} \geq 10^{-3.0} h^3{\rm Mpc}^{-3}/{\rm mag}^2$, 
where $p_{\rm g} \equiv {\rm d}^2n_g/{\rm dM_r/d(g-r)}$ is the number density 
of galaxies per magnitude per colour.
The middle panel shows the corresponding distribution from the best-fitting 
CCMD model. In the right panel, the colour scale shows the fractional difference in 
galaxy number density distribution between model and observation, 
$(p_\textrm{g,model}-p_\textrm{g,data})/p_\textrm{g,data}$, while the magenta
and black contours are the observed and model distribution of galaxy abundance.
}
\label{fig:cmdcolourscale}
\end{figure*}

\begin{figure}
\includegraphics[width = \columnwidth]{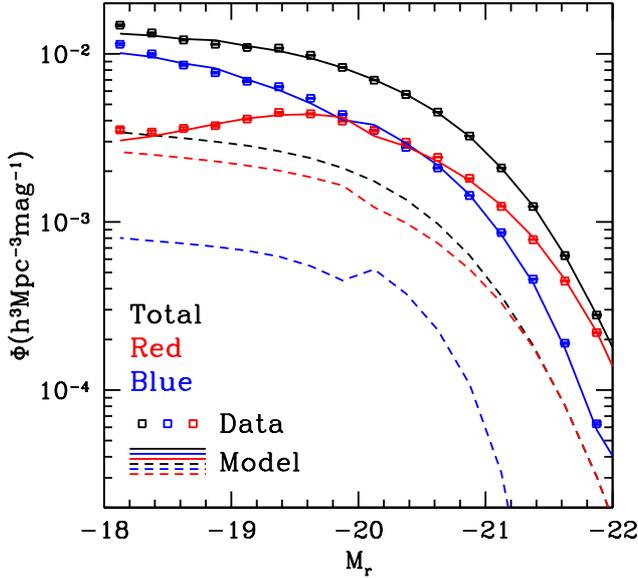}	
\caption{
Comparison of observed and model galaxy LFs. The black data
points show the observed LFs of all the galaxies, which is
further separated into that of blue galaxies (blue) and that of red galaxies 
(red).  The solid curves are the
corresponding LFs from the best-fitting CCMD model, and the dashed curves are
the predicted contributions from satellites.
A luminosity-dependent colour cut is adopted to divide galaxies into blue 
and red populations (\citealt{Zehavi11}; see text \S~\ref{subsec:obsvsmodel} for detail).
}
\label{fig:lf}
\end{figure}

\begin{figure*}
\includegraphics[width=\textwidth]{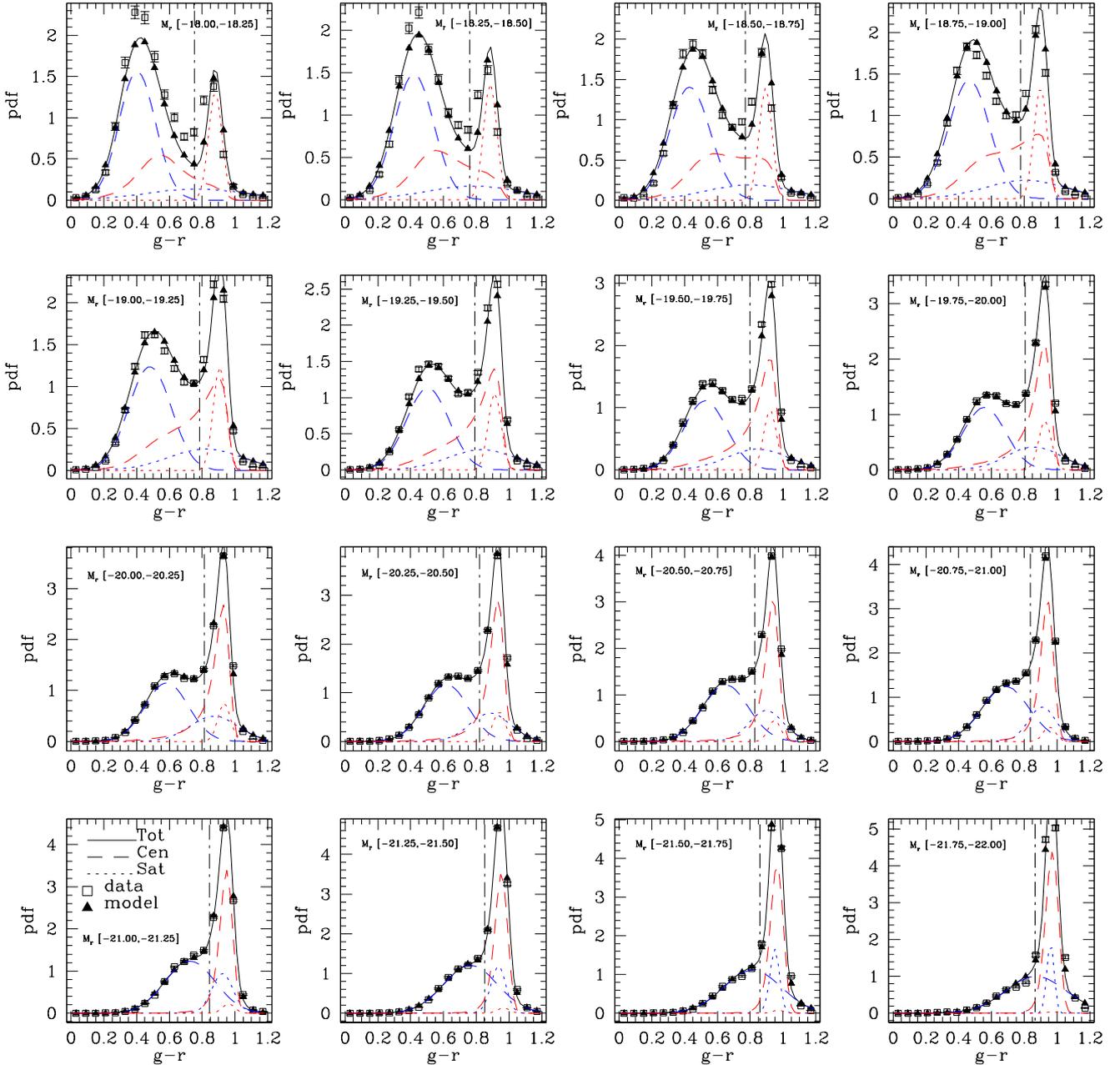}
\caption{
Comparison of luminosity-dependent galaxy colour distribution from observation
and model. In each panel, the squares are the observed probability distribution 
function (PDF) of galaxy colour for a given luminosity bin (indicated inside each panel).
The solid curve is the PDF from the best-fitting model, while the triangles
are the model PDF integrated over the same colour bins as used in the observation. 
The blue/red dashed  and dotted curves are the contributions from central 
and satellite galaxies of the pseudo-blue/pseudo-red populations in the model 
(see text for detail). In each panel, the vertical line indicates the 
luminosity-dependent colour cut \citep{Zehavi11} to divide galaxies into
blue and red populations in observation.
}
\label{fig:ngdm}
\end{figure*}

\begin{figure*}
\includegraphics[width = \textwidth]{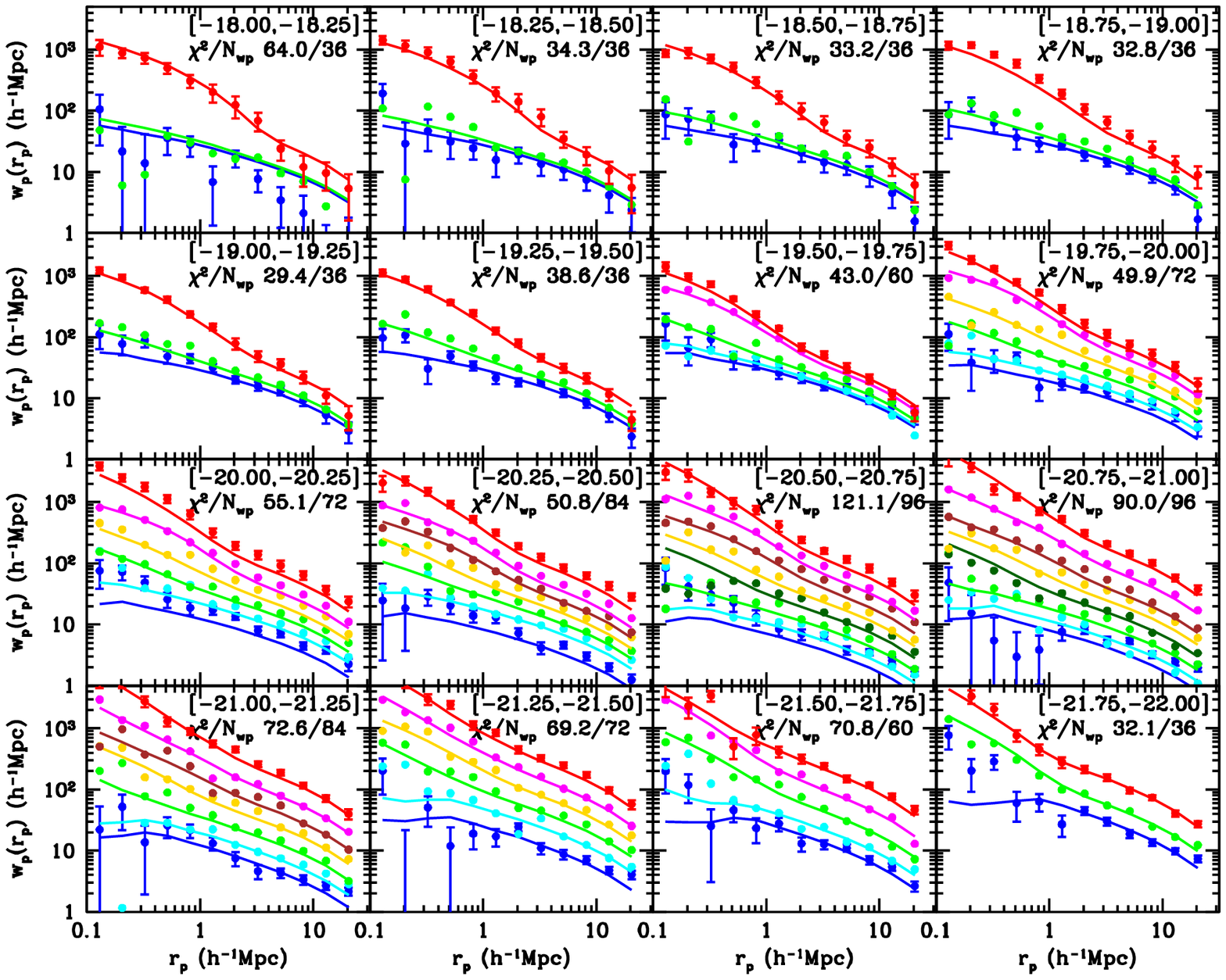}
\caption{
Comparison of the projected 2PCFs between observation and model.
Each panel shows the colour-dependent projected 2PCFs for a given luminosity
bin (marked in the top-right corner). The filled circles represent the 
measurements from the observational data and the solid curves stand for those
from the best-fitting CCMD model. For clarity, offsets are added to
both the data and model to separate galaxy sub-samples of different colours 
and error bars are only plotted
for the bluest and reddest samples.
The value of $\chi^2$ computed using the $\wpp$ data in the panel is labelled,
together with the number of $\wpp$ data points, which is meant to aid the 
comparison between data and model (see text \S~\ref{subsec:obsvsmodel} for detail).
}
\label{fig:wpcomparison}
\end{figure*}

As the model is based on fitting the number densities and 2PCFs of 
the 79 galaxy samples in fine bins of colour and magnitude across the 
galaxy CMD, we expect that the bimodal distribution in the CMD is well
captured by the model. In Fig.~\ref{fig:cmdcolourscale}, we compare 
the galaxy CMD from observation and that from the best-fitting model.
To resolve the key features in the CMD, we use much finer colour bins 
(0.06 mag) than the ones adopted in the modelling, which put the model 
under a more stringent test.  The model passes
such a check and reasonably reproduces the overall CMD. The
right panel shows the fractional differences between the model and data.
The model works well for luminous galaxies, especially those more luminous
than $L^\ast$. These galaxy samples have large survey volumes and thus small 
fractional uncertainties in their 2PCF measurements. Also for these samples, 
we are able to construct sub-samples with finer colour bins. Both factors 
contribute to the tight constraints on the CCMD model in the corresponding 
regions of the parameter space. For faint galaxies, 
especially those fainter than $\Mr\sim -19$, the difference between model and
data increases. Besides the small survey volume and relatively large 
uncertainties in the 2PCF and number density measurements, the coarse 
colour bins (e.g. 3 colour subsamples at fixed magnitude) also contribute
to the increased difference. Fitting the number densities of the 
coarse-colour sub-samples well does not guarantee that the model can do 
well in predicting those for the finer-colour-bin sub-subsamples. In our case,
the worst match is in the green valley region of the faint samples, with
the difference increasing to about 30 per cent.

The model reproduces the CMD reasonably well, implying that the
colour-dependent galaxy LF inferred from the the model
should agree with the data. Fig.~\ref{fig:lf} shows such a comparison between 
the model and observed galaxy LF, which is further 
decomposed into those from blue and red galaxies, respectively. A 
luminosity-dependent colour cut as in \citet{Zehavi11} is adopted for the 
division between red and blue galaxies. As expected, the CCMD modelling 
results nicely follow the data. Only at the very faint end ($\Mr \sim -18$)
and the very luminous end ($\Mr \lesssim -22$; not shown), the agreement degrades. 
The former reflects the loose constraints in the corresponding regions
of the CCMD parameter space caused by the small survey volumes and coarse
colour bins used for the modelling. The latter corresponds to samples 
outside of the luminosity range of modelled ones such that the model curves
there are actually from extrapolation. 
It is encouraging that the extrapolated results are still close to the observation.

The black dashed curve in Fig.~\ref{fig:lf} shows the satellite contribution to the LF predicted by the best-fitting CCMD, which is decomposed into those from red and blue galaxies (red and blue dashed curves, respectively) 
with the luminosity-dependent colour cut of \citet{Zehavi11}. The satellite LF is dominated by the red galaxies
over the whole luminosity range, with the fraction of red satellites increasing 
from $\sim$80 per cent at $\Mr \sim -18$ to 100 per cent at $\Mr\sim -22$. For red
galaxies, the majority of faint ones ($\sim$90 per cent) are satellites and the 
fraction of satellites decreases with increasing luminosity. For blue galaxies, 
they are dominated by central galaxies over the whole luminosity range. 

\begin{figure*}
\includegraphics[width = \textwidth]{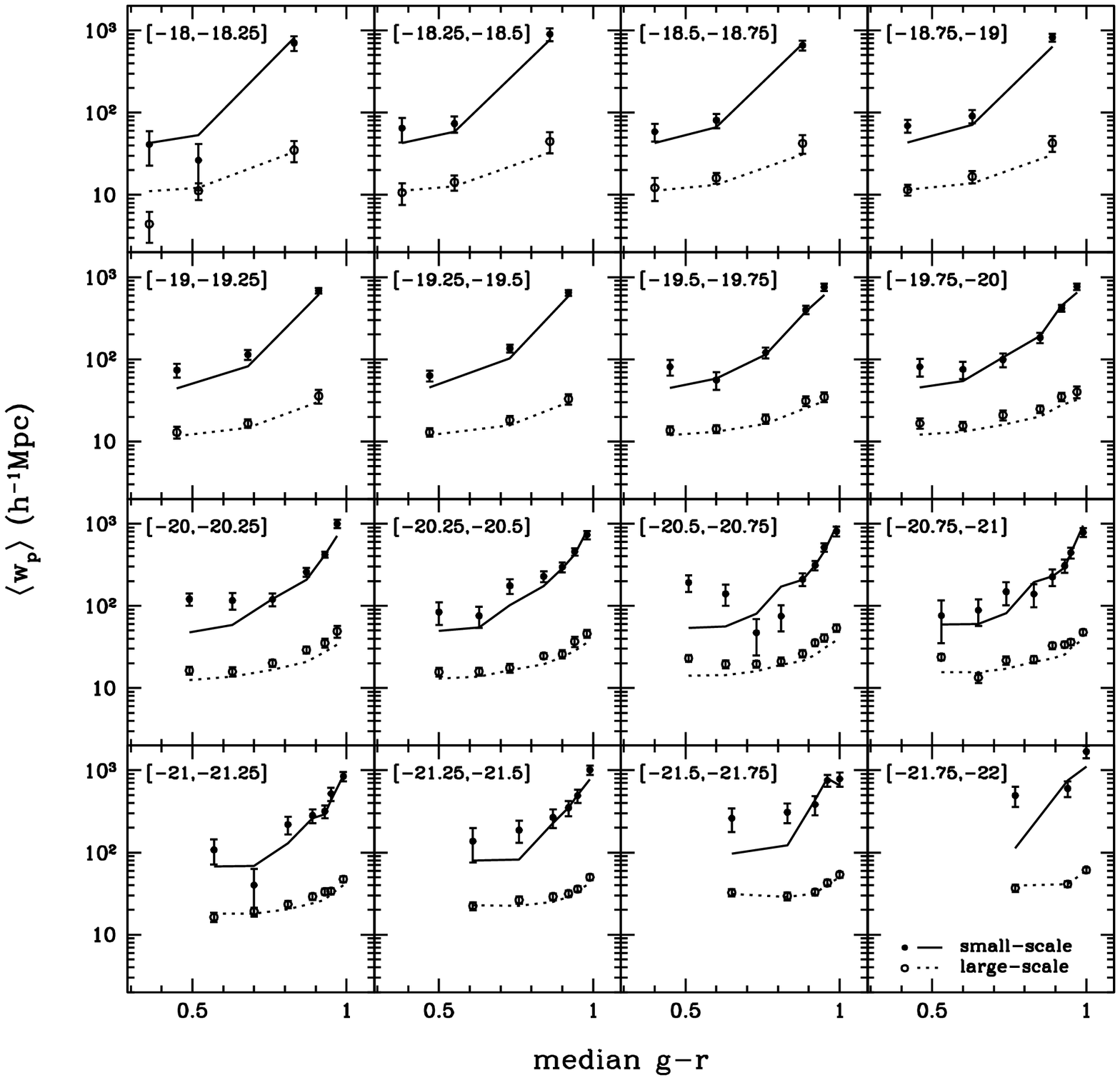}
\caption{
Similar to Fig.~\ref{fig:wpcomparison}, but on the comparison of the average projected
2PCFs between observation and model.
Each panel shows the average projected 2PCFs as a function of median colour of
the galaxy sub-samples for a given luminosity bin (marked in the top-left corner). 
The filled and open circles represent the average values of the observed $\wpp(\rp)$
over the $\rp$ ranges of 0.1--1$\hinvMpc$ (one-halo regime) and 2.51--10$\hinvMpc$
(two-halo regime), respectively. The solid and dotted curves correspond to those from the best-fitting CCMD model.
}
\label{fig:wplumincol}
\end{figure*}

\begin{figure}
\includegraphics[width = \columnwidth]{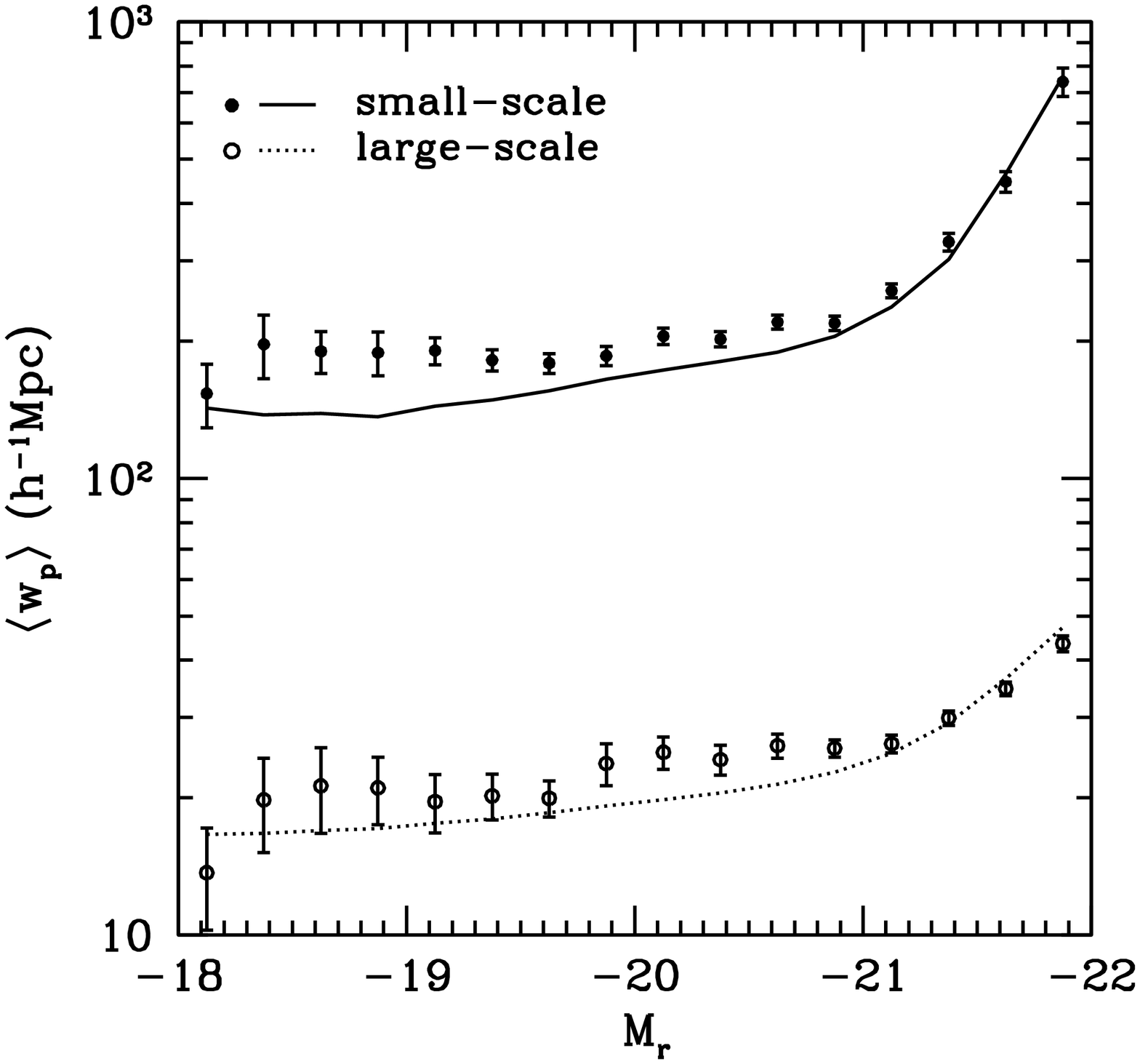}
\caption{
Similar to Fig.~\ref{fig:wplumincol}, but on the comparison of the average 
projected 2PCFs between observation and model for luminosity-bin samples.
The filled and open circles represent the average values of the observed $\wpp(\rp)$
over the $\rp$ ranges of 0.1--1$\hinvMpc$ (one-halo regime) and 2.51--10$\hinvMpc$
(two-halo regime), respectively. The solid and dotted curves correspond to those from the best-fitting CCMD model. See the text for the explanation of the apparent 
difference between the data and the model below $L^\ast$.
}
\label{fig:wplumin}
\end{figure}

We can also take a close look at the luminosity-dependent colour distribution of
galaxies, which is shown in Fig.~\ref{fig:ngdm}. The solid curve in each panel 
(representing a fixed magnitude bin) is computed from the best-fitting model 
by accounting for the contributions from haloes of all masses. 
The model also gives the contributions from central/satellite galaxies of 
different pseudo-colour populations 
(dashed and dotted curves; which will be discussed in \S~\ref{sec:ccmd_constraints}). 
The open squares are the data points from observation, while the solid triangle
points are from the model with the finite colour bin size effect
taken into account.
At low luminosity, the colour distribution profile appears to be double-peak, 
with a broad blue peak and a narrow red peak. For the two lowest luminosity bins,
the model tends to underestimate the fraction of galaxies near the blue peak and
overestimate that redder than the red peak. 
As the luminosity increases, the broad blue peak gradually merger with the red 
one. For luminous bins, the model reproduces the colour distribution remarkably
well, including the overall shape and the tail distribution. In each panel, the 
vertical line marks the luminosity-dependent colour cut adopted in \citet{Zehavi11} 
to divide galaxies into blue and red populations in observation.

With the colour and luminosity distributions well reproduced by 
the best-fitting model, we turn to galaxy clustering. In 
Fig.~\ref{fig:wpcomparison}, the comparison is done between 
the observed and model projected 2PCFs at each luminosity bin as a 
function of galaxy colour (different points and curves in each panel).
In each panel, 
the value of $\chi^2$ from comparing the model with the data
for all the samples in the corresponding magnitude bin is shown, together
with the number of data points (denoted as `${\rm N_{\rm wp}}$').
Although this does not represent the exact goodness of fit for $\wpp$ 
in each panel (as it neglects the number of parameters),
the values can give a sense of how well the model works in each luminosity bin.

As a whole, the model fits the 2PCFs well for galaxy samples covering
a range of $\sim$40 in luminosity (or 4 mag in magnitude) and across the 
full colour range (about 1.2 mag). Noticeably, the fit appears to be poor
for samples with the lowest luminosity ($-18.25<\Mr<-18.00$; the 
bottom-left panel), which has $\chi^2=64$ for 36 data points. A close 
inspection reveals that for the blue sample the model predicts an 
amplitude too high compared to the data points on scales above 
$\sim$3$\hinvMpc$. 
The model also predicts lower $w_{\rm p}$ at small scales 
for faint samples with $\Mr>-19.5$ (except for the faintest sample; also see Fig.~\ref{fig:wplumin}).
We note that the samples with the lowest luminosity 
suffer the most from the small survey volume and thus have the largest
sample variance \citep[e.g.][]{Zehavi05,Zehavi11,Xu16}. The other trend we notice 
is that for the bluest galaxy samples with high luminosity (more luminous 
than $\sim L^\ast$ or -20.44 mag) the model tends to under-predict the 
clustering amplitude on scales smaller than $\sim$0.5$\hinvMpc$. Although 
given the large measurement uncertainties the trend appears to be weak, 
it systematically shows up in almost all the high luminosity bins. The model
curve is flattened towards small scales, while the data seem to have an 
inflection and have a steep increase. This indicates that there may be
too few one-halo galaxy pairs in the model. Further investigations are 
needed to study the clustering of the luminous blue galaxies to determine whether
the trend is real and to discuss its implications.

Since offsets are added in the projected 2PCFs in Fig.~\ref{fig:wpcomparison}, the
luminosity and colour dependence in the 2PCF amplitude is not easily
revealed. To show 
the trend, we compute projected 2PCF amplitudes averaged over $\rp$ ranges of 
0.1--1$\hinvMpc$ and 2.51--10$\hinvMpc$, respectively, and plot their dependence 
on luminosity and (median) colour of galaxy samples in 
Fig.~\ref{fig:wplumincol}. In detail, we compute the arithmetic mean of $N$ values 
of $\wpp$,  $\langle \wpp \rangle = \sum_{i=1}^{N} w_{{\rm p},i}/N$, and derive the
error bars from the variance in the mean, $\sigma^2=\sum_{i,j=1}^N C_{ij}/N^2$, 
where $C_{ij}$ is the covariance matrix of $\wpp$. The small-scale and large-scale
averages represent the mean clustering amplitudes in the one-halo and two-halo regime.

At fixed colour, the points across different panels show the luminosity dependence, 
while in each panel the colour dependence at fixed luminosity is shown.
We see that the clustering amplitude at both small and large scales 
has a stronger dependence on colour than luminosity. 
At fixed luminosity, the small-scale clustering amplitude
shows a stronger dependence on colour than the large-scale one, indicating the
strong dependence of satellite fraction on galaxy colour (see \S~\ref{sec:derived}
for more discussions). Overall, the CCMD reproduces the luminosity and colour  
dependent clustering amplitude (note that the data points in each panel are 
correlated). As with Fig.~\ref{fig:wpcomparison}, the relatively
large deviations between data and model are seen in the one-halo amplitude for the
luminous blue galaxies (above $L^\ast$) and the two-halo amplitude for the faint blue
galaxies (in the $-18.25<\Mr<-18$ panel). 

The CCMD model also allows us to predict the luminosity-only dependence of the
2PCF. The 2PCF of the full sample of galaxies at a fixed luminosity encodes more
information than the individual auto-correlation functions of the colour sub-samples, 
as it has contributions from cross-correlations among the colour sub-samples.
Therefore, the predicted 2PCF of the full sample from the best-fitting model 
involves extrapolations. Similar to Fig.~\ref{fig:wplumincol}, we compute the
arithmetic mean $\wpp$ on small and large scales and compare the data and 
the CCMD model prediction in Fig.~\ref{fig:wplumin}. The model successfully 
captures the trend that the clustering amplitude increases slowly with luminosity 
below $L^\ast$ but steeply above it \citep[e.g.][]{Zehavi05a,Zehavi11}. Towards the
luminous end, a steeper change is seen in the one-halo regime than in the two-halo
regime. Around $L^\ast$ ($-21\lesssim \Mr \lesssim -20$), the model prediction lies 
below the data in both regimes with a wiggle feature shown in the data. This 
likely reflects the effect of the Sloan Great Wall on the measurements (see the test 
in \citealt{Zehavi11}). For luminosity fainter than $\Mr\sim -19.5$, the model curves
are also below the data (except for the faintest bin), which is systematically seen 
in previous HOD or CLF modelling results of luminosity-bin samples (e.g. left panel 
of fig.4 and fig.13 in \citealt{Zehavi11} and left panels of fig.2 in \citealt{Cacciato13}). This is consistent with the trend seen in Fig.~\ref{fig:wpcomparison}, in all colour sub-samples. Note that the fits
shown in Fig.~\ref{fig:wpcomparison} are good, once the covariance of the data points
is accounted for. The apparent difference between the data and the model can be a
manifestation of the sample variance effect on the measurements for faint galaxy 
samples with small survey volumes. It has been shown that sample variance has little
effect in the modelling results (see e.g. fig.15 of \citealt{Zehavi05}).

\subsection{CCMD Constraints}  
\label{sec:ccmd_constraints}

\begin{figure*}
\includegraphics[width=\textwidth]{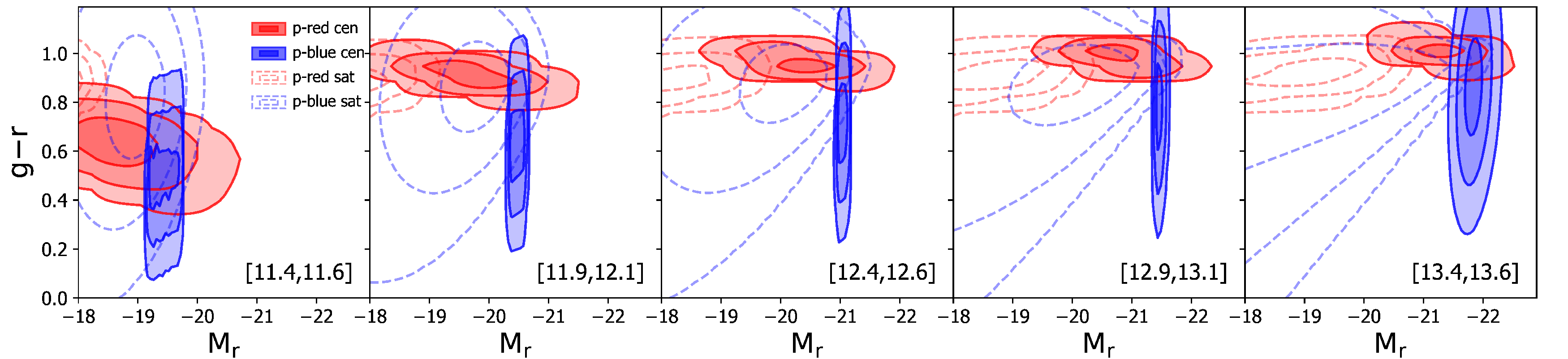}
\caption{
CCMD as a function of halo mass from the best-fitting model. 
Solid and dotted blue contours
are for the pseudo-blue central and satellite 
galaxy components. Solid and dotted red contours are the same, but for the
pseudo-red population. In each component, the contour levels are $\exp(-1/2)$,
$\exp(-4/2)$, and $\exp(-9/2)$ times the peak value for this component 
(corresponding to the inclusion of 39, 86, and 99 per cent of galaxies in the
component for a 2D Gaussian distribution). That is,
no overall normalisation is applied for the contour levels of different 
components (see text for detail).
The halo mass range in terms of 
$\log[\Mh/(\hinvMsun)]$ is labelled in each panel.
}
\label{fig:halomassctur}
\end{figure*}

\begin{figure*}
\includegraphics[width=\textwidth]{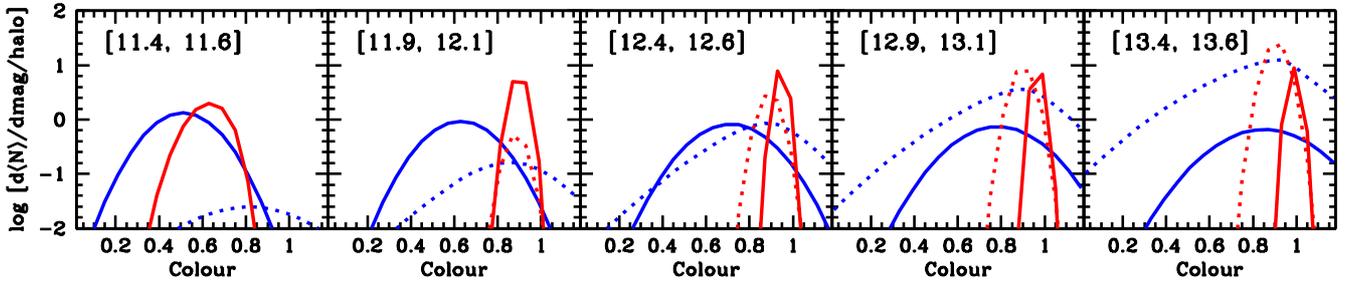}
\caption{
Conditional Colour Function (CCF) 
as a function of halo mass from the best-fitting model.
Solid and dotted blue curves are for the pseudo-blue central and satellite 
galaxy components. Solid and dotted red curves 
are the same, but for the pseudo-red population. The CCF of each component is 
computed for galaxies more luminous than $\Mr=-18$.
The halo mass range in terms of 
$\log[\Mh/(\hinvMsun)]$ is labelled in each panel.
}
\label{fig:ccfhalomass}
\end{figure*}

\begin{figure*}
\includegraphics[width=\textwidth]{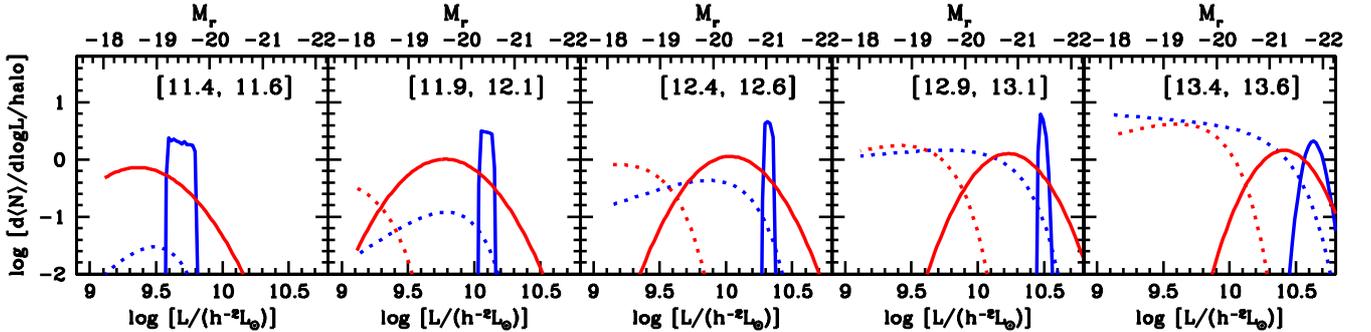}
\caption{
CLF as a function of halo mass from the best-fitting model.
Solid and dotted blue curves are for the pseudo-blue central and satellite 
galaxy components. Solid and dotted red curves 
are the same, but for the
pseudo-red population.
The halo mass range in terms of 
$\log[\Mh/(\hinvMsun)]$ is labelled in each panel.
}
\label{fig:clfhalomass}
\end{figure*}

\begin{figure}
\includegraphics[width =\columnwidth]{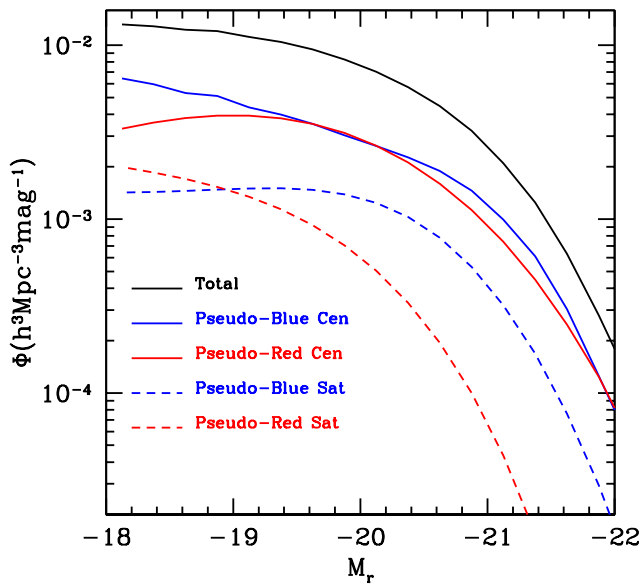}	
\caption{
Decomposition of galaxy LF into model components.
The LF from the best-fitting model is shown as the thick
black curve, which is decomposed into contributions from the pseudo-blue
central (thin solid blue) and satellite (thin dashed blue) components and
the pseudo-red central (thin solid red) and satellite (thin dashed red) 
components.
}
\label{fig:lf_model}
\end{figure}

\begin{figure*}
\includegraphics[width=\textwidth]{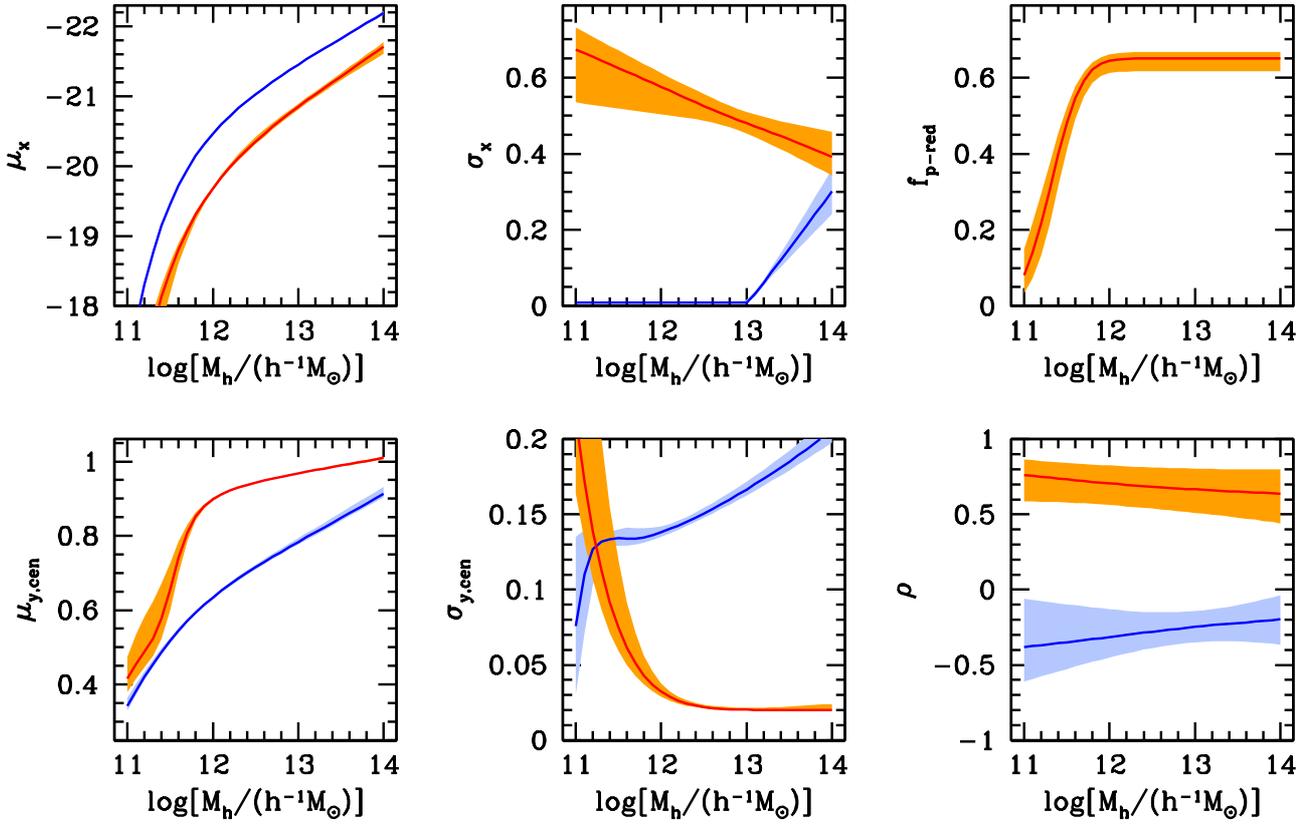}
\caption{
CCMD quantities as a function of halo mass for central galaxies.
From left to right and top to bottom, the panels show the halo mass dependence 
of the mean magnitude, the standard deviation of magnitude, the fraction 
of pseudo-red central galaxies, the mean colour, the standard deviation of colour,
and the correlation coefficient between colour and magnitude (see \S~\ref{subsec:cenpara}
for physical meanings or definitions of these parameters).
Blue and red curves are for pseudo-blue and pseudo-red central galaxies from the
best-fitting CCMD model, respectively. The shaded region around each curve 
represents the 1$\sigma$ range determined from the MCMC chain.
}
\label{fig:msmscen}
\end{figure*}

\begin{figure*}
\includegraphics[width=1.0\textwidth]{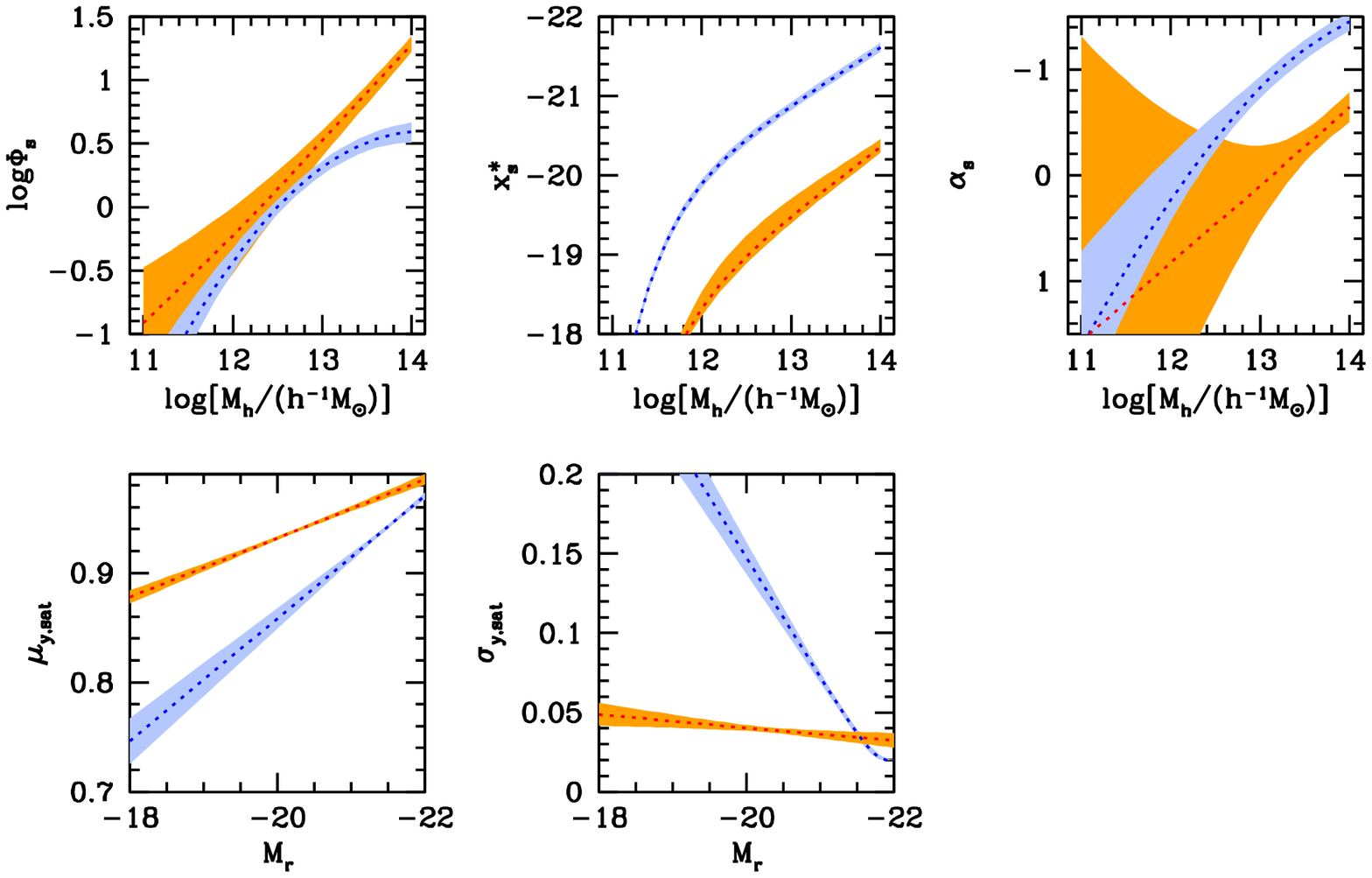}
\caption{
CCMD quantities for satellite galaxies.
From left to right, the top panels show the halo mass dependence 
of the normalisation, the characteristic magnitude, the faint-end slope of 
the satellite CLF, and the bottom panels show the luminosity-dependent mean 
colour and standard deviation of the colour for satellite galaxies 
(see \S~\ref{subsec:satpara}
for physical meanings or definitions of these parameters).
Light blue and orange curves are for pseudo-blue and pseudo-red satellite 
galaxies from the
best-fitting CCMD model, respectively. The shaded region around each curve 
represents the 1$\sigma$ range determined from the MCMC chain.
}
\label{fig:msmssat}
\end{figure*}

After discussing the reasonable fits to the projected 2PCFs and colour-magnitude
distributions of galaxies from the CCMD model, we turn to the CCMD model
itself and present the constraints on CCMD parameters and relations.

Fig.~\ref{fig:halomassctur} shows the colour-magnitude distributions of 
galaxies in haloes of different masses from the best-fitting model, which
demonstrates the CCMD model's spirit of de-projecting the global CMD along 
the halo mass axis. In each panel, the two sets of solid (dotted) contours are 
the distributions of the pseudo-blue and pseudo-red central (satellite) galaxies.
As mentioned in \S~\ref{sec:para}, the division of pseudo-blue and pseudo-red
galaxies are motivated by the bimodal colour distribution, which can be 
approximately described as the superposition of two Gaussian components at
fixed luminosity. 
Therefore the pseudo-blue and pseudo-red components do not need to have 
distinct colours and they can overlap in certain colour ranges.

For each pseudo-colour population, central galaxies (solid contours) appear to 
be more luminous than satellite galaxies (dotted contours), reflecting the 
luminosity gap phenomenon \citep[e.g.][]{Yang08}. 
In low mass haloes (e.g. with mass of a few times
$10^{11}\hinvMsun$; leftmost panel of Fig.~\ref{fig:halomassctur}), the 
pseudo-blue and pseudo-red central galaxies overlap largely in colour and to a 
lesser extent in luminosity. In high mass haloes (above $10^{12}\hinvMsun$), on
average the pseudo-blue central galaxies are more luminous (about 0.5 magnitude 
in median luminosity) than the pseudo-red central
galaxies, and they are bluer on average as expected (about 0.1--0.3 magnitude, 
decreasing slightly with increasing halo mass). 
While the pseudo-blue central galaxies have a smaller 
spread in luminosity in high mass haloes, the pseudo-red central galaxies have a
narrower colour distribution weakly dependent with halo mass above $10^{12}\hinvMsun$. 
In each central component, galaxy colour does not appear to have a 
strong correlation with luminosity
(see more discussions later in this section). 
Note that in Fig.~\ref{fig:halomassctur}, the contours are drawn for each
individual component, and the contour levels from different components are not
meant to be compared. Even though there is the pseudo-red (pseudo-blue) central
component in low (high) mass haloes, their total occupation fraction is small 
(see below).

To have a more detailed view on the colour distributions of pseudo populations
as a function of halo mass, we project the CCMD in Fig.~\ref{fig:halomassctur} 
onto the colour dimension to form the conditional colour distribution (CCF) in Fig.~\ref{fig:ccfhalomass} for galaxies more luminous than $\Mr=-18$. 
For the central components, 
both the pseudo-blue and pseudo-red populations have redder mean colour 
in more massive halo, and the dependence on halo mass becomes weak above
$10^{12}\hinvMsun$. The CCF for pseudo-red central galaxies is narrow with an 
almost constant width in haloes above $10^{12}\hinvMsun$, while that for the 
pseudo-blue central galaxies is broad with a increasing width with halo mass 
(also see the bottom middle panel in Fig.~\ref{fig:msmscen}). 
For the satellite components, like the central ones, the colour distribution of 
the pseudo-red component is much narrower than that of the pseudo-blue one. In 
haloes above $10^{12}\hinvMsun$, the colour distribution of each pseudo-colour 
component has little change with halo mass and the median colour or the colour
at the distribution peak is around $g-r\sim 0.9$.
Compared with the central galaxies, the median colour of pseudo-red satellites is 
redder in low-mass haloes (leftmost panel of Fig.~\ref{fig:halomassctur}, where 
the pseudo-red satellite population does not show up because of low normalisation)
and becomes slightly bluer in high-mass haloes. For the pseudo-blue population,
the median colour of satellites is much redder than that of central galaxies, and
the difference decreases with increasing halo mass.

In addition, the CCF plot (Fig.~\ref{fig:ccfhalomass}) allows a comparison of the 
normalisation amplitudes of the four pseudo-colour components. It can be clearly
seen that the number of satellites keeps increasing with increasing halo mass, from
negligibly small in low mass haloes (the leftmost panel) to outer-numbering the
central galaxies by more than an order of magnitude in high mass haloes (the
rightmost panel). In haloes above $10^{12}\hinvMsun$, the relative fraction of
pseudo-red and pseudo-blue central galaxies does not change much with halo mass
(also see the top right panel in Fig.~\ref{fig:msmscen}). The CCF also suggests the
constitution of the colour bimodality. The satellite colour distribution, consisting
of a narrow pseudo-red and a diffuse pseudo-blue contribution with similar median
colour, does not show a bimodal pattern. In haloes above $10^{12}\hinvMsun$, the 
median colours of pseudo-blue and pseudo-red central galaxies are well separated,
forming a bimodal distribution. The colour bimodality shown in observations (Fig.~\ref{fig:ngdm}) is caused by the superposition of different galaxy 
components in haloes of different masses. We emphasise that 
Fig.~\ref{fig:ccfhalomass}, along with Fig.~\ref{fig:ngdm} and
Fig.~\ref{fig:halomassctur}, demonstrates that the pseudo-colour components do 
not represent observational colour components and they can well overlap in colour. 

Marginalising the CCMD over colour reduces to the CLF. 
In Fig.~\ref{fig:clfhalomass}, the CLFs for five halo mass ranges are shown,
with the decomposition into the four
pseudo-colour components. As inferred from
Fig.~\ref{fig:halomassctur}, the median luminosity of the pseudo-blue central
galaxy population is higher than that of the pseudo-red one and its CLF has 
a more narrow distribution. 
Particularly in low mass haloes, the luminosity of the pseudo-blue
central galaxies show a tight correlation with halo mass, with the scatter 
reaching the floor value set by the model (top-middle panel in 
Fig.~\ref{fig:msmscen}), and the width here is mainly determined 
by the size of the halo mass bin. 
For haloes with the lowest mass in Fig.~\ref{fig:clfhalomass}, the luminosity
distribution is wider, because of the steep slope in the luminosity-halo 
mass relation (top-left panel in Fig.~\ref{fig:msmscen}) around that mass range.
For haloes with the highest mass in Fig.~\ref{fig:clfhalomass},  
the luminosity distribution is also wider, but this mainly reflects the large
scatter in luminosity at fixed halo mass (top-middle panel in Fig.~\ref{fig:msmscen}).
The amplitude of satellite CLF for either the pseudo-blue or pseudo-red
component increases with increasing halo mass, and that for the 
pseudo-blue satellites shifts towards the more luminous end. The faint-end 
of the CLF for the pseudo-blue component becomes steeper at higher halo 
mass, while the change in the pseudo-red component is relatively weaker.
A comparison with the CLF derived from a group catalogue is presented in 
\S~\ref{sec:cmppre}. 

In Fig.~\ref{fig:lf_model}, the global LF is decomposed into contributions 
from the various model components. Over the luminosity range considered here, 
central galaxies dominate the contribution to the LF at any given luminosity.
The component contributions from pseudo-blue and pseudo-red central galaxies 
are comparable, except towards the faint end where the component LF from 
pseudo-blue central galaxies reaches a factor of two higher than that from 
pseudo-red central galaxies. For the LF contribution from satellites is 
dominated by pseudo-blue satellites above $\Mr \sim -19$ and is taken over 
by pseudo-red satellites towards the faint end. It is worth emphasising 
that at low halo masses the colours of pseudo-red centrals are only
slightly shifted from those of pseudo-blue centrals 
(Fig.~\ref{fig:halomassctur}), so the comparable contributions of the
two components are consistent with the observation that most low mass
centrals are blue (Fig.~\ref{fig:lf}). 
Conversely, the mean colours of the pseudo-blue and
pseudo-red satellite populations are similar at high halo mass, so the
dominance of the pseudo-blue contribution at the bright end is driven
by the tail of pseudo-blue satellites in its broader Gaussian, where these
pseudo-blue galaxies are supposed to be classified as red galaxies in observation,
indicated by the dashed lines in Fig.~\ref{fig:lf}.
The contributions of blue and red galaxies to the 
luminosity function (as opposed to pseudo-blue and pseudo-red ones here) 
were shown previously in Fig.~\ref{fig:lf}.

The overall features and trends in the CCMD and CLF can be more clearly 
understood by expressing the main CCMD quantities (\S~\ref{subsec:cenpara} 
and \S~\ref{subsec:satpara}) as a function of halo mass. In fact, these are 
the fundamental relations we are ultimately after with the CCMD model.
The CCMD relations related to central galaxies are shown in 
Fig.~\ref{fig:msmscen}, as a function of halo mass. In each panel, solid 
curves correspond to the best-fitting model and the shaded regions enclose 
the central 68.3 per cent distribution. In the top-left panel, the median 
luminosity of the two pseudo-colour central galaxy populations has a similar
dependence on halo mass, an exponential cutoff towards the low mass end 
and a power-law form towards the high mass end. The power-law index at the 
high mass end is 0.28 (0.35) for the pseudo-blue (pseudo-red) central
galaxies. The offset between the two curves reflects that 
at fixed halo mass
pseudo-blue central galaxies are on average more luminous or at the same
$r$-band median luminosity pseudo-red central galaxies reside in more massive 
haloes. The scatter in the central galaxy luminosity for the pseudo-red
galaxies monotonically decreases with increasing halo mass (top-middle panel), 
reaching about 0.4 mag in haloes of $10^{14}\hinvMsun$. 
For pseudo-blue central galaxy, the luminosity is tightly correlated with halo 
mass below $10^{13}\hinvMsun$ (nearly zero scatter, reaching the floor 0.01 set 
in our model)\footnote{
We perform tests by increasing the floor to see what drives the small scatter in 
the luminosity of the pseudo-blue central galaxies at fixed halo mass. Around 
$10^{13}\hinvMsun$, if the scatter increases, more pseudo-blue galaxies would be populated into haloes of lower mass and the model tends to under-predict the 
clustering amplitude of luminous galaxies. At low mass, we find that an increased
scatter would not noticeably change the 2PCFs of low-luminosity galaxies, while 
their number densities have a substantial change in comparison to their small 
error bars, leading
to a large change in $\chi^2$. We find that increasing the scatter to about 0.04 
results in a $\chi^2$ change of $\sim$40 (from number density), the 1-$\sigma$ range
of the $\chi^2$ distribution of the model. We can treat 0.01--0.04 as the possible allowed range of the scatter and for simplicity we take the value 0.01 to present 
our results. The tests show that the pseudo-blue central galaxies indeed have a much 
smaller scatter in luminosity than the pseudo-red ones at fixed halo mass. This is 
also supported by the shape of the central galaxy CLF derived from a group 
catalogue (see \S~\ref{sec:cmppre} and Fig.~\ref{fig:clfcompare}).
} 
and then the scatter increases to become similar as the pseudo-red central 
galaxies in cluster-size haloes. The fraction of central galaxies being pseudo-red 
increases from about 10 per cent in $10^{11}\hinvMsun$ haloes to about 65 per 
cent in $10^{12}\hinvMsun$ haloes and flattens towards the high mass end.
Note that the luminosity-halo mass relations derived here is for pseudo-colour
components. These are different from observationally inferred relations by 
separating galaxies into blue and red populations based on a luminosity dependent
colour cut, e.g. the ones in \citet{Surhud11} based on satellite kinematics.
Predictions for such relations can be computed from the model.

The mean colour of the pseudo-blue central galaxies increases steadily from 
$\sim$0.35 to $\sim$0.9 over the halo mass range of $10^{11}$--$10^{14}\hinvMsun$
(bottom-left panel in Fig.~\ref{fig:msmscen}).
For the pseudo-red central galaxies, the mean colour has a steep rise of
$\sim$0.5mag within the range of $10^{11}$--$10^{12}\hinvMsun$, and then
slowly increases towards higher halo mass ($\sim$0.1mag over two orders of
magnitude in halo mass). For the scatter in colour, the two components have
different trends (bottom-middle panel) --- while that of the pseudo-blue central
galaxies keeps increasing, that of the pseudo-red central galaxies has a sharp 
decrease in haloes of mass from $10^{11}\hinvMsun$ to $10^{12}\hinvMsun$ and 
then reaches a plateau of
$\sim$0.02mag towards higher halo mass. This narrow colour distribution of
pseudo-red central galaxies suggests that they make a large contribution to the
tight red sequence. We note that for the pseudo-red central galaxies, the major
changes in the colour distribution (i.e. mean and scatter) occur at halo mass
$10^{12}\hinvMsun$, coinciding with a characteristic mass in many aspects of galaxy
formation, such as the transition from cold-mode to hot-mode accretion 
\citep[e.g.][]{Birnboim03,Keres05}.

The bottom-right panel of Fig.~\ref{fig:msmscen} shows the correlation between
the colour and the magnitude of the pseudo-colour central galaxies
as a function of halo mass. Interestingly
a positive (negative) correlation exists for the pseudo-red (pseudo-blue) 
components over the halo mass range probed here. It means that in haloes of
fixed mass pseudo-red central galaxies appear slightly bluer at higher 
luminosity (more negative in magnitude), and the opposite for the pseudo-blue 
central galaxies. In the CCMD plot (Fig.\ref{fig:halomassctur}), given the
correlation the orientation angle $\theta$ of the contours with respect to the 
$x$-axis (magnitude) is determined by 
$\tan 2\theta = 2\rho\sigma_{\rm x}\sigma_{\rm y}/(\sigma_{\rm x}^2+\sigma_{\rm y}^2)$, 
with $\sigma_{\rm x}$
and $\sigma_{\rm y}$ the scatters in magnitude and colour, respectively. 
In most cases we are in the regime that one of $\sigma_{\rm x}$ and $\sigma_{\rm y}$ 
is much larger than the other (middle panels of Fig.~\ref{fig:msmscen}) and hence we 
have either $\theta \sim 0$ or $\theta \sim \pi/2$, which explains why the 
correlation does not show up prominently in Fig.~\ref{fig:halomassctur}.

Fig.~\ref{fig:msmssat} shows the CCMD quantities related to the satellite
galaxies. The normalisation $\phi_s^\ast$ of the satellite CLF for the
pseudo-red component is higher than that of the pseudo-blue component and the
difference reaches a factor of $\sim$5 in cluster-size haloes (top-left panel).
The characteristic luminosity $x_{\rm s}^\ast$ of the pseudo-red satellites
is lower than that of the pseudo-blue ones (top-middle panel), e.g. about 1.2mag 
in massive haloes. At the faint end, the CLF of the pseudo-red satellites has a
shallower slope than the pseudo-blue satellites (top-right panel of 
Fig.~\ref{fig:msmssat}). Below $10^{13}\hinvMsun$, the faint end slope for the
pseudo-red satellites is not well constrained in the model. On average satellites
become redder with increasing luminosity, and the dependence is weaker for the
pseudo-red component ($\sim$0.1mag in colour over 4 mag in luminosity; 
bottom-left panel). The scatter in the colour of the pseudo-red satellites show
a weak dependence on luminosity, decreasing from 0.05mag to 0.03mag over 4 mag 
in luminosity. However, the scatter for the pseudo-blue satellites is
a relatively steep function of luminosity, reaching a value similar to those of 
the pseudo-red satellites at the bright end but a few times higher at the faint 
end.

Finally, for the luminosity gap (the difference between the median luminosity of 
central galaxies and the characteristic luminosity of satellite galaxies), the
best-fitting model gives 0.64mag for pseudo-blue galaxies and 1.43mag for pseudo-red
galaxies.

\subsection{Derived Quantities and Relations}
\label{sec:derived}

\begin{figure*}
\includegraphics[width=\textwidth]{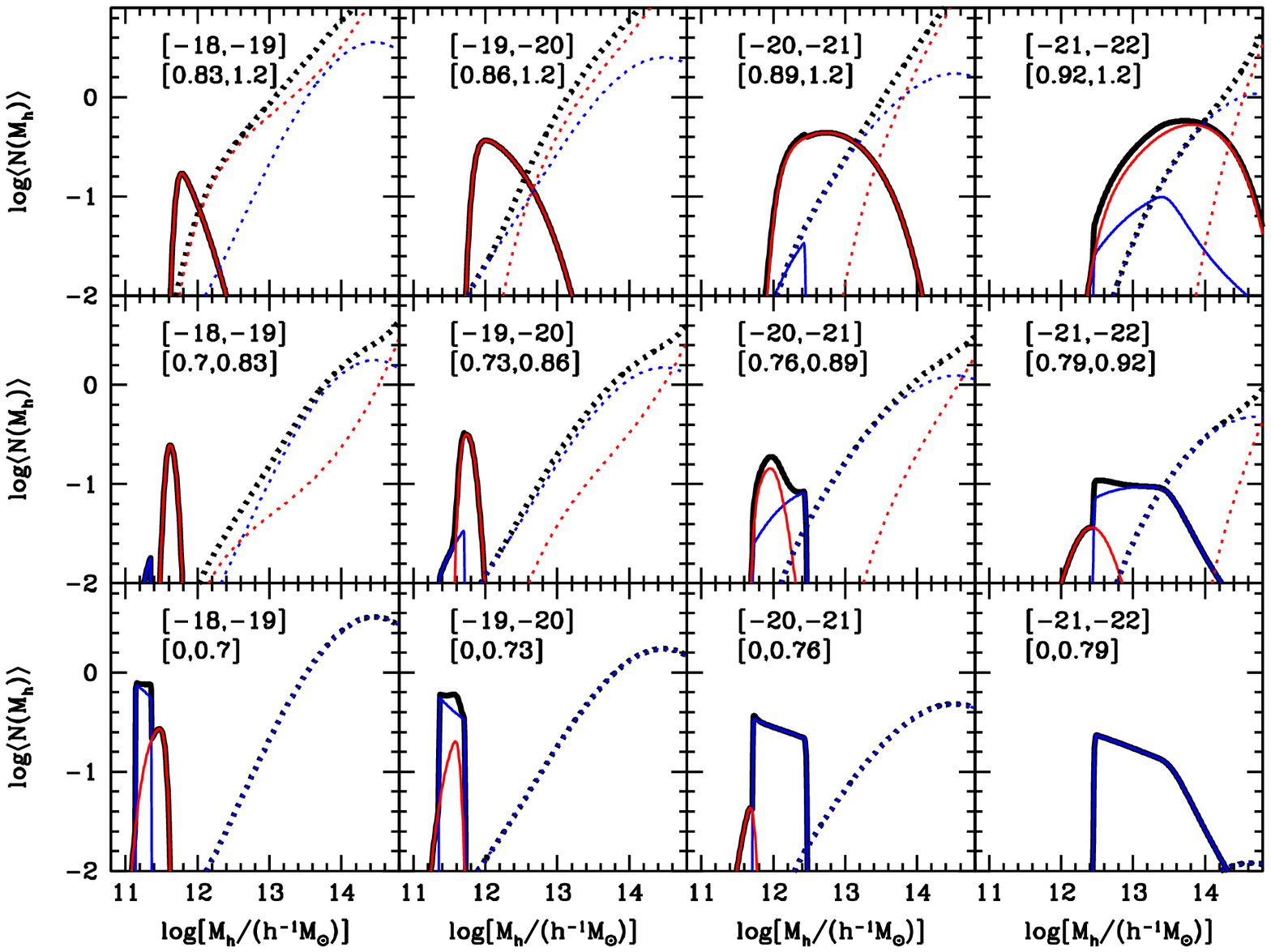}
\caption{
Dependence of the mean occupation function on galaxy colour and luminosity.
Each column corresponds to a fixed magnitude bin, with three colour sub-samples 
of galaxies (blue, green, and red sub-sample from bottom to top), with the 
magnitude and colour ranges labelled in the panel.
In each panel, the thick solid (dotted) black curve is the mean occupation 
function of central (satellite) galaxies in 
the given colour and magnitude bin, derived from the best-fitting
CCMD model. The thin blue and red curves correspond to those of the pseudo-blue
and pseudo-red galaxies (solid for central and dotted for satellite galaxies).
}
\label{fig:mof}
\end{figure*}

The CCMD modelling connects galaxy colour and luminosity with dark matter haloes. 
We present a few quantities and relations derived from such a connection.

First, the CCMD model enables us to compute the halo occupation function for
a galaxy sample defined by arbitrary cuts in colour and luminosity. In fact, this 
is how we do the modelling in the first place by figuring out the halo occupation
function for each galaxy sample. In Fig.~\ref{fig:mof}, we show the mean halo
occupation functions for galaxies in four magnitude bins with each divided into
three colour bins (roughly representing the blue, green,  and red samples). 
In each panel, the mean occupation functions are decomposed
into the four pseudo-blue/pseudo-red central/satellite components. At each 
fixed magnitude bin, as the sample becomes redder, the relative contribution from 
the pseudo-red components increases. The mean occupation function of the 
pseudo-red central galaxies is smooth, while that of the pseudo-blue central
galaxies shows relatively sharp, step-like features, a manifestation of the
tight correlation in the model between luminosity of pseudo-blue central galaxies 
and halo mass in low mass haloes (Fig.~\ref{fig:msmscen}). The total mean 
occupation function of central galaxies at a fixed magnitude bin is in the form 
of a bump, which can usually be approximated by a Gaussian profile. For that of 
the satellites, it is approximately in a form of power law rolling off
towards low mass, except for the bluest sample, which turns over towards 
the high mass end.

\begin{figure*}
\includegraphics[width=\textwidth]{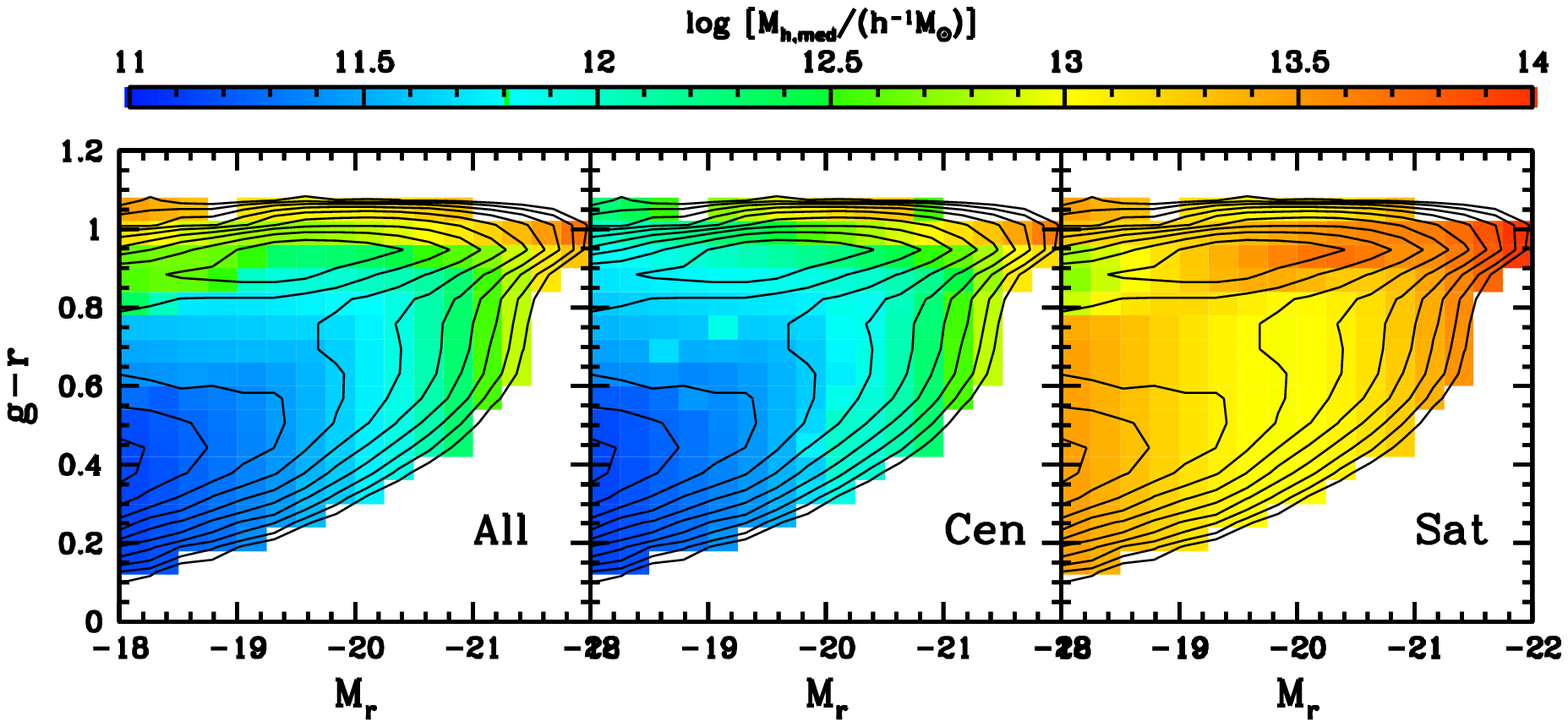}
\caption{
Dependence of median host halo mass on galaxy colour and luminosity from 
the best-fitting CCMD model.
In each panel, colour-coded is the median halo mass $M_{\rm h, med}$ for 
galaxy samples defined in fine colour and magnitude bins. From left to right,
the panels show the median host halo mass distribution for all galaxies, 
central galaxies, and satellite galaxies, respectively. Contours of galaxy 
number density are overlaid in each panel to show the bimodality in the overall
galaxy population. 
}
\label{fig:mdmctur_3panels}
\end{figure*}

With the mean occupation function of galaxies in each colour and magnitude bin,
we can figure out characteristic quantities describing the occupation of galaxies.
In Fig.~\ref{fig:mdmctur_3panels}, we show an example of such quantities, 
the median mass of host haloes as a function of galaxy colour and magnitude.
The left panel is the median halo mass for all the galaxies in each colour and
magnitude bin. In general, the more luminous and redder the galaxy population is,
the higher the median halo mass becomes. The majority of blue cloud galaxies 
have median host halo mass in the range of $10^{11}$--$10^{12}\hinvMsun$. 
The median halo mass for galaxies at the ridge of the red sequence increases from 
$\sim 10^{12}\hinvMsun$ to a few times $10^{13}\hinvMsun$ over a range of 
four magnitudes. At each luminosity, the reddest galaxies appear to have the
highest median halo mass. 

The middle and right panels in Fig.~\ref{fig:mdmctur_3panels} show the median
halo mass for central and satellite galaxies, respectively. The pattern for the
central galaxies is similar to that of the total, but the trend along the colour
direction becomes weaker. For satellites, the majority of them have median halo mass
above $10^{13}\hinvMsun$, including those in the blue cloud. At fixed colour
and luminosity, satellite galaxies on average reside in more massive haloes than
central galaxies. For instance, the faint red 
central galaxies resides in haloes of a few times $10^{11}\hinvMsun$, while faint 
red satellites are in haloes of a few times $10^{12}\hinvMsun$ to a few times
$10^{13}\hinvMsun$ \citep[e.g.][]{Xu16}. From the three panels, we can also tell that 
reddest galaxies are dominated by satellites in relatively massive haloes.

\begin{figure}
\includegraphics[width=\columnwidth]{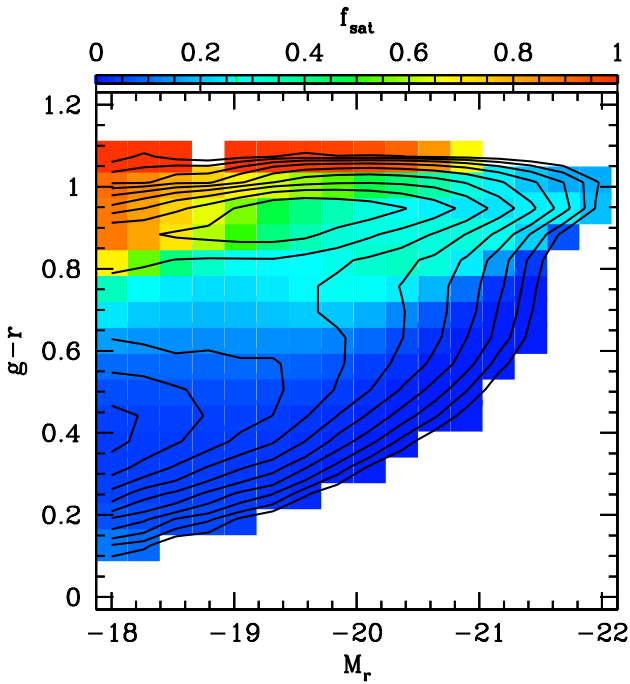}
\caption{
Dependence of satellite fraction on galaxy colour and luminosity from 
the best-fitting CCMD model.
Colour-coded is the satellite fraction $f_{\rm sat}$ for 
galaxy samples defined in fine colour and magnitude bins.  
Contours of galaxy 
number density are overlaid to show the bimodality in the overall
galaxy population.  
}
\label{fig:satcmd}
\end{figure}

\begin{figure*}
\includegraphics[width=1.0\textwidth]{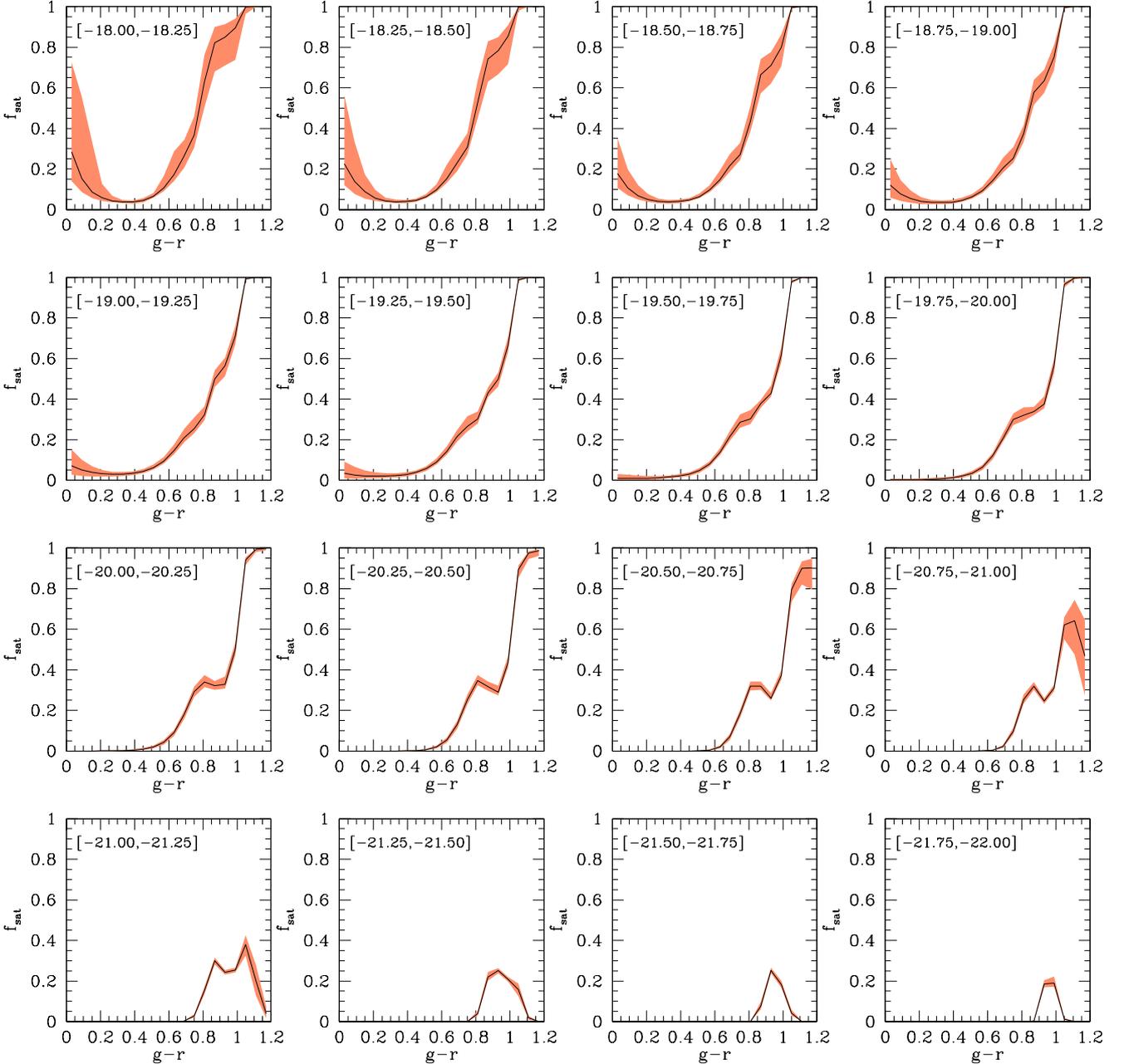}
\caption{
Dependence of satellite fraction on colour for galaxies in different luminosity 
bins. 
Galaxy luminosity increases from left to right and top to bottom panels, as 
labelled in the top-left corner in each panel. 
In each panel, the solid curve is the satellite fraction from the best-fitting
CCMD model and the shaded region indicate the 1$\sigma$ range.
}
\label{fig:fsatL}
\end{figure*}

In Fig.~\ref{fig:satcmd}, we show the fraction of satellites as a function of 
galaxy colour and magnitude. The gradient is mainly along the colour direction,
and redder galaxies are more likely to be satellites. For a detailed inspection,
Fig.~\ref{fig:fsatL} shows satellite fraction as a function of galaxy colour in 
several magnitude bins. \citet{Zehavi11} model the clustering of galaxies in 
the $-20<\Mr<-19$ magnitude bin with six fine bins in colour, using a simple
HOD model (their fig.20). The satellite fractions derived here are in broad 
agreement with theirs. The satellite fraction 
for blue cloud galaxies is generally below 10 per cent. The red galaxies beyond 
the red sequence ridge are mainly satellites. For faint red galaxies, the 
satellite fraction is high. For example, for galaxies with $-19<\Mr<-18$ 
and $g-r\gtrsim 0.75$, the average satellite fraction is above 50 per cent,
increasing with colour. This is consistent with the HOD modelling results
in \citet{Xu16}.\footnote{
If the stronger clustering of the fainter, redder galaxies were caused by 
assembly bias effect, the satellite fraction would be reduced. As no
assembly bias is incorporated into the CCMD model, the model tends to
populate more satellites in massive haloes to match the observed 
clustering. As mentioned before, our interpretation of the results 
assumes no assembly bias and discussions on assembly bias effect can be
found in section~\ref{sec:dis}.
}
In the tail of the colour distribution, e.g. redder than $g-r=1$,
galaxies are dominated by satellites, especially at the faint end. They are
satellites in massive haloes (with median mass of a few times $10^{13}\hinvMsun$;
right panel of Fig.~\ref{fig:mdmctur_3panels}). \citet{Pasquali10} show that 
low-stellar-mass satellites in massive haloes are generally older and more metal rich
than central galaxies of similar stellar mass. However, such an average trend is not
able to explain why those faint satellites are so red. They are even as red as the
most luminous galaxies in the CMD here ($\Mr\sim -22$). It is likely that they have 
high metallicity, comparable to the massive galaxies. This means that they 
substantially deviate from the mass-metallicity relation, which is possible 
if they represent the tail of the heavily stripped galaxies. As shown by 
\citet{Puchwein10} with a high-resolution simulation, satellites approaching closely 
to the cluster centre can be substantially stripped (e.g. with more than 90 per 
cent of their stellar mass lost; see their fig.16). 
A metallicity study of those galaxies with spectroscopic observation 
can help verify the above picture and understand their nature.

We note that, in Fig.~\ref{fig:fsatL}, there appears to be an upturn in
the satellite fraction towards the blue end for galaxy samples with $\Mr>-19$. One possible reason is that galaxies at the
blue end come from the tail of a narrow distribution of pseudo-blue central galaxies and that of a broad distribution of pseudo-blue and 
pseudo-red satellite galaxies (see Fig.~\ref{fig:ngdm}). A slight change
in the satellite distribution would lead to a large change in the
satellite fraction. Also as we approach the tail of the distribution, the results may be sensitive to 
the functional forms used in the model parameterization. 
Anyway, at the faint blue end, since there is not
much constraining power from the clustering, the uncertainty in the
satellite fraction is large and the results are still consistent with 
a low satellite fraction. 

\begin{figure*}
\includegraphics[width=\textwidth]{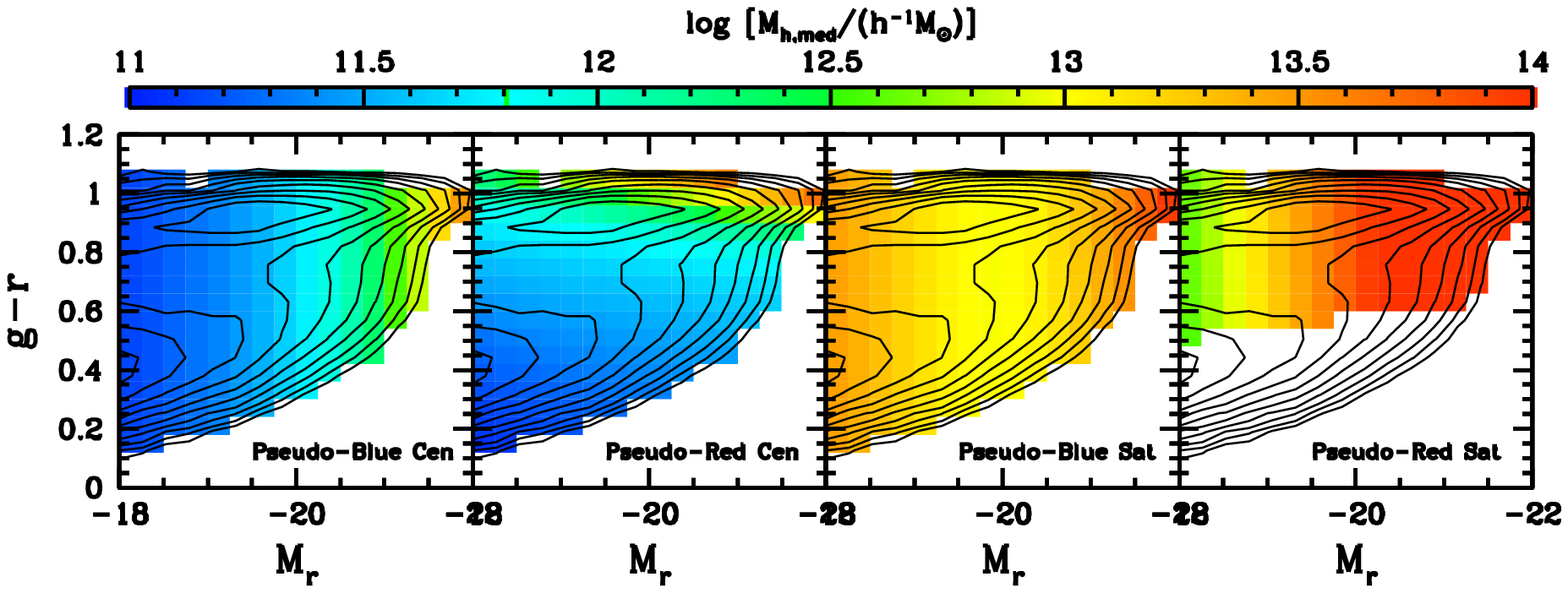}
\caption{
Dependence of median host halo mass on galaxy colour and luminosity for each 
pseudo-colour model population from the best-fitting CCMD model.
In each panel, colour-coded is the median halo mass $M_{\rm h, med}$ for 
galaxy samples defined in fine colour and magnitude bins. From left to right,
the panels show the median host halo mass distribution for pseudo-blue 
central galaxies, 
pseudo-red central galaxies, pseudo-blue satellites, and pseudo-red satellites, 
respectively. Contours of galaxy 
number density are overlaid in each panel to show the bimodality in the overall
galaxy population. 
Note that the `blue cloud' region in the rightmost panel has no value 
assigned, as almost no pseudo-red satellites exist there.
}
\label{fig:mdmctur_4panels}
\end{figure*}

Finally, the median halo mass trends seen in Fig.~\ref{fig:mdmctur_3panels} 
motivate us to study those for each individual pseudo-colour population, 
as the CCMD model naturally separates them. Such a separation may be more 
illuminating. Indeed, we find clear and interesting trends. 
Fig.~\ref{fig:mdmctur_4panels} displays the median halo mass for 
each of the pseudo-blue/pseudo-red central/satellite components. 
For pseudo-blue central galaxies (leftmost panel), the median halo mass 
gradient is along the luminosity direction, with more luminous pseudo-blue 
central galaxies residing in more massive 
haloes. In striking contrast, for pseudo-red central galaxies (second to the left
panel), the median halo mass gradient is almost completely 
along the colour direction, with redder pseudo-red galaxies found 
in more massive haloes. The trends can also be inferred from Fig.~\ref{fig:mof}.
The orthogonal dependences of median halo mass on
luminosity and colour for the two central galaxy components are remarkable. 
The results suggest that while halo mass determines the luminosity of 
pseudo-blue central galaxies, 
it mainly affects the colour of pseudo-red central galaxies. 
This seems to indicate that the pseudo-colour populations in the CCMD model 
have a physical origin, and we 
leave more discussions on the implications to \S~\ref{sec:dis}.

For pseudo-blue satellites (second to right panel in 
Fig.~\ref{fig:mdmctur_4panels}), the median halo mass does not depend on 
colour and shows only a weak 
dependence on luminosity. It is about $10^{13}\hinvMsun$ for pseudo-blue 
satellites with luminosity around $L^\ast$ and becomes higher for fainter 
as well as more luminous satellites. 
The median halo mass for pseudo-red satellites (rightmost panel), 
on the contrary, 
shows a clear dependence on luminosity, higher for more luminous satellites. 
Note that in this panel, no value is assigned in the `blue cloud' region, as
there exist almost no pseudo-red satellites. The 
median halo mass in satellites can also be inferred 
from Fig.~\ref{fig:mof} -- while 
the shape of the mean occupation function of the pseudo-blue satellites 
does not vary significantly with either colour or luminosity, 
that of the pseudo-red satellites
becomes steeper for more luminous satellites. This is consistent with 
the pseudo-blue satellites being more widely distributed in colour and luminosity 
across a large halo mass range than the pseudo-red galaxies, as shown 
in Fig.~\ref{fig:halomassctur}. 
Similar to the pseudo-colour central components, the distinct difference between 
the two pseudo-colour satellite components and the clear trend seen with 
the median halo mass imply that there may be a physical origin of the 
separation into those components (see more discussions in \S~\ref{sec:dis}).

\subsection{Comparison of the Derived 
Galaxy-Halo Relations with Previous Work}
\label{sec:cmppre}

\begin{figure}
\includegraphics[width = \columnwidth]{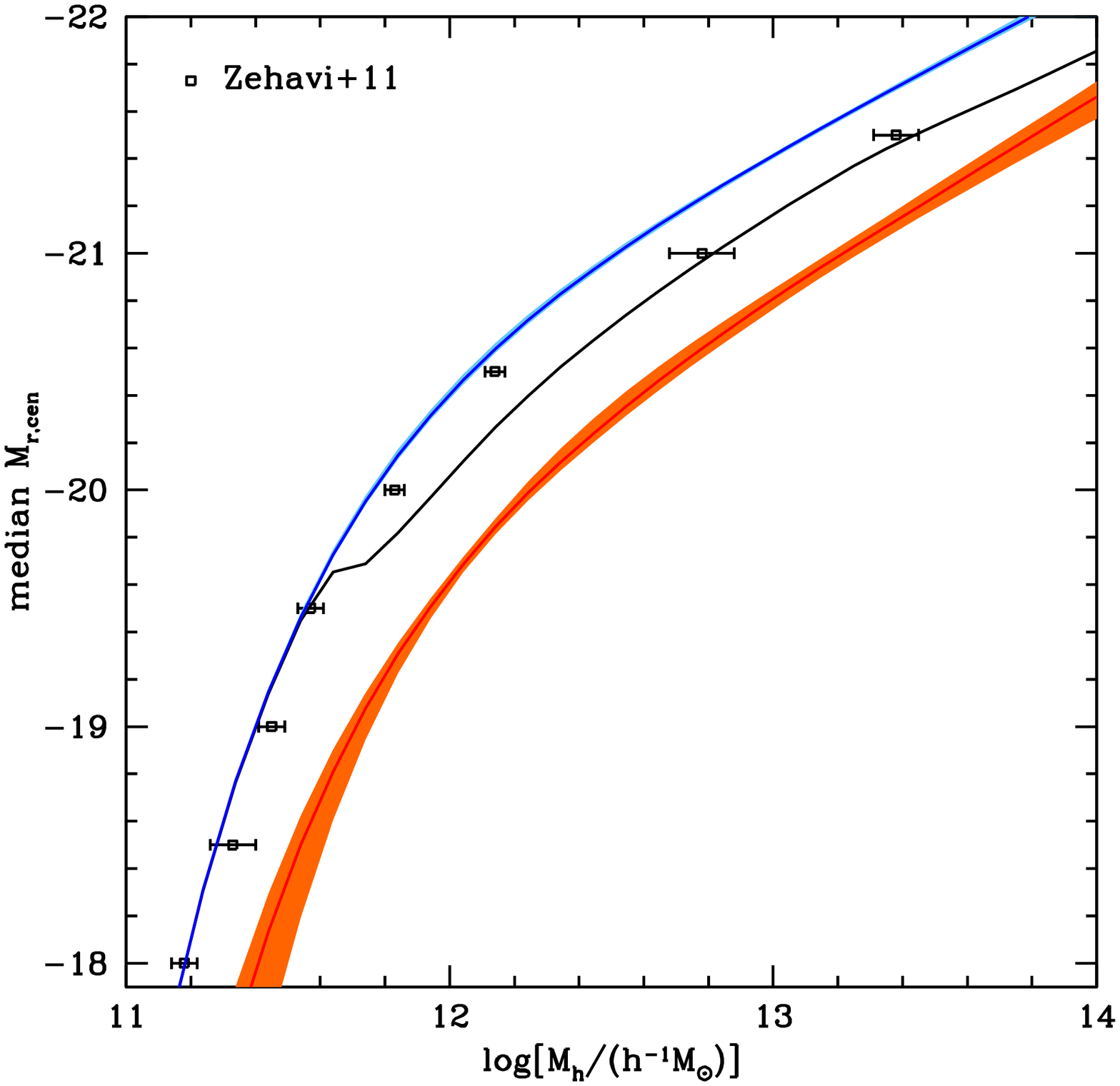}
\caption{
Comparison of the relation between the median central galaxy luminosity 
and halo mass. The blue and red curves are the relations for pseudo-blue and 
pseudo-red central galaxies from
the best-fitting CCMD model (same as in the top-right panel of 
Fig.~\ref{fig:msmscen}), and the
black curve is the overall median central galaxy luminosity 
as a function of halo mass. 
The data points are from \citet{Zehavi11}, from HOD modelling of the 
projected 2PCFs of luminosity-threshold samples. See details in the text \S~\ref{sec:cmppre}.
}
\label{fig:luminmh}
\end{figure}

\begin{figure}
\includegraphics[width = \columnwidth]{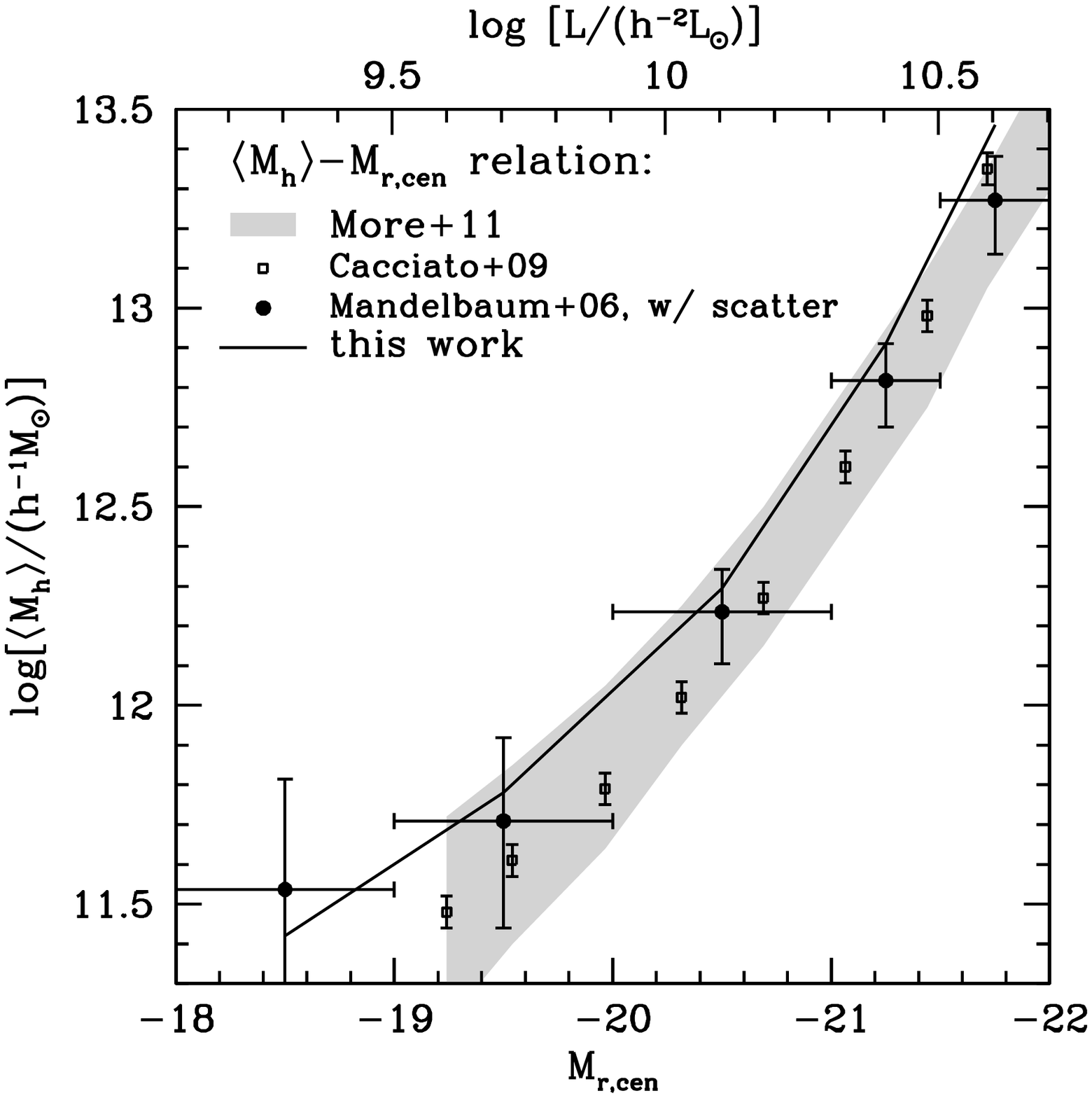}
\caption{
Comparison of the relation between the mean halo mass and
central galaxy luminosity.
The curve is the mean halo mass as a function of central galaxy luminosity 
from the best-fitting CCMD model.
The different sets of data points show the same relation inferred 
from different probes, including galaxy-galaxy weak lensing \citep{Mandelbaum06}, 
satellite kinematics \citep{More11}, and CLF modelling of galaxy clustering 
and galaxy lensing \citep{Cacciato09}, as labelled in the legend.
All vertical error bars, including the boundaries of shaded region, 
represent the 95 per cent 
confidence levels, and the horizontal error bars on filled circles 
indicate the magnitude bins adopted in the galaxy lensing measurements 
\citep{Mandelbaum06} and used in the calculation 
of the CCMD result. The lensing mass is derived by combining those for the 
late- and early-type central galaxies in \citet{Mandelbaum06} and 
corrected to account for the scatter
in the mass-luminosity relation. See the text \S~\ref{sec:cmppre} for detail.
}
\label{fig:luminmeanmh_compare}
\end{figure}

Our CCMD model presents a new way to describe the relation between galaxies 
and dark matter haloes. It extends the CLF framework by adding a dimension of 
galaxy colour. The modelling results can be used to derive the CLF or HOD for 
a galaxy sample defined by cuts in luminosity and colour. We compare the 
results with the HOD and CLF derived based on previous study. 
To be consistent with the halo definition used in previous work, we apply 
the correction mentioned in \S~\ref{subsec:simubased} to convert the halo mass 
in our results to $M_{\rm 200b}$.

\citet{Zehavi11} perform HOD modelling of the projected 2PCFs of galaxy 
samples defined by different luminosity thresholds, using the SDSS DR7 data. 
For each sample, one of the HOD parameters is $M_{\rm min}$, which is the mass 
scale of haloes that on average half of them host central galaxies above the 
luminosity threshold. If at fixed halo mass central galaxy
luminosity follows a log-normal distribution 
(i.e. normal distribution in terms of magnitude)
and the median luminosity has a power-law 
dependence on halo mass, the relation between the threshold luminosity 
and $M_{\rm min}$ would be just that between the median central galaxy 
luminosity $\langle M_{\rm r,cen}\rangle$ and halo mass $\Mh$ \citep{Zheng07b}. 
Note that here, for a log-normal luminosity distribution, 
the median luminosity corresponds to the mean magnitude. 
The data points in Fig.~\ref{fig:luminmh} show the 
$\langle M_{\rm r,cen}\rangle$--$\Mh$ relation derived in such a way in \citet{Zehavi11}.
The relation is also consistent with that inferred from satellite kinematics in
\citet{Surhud09}.

The blue (red) curve is the median luminosity of pseudo-blue (pseudo-red) 
central galaxies as a function of halo mass from the best-fitting CCMD model 
(same as in the top-left panel of Fig.~\ref{fig:msmscen}). 
As the CCMD model has the full distribution of the central galaxy 
luminosity, which is the superposition of those of pseudo-red and 
pseudo-blue central galaxies weighted by the relative fraction 
(top-right panel of Fig.~\ref{fig:msmscen}), we can obtain the median central 
galaxy luminosity at each halo mass. The black curve is the
$\langle M_{\rm r,cen}\rangle$--$\Mh$ relation from the best-fitting 
CCMD model. An inflection feature occurs around $\log[\Mh/(\hinvMsun)]=11.6$, 
the mass scale where pseudo-blue and pseudo-red have equal fraction 
(top-right panel of Fig.~\ref{fig:msmscen}). 
The feature is caused by the narrow luminosity distribution of 
the pseudo-blue central galaxies (top-middle panel of Fig.~\ref{fig:msmscen}). 
With a test of increasing the floor set for the scatter in 
the pseudo-blue central galaxy luminosity, we find that the inflection 
feature becomes smoothed, with little effect on other parts of the curve. 
We note that this inflection will also go away if we use the mean luminosity.
While the curve follows well the
overall trend seen in \citet{Zehavi11}, it does not exactly match the points 
at the low mass end. This is not surprising, as the distribution of central 
galaxy luminosity is not a single log-normal distribution and the relation 
between median central luminosity and halo mass is not a pure power law. 
That is, the two conditions to interpret the relation between
threshold luminosity and $M_{\rm min}$ as the 
$\langle M_{\rm r,cen}\rangle$--$\Mh$ relation in 
\citet{Zehavi11} are not satisfied in the CCMD model. 
From the view point of the CCMD
model, the central galaxy luminosity distribution from the contributions of
the pseudo-blue and pseudo-red population does not follow a single 
log-normal distribution, and the commonly adopted form of the central galaxy 
mean occupation function for luminosity-threshold samples 
\citep[e.g.][]{Zheng05,Zheng07b,Zehavi11} should be only treated
as a good approximation. The comparison here between the HOD and 
CCMD modelling results is not exact, but the agreement is reasonable anyway.

The relation between central galaxy luminosity and halo mass can 
also be expressed in terms of mean halo mass as a function of galaxy luminosity, 
i.e. the $\langle\Mh\rangle$--$M_{\rm r,cen}$ relation. 
In Fig.~\ref{fig:luminmeanmh_compare}, we conduct a comparison between 
the relation from the best-fitting CCMD model with those yielded from various other methods,
which include galaxy-galaxy weak lensing, satellite kinematics, and CLF 
modelling of galaxy clustering and weak lensing. \citet{Mandelbaum06} construct 
luminosity bin SDSS galaxy samples and perform weak lensing measurements to 
infer the mean halo mass of early- and late-type central galaxies, respectively. 
We compute the mean halo mass for all central galaxies 
based on those for early- and late-type galaxies and their relative fraction
(see table 4 and table 2 in \citealt{Mandelbaum06}). 
The lensing measurement result is plotted as the filled 
circles with vertical error bars marking the 95 per cent confidence level 
and horizontal error bars indicating the magnitude bins. Note that to 
convert from the measured excess 
surface density of mass to mean halo mass, 
there are additional corrections to account for 
the halo mass distribution and the scatter in the $\Mh$--$M_{\rm r,cen}$ relation 
\citep{Mandelbaum05}. In \citet{Mandelbaum06}, the result has been corrected, 
assuming no scatter in the $\Mh$--$M_{\rm r,cen}$ relation. 
To be more realistic, here we make 
a further correction to include the effect of the scatter using a model considered in
\citet{Mandelbaum05} (see their table 1). In detail, with respect to the 
mean halo mass in \citet{Mandelbaum06}, we add upward corrections of 
0 dex, 0.08 dex, 0.15 dex, and 0.14 dex
in halo mass for $\Mr$ magnitude bins [-19,-18], [-20,-19], [-21,-20], 
and bins with $\Mr$ more luminous than -21, respectively. 
The solid curve is the result calculated from the
best-fitting CCMD model with the same magnitude bins as in \citet{Mandelbaum06}, 
which shows an excellent agreement with that inferred from galaxy lensing. 
For the most luminous bin, the lensing result is slightly lower than the 
CCMD result, which may be related to the details 
in modelling the scatter in the $\Mh$--$M_{\rm r,cen}$ relation and the 
slightly different cosmological model adopted in \citet{Mandelbaum06} for 
modelling the lensing signal. A more
direct comparison would be between the excess surface density predicted 
by the CCMD model and the lensing measurement, which we reserve for future work.

The shaded region in Fig.~\ref{fig:luminmeanmh_compare} 
represents the constraints (95 
per cent confidence level) on the $\langle\Mh\rangle$--$M_{\rm r,cen}$ 
relation from satellite kinematics in \citet{More11}. 
While the constraints are consistent with the
CCMD result, the mean halo mass inferred from satellite kinematics is 
systematically lower (by about 0.2 dex). 
The reason is not clear, and it is possibly related to some
additional effects from satellite velocity bias \citep[e.g.][]{Guo15} and the 
misidentification of a small fraction of central galaxies as satellites. 
The open squares show the $\langle\Mh\rangle$--$M_{\rm r,cen}$ relation 
derived from CLF modelling of galaxy clustering and lensing measurements 
by \citet{Cacciato09}, which is presented in 
\citet{More11}. As the CLF modelling in \citet{Cacciato09} makes 
use of the information from the group catalogue \citep[e.g.][]{Yang07}, 
the $\langle\Mh\rangle$--$M_{\rm r,cen}$
relation is almost identical to that from the group 
catalogue \citep[see][fig.10]{More11}.
The mean halo mass from \citet{Cacciato09} is also systematically lower 
than the CCMD result,
especially for faint galaxies in low mass haloes (e.g. below
$\log[\langle\Mh\rangle/(\hinvMsun)]\sim 13$). The discrepancy may be 
related to the underestimation of the halo mass from the group catalogue 
for haloes hosting blue
central galaxies (\citealt{Lin16} and discussions below).

\begin{figure*}
\includegraphics[width = \textwidth]{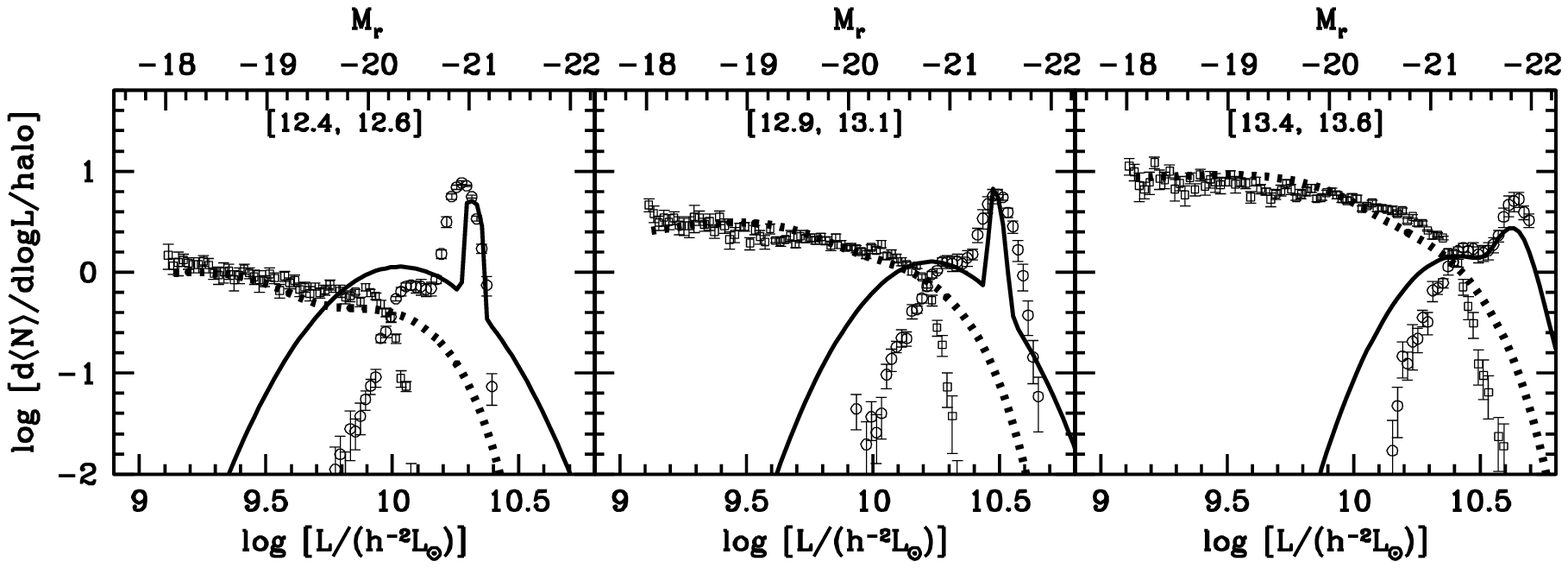}
\caption{
Comparison between the CLF from a group catalogue and that from the 
best-fitting CCMD model. The halo mass range in terms of 
$\log[\Mh/(\hinvMsun)]$ is labelled in each panel.
The circles and squares are the CLFs of central and satellite galaxies 
from the SDSS DR7 group catalogue, constructed in the same way 
as in \citet{Yang08}. The error bars are estimated based on 10 jackknife samples. 
The solid and dotted curves are those from the best-fitting CCMD model.
The differences between group-based CLF and CCMD-derived CLF are related 
to the way of assigning
halo mass to groups in the group catalogue. See the text \S~\ref{sec:cmppre} for details.
}
\label{fig:clfcompare}
\end{figure*}

As the CCMD can be readily inferred with a group catalogue, 
it is interesting to compare our 
CCMD modelling results with those from a group catalogue. 
In particular, this potentially 
allows the comparison of the full distribution of galaxy luminosity and colour.
For such a purpose, we make use of the 
group catalogue\footnote{\url{http://gax.sjtu.edu.cn/data/Group.html}}
based on the SDSS DR7 data and constructed in the same way as 
in \citet{Yang08}, who used the DR4 galaxy sample. 
For groups in a given halo mass
bin, the CLF is computed, separating into contributions from central 
and satellite galaxies. 
To account for the effect of flux limit for galaxies and the completeness of groups, 
at each luminosity bin, the average number of galaxies per halo is 
computed with groups in the volume 
where both galaxies down to that luminosity and groups in the given 
halo mass bin are complete.
Fig.~\ref{fig:clfcompare} shows the CLFs from the group catalogue 
in three halo mass bins, in 
comparison with those from our best-fitting CCMD model. 

For the central galaxies, the CLF from the group catalogue shows a clear 
double-peak profile, similar to the superposition of two Gaussian profiles 
(in magnitude). This is exactly in line with the CCMD model, 
where the central galaxies are composed of the pseudo-blue and pseudo-red
populations, each with a Gaussian distribution. In the group-based central 
CLF, the component with the blue peak has a much narrower distribution than 
the one with the red peak, meaning that the blue component has a tighter 
correlation between central galaxy luminosity and halo mass.
This is again consistent with the constraints from the CCMD model. 
However, quantitatively there are clear differences between the group-based 
CLF (open circles) and the CCMD modelling
result (solid curves). The group-based central CLF is more concentrated in luminosity,
the peak positions shift with respect to the CCMD model, and the relative 
contribution of the two components may not agree perfectly with the CCMD model. 

The differences can be largely explained by the way halo masses are 
assigned to the groups. In the group catalogue we use, (apparent) halo mass 
is determined from the total $r$-band galaxy
luminosity (down to $\Mr=-19.5$) of each group, i.e. through matching the 
abundances of groups ranked by total luminosity and that of haloes 
ranked by halo mass. To have a conceptual understanding, suppose that we work 
in the regime where the total galaxy luminosity of each halo 
is dominated by the central galaxy (approximately the case for haloes below 
$\sim 10^{12.5}\hinvMsun$). 
Central galaxy luminosity is then mapped directly onto halo mass. The 
scatter in central galaxy luminosity in groups of a given halo mass bin 
simply reflects the size of the mass bin. The intrinsic scatter of central 
galaxy luminosity at fixed halo mass does not show up in the CLF,
which leads to the narrower profile of the group-based central galaxy CLF. 
Such abundance matching also shifts the median galaxy luminosity and 
halo mass relation, resulting in the apparent offset of the CLF profile 
at a given halo mass. By performing a test of the effect 
of group finding on the conditional stellar mass function (CSMF), 
\citet{Reddick13} find that the central CSMF from the group catalogue 
has a reduced width and shows a peak offset (see their fig.20,
squares versus diamonds), in line with our simple analysis. In addition, 
the halo mass in the group catalogue can be different from the truth 
and the difference depends on the type of central
galaxies. For example, through galaxy lensing measurements, 
\citet{Lin16} show that the halo 
masses of groups of the same apparent halo mass (around $10^{12}\hinvMsun$) 
in the \citet{Yang07} group catalogue can differ by a factor of 1.9--3.4, 
with higher masses for those hosting 
central galaxies of early type or low specific star formation rate 
(see their fig.1 and fig.2). Such a mass difference is quite consistent 
with that between the halo masses of the pseudo-blue 
and pseudo-red central galaxies of the same median luminosity in the CCMD 
model (top-left panel of Fig.~\ref{fig:msmscen}). The difference between the 
apparent halo mass in the group catalogue and the true halo mass 
can also cause the shift in the peaks of the central 
galaxy CLF and affect the relative fraction of the red and blue components.

In Fig.~\ref{fig:clfcompare}, the group-based satellite CLF (squares) and 
that from the CCMD model (black dotted curve) appear to have a better agreement 
than the central CLFs. The big difference is that the group-based satellite 
CLF has a steeper cutoff at the high luminosity end than that from the CCMD model. 
This in fact can be explained in the same way as why the group-based 
central galaxy CLF is narrower -- with the apparent halo mass largely 
determined by the central galaxy luminosity by ignoring the scatter 
between luminosity and halo mass, the luminosity gap between the central 
galaxy and satellites is enhanced. This phenomenon also 
shows up in the test of halo finding effect in \citet{Reddick13}, as seen in their 
fig.20 (also see \citealt{Campbell15}). 
In reality, even if the luminosity gap is independent of central galaxy 
luminosity, the intrinsic scatter in central galaxy luminosity can lead to 
a much softer high-luminosity cutoff in the
average satellite CLF, i.e. a smearing effect. Towards the faint end, 
the group-based satellite CLF and the CCMD one show a remarkable agreement. 
We speculate that given the relatively shallow
profile of the faint end satellite CLF,  any smearing effect 
(e.g. caused by scatter in central galaxy luminosity or by the inaccurate 
halo mass in the group catalogue) would not change the
profile substantially.

With the group catalogue, more comparisons could be done, such as the 
colour distribution of central and satellite galaxies 
\citep[e.g.][]{vandenBosch08} or even the CCMD. However,
from the above CLF comparison, it is clear that any comparison would not 
be direct and the interpretation of any differences would not be straightforward. 
Many factors can affect the comparisons, including the group finding algorithm, 
the purity and completeness of groups,
the way halo mass being assigned, and the definitions of central and 
satellite galaxies. The best way to do the comparisons is to construct 
a mock galaxy catalogue from our CCMD 
modelling results and apply the specific group finding code to find groups 
in the mock \citep[e.g.][]{Campbell15}. Then with
the SDSS groups and the mock ones, we can do direct and fair comparisons 
on the distribution of galaxy properties (luminosity and colour). 
The mock catalogue can also help understand and
quantify the systematics in finding groups and in using the group catalogue 
for various studies. We reserve more comparisons and such a full investigation 
for future work, with the help of CCMD-based mock galaxy catalogues. 
The importance of fully modelling the redshift-space group identification
procedure for HOD-type clustering analysis is emphasised by \citet{Sinha17}. 

\section{Summary and Discussion}
\label{sec:dis}

The bimodal distribution in the galaxy CMD is a main observational feature 
in properties of galaxies. It encodes important information on galaxy formation 
and evolution. In this paper, we study the luminosity and colour distribution 
of galaxies by developing a CCMD model and by modelling the luminosity and 
colour dependent galaxy clustering, which enables us to connect
galaxy luminosity and colour with dark matter haloes and to deproject the 
global galaxy CMD into a halo mass dependent distribution with a decomposition 
into contributions from central and satellite galaxies. This model, and its
calibration against clustering measurements for many finely binned samples
of SDSS galaxies, substantially extends our previous studies of the
luminosity and colour dependence of the galaxy correlation function
\citep{Zehavi05,Zehavi11}.

The CCMD parameterization is mainly motivated by the fact that the colour 
distribution of galaxies at fixed luminosity can be well described by the 
superposition of two Gaussian components, which guides us to introduce the 
pseudo-blue and pseudo-red model populations. Each population is further 
divided into the central and satellite galaxy components. At fixed halo
mass, the magnitude and colour of each pseudo-colour central population 
follows a 2D Gaussian distribution. For satellite galaxies, at fixed halo mass, 
their luminosity distribution is 
parameterized as a Schechter-like function and their colour follows a 
luminosity-dependent Gaussian distribution. The characteristic quantities of 
the central and satellite CCMD are
further parameterized as a function of halo mass, motivated by previous work 
on HOD and CLF modelling. The quantities include the median and scatter of 
the central galaxy magnitude and those of the colour and the correlation 
between central galaxy magnitude and colour for each pseudo-colour central
component, and the normalization, faint end slope, and characteristic 
luminosity of the satellite CLF for each pseudo-colour satellite component. 
Because of the width of the pseudo-blue and pseudo-red distributions,
the two populations can overlap in true colour, sometimes substantially
(see Figs.~\ref{fig:ngdm} and ~\ref{fig:halomassctur}). 
The CCMD model employs a global parameterization in the sense that it 
describes the overall distribution of galaxy luminosity and colour as a
function of halo mass, not specific to any particular galaxy sample. 
Given a set of CCMD parameters, the mean occupation function 
can be readily calculated for a sample of galaxies with any 
luminosity and colour cuts, and then the clustering statistics of the 
sample can be computed with the halo population of a cosmological model.

We apply the CCMD formalism to model the abundances and projected 2PCF 
measurements of galaxy samples defined by fine luminosity and colour bins, 
constructed from the SDSS DR7 data. The model is able to well reproduce 
the data, which include the abundances and projected 2PCF measurements
of 79 galaxy samples covering the galaxy CMD in the range 
of $-22<\Mr<-18$ and $0<g-r<1.2$, with 13 data points per sample. 
The model seems to have a large number of free parameters, 59 in total.
However, the number of parameters of this global CCMD model 
is in fact much less than that in a
model where each sample is parameterized separately, as done in previous work 
(e.g. 5 parameters per sample). In our application here, 
each sample only has $\sim 0.7$ parameter on average. The 
luminosity and colour dependent clustering makes it possible to infer 
the halo mass dependent CCMD, and the small-scale clustering provides 
sufficient constraining power to separate the 
central and satellite components.

The main results from the CCMD modelling of SDSS galaxy clustering include:
\begin{itemize}
    \item[(1)] For central galaxies, the CCMDs of the two pseudo-colour 
    populations are distinct and almost orthogonal to each other. 
    At fixed halo mass, the pseudo-blue central galaxies 
    have a narrow luminosity distribution but spread in colour, 
    while the pseudo-red central galaxies have a narrow colour distribution 
    but spread in luminosity. The mean colours of the two components become
    comparable at halo mass $\sim 10^{11.5}\hinvMsun$.
    
    \item[(2)] For each pseudo-colour central component, below halo mass of 
    $\sim 10^{12}\hinvMsun$, the median luminosity sharply increases with 
    halo mass. At the high mass end, the dependence approaches a power law 
    with an index of 0.28 (for pseudo-blue galaxies) or 
    0.35 (for pseudo-red galaxies). At fixed halo mass, pseudo-blue
    central galaxies are more luminous in $r$-band than pseudo-red ones, 
    by about one (half) magnitude at the low (high) mass end. 
    Or, for pseudo-blue and pseudo-red central galaxies of
    similar luminosity (in the sense of median luminosity), the latter 
    reside in more massive haloes, with mass about 0.2 (1) dex higher
    at the faint (luminous) end.
    
    \item[(3)] The fraction of central galaxies in the pseudo-red 
    component increases with halo mass and approaches a plateau of $\sim$65 
    per cent above $10^{12}\hinvMsun$, where the colour distribution of this 
    component also becomes narrow (about 0.02mag in scatter).

    \item[(4)] Based on the median mass of the host haloes, 
    for pseudo-blue central galaxies it is the luminosity that strongly 
    correlates with halo mass, more luminous 
    in more massive haloes, while for pseudo-red central galaxies, 
    it is the colour, redder in more massive haloes.
    
    \item[(5)] For satellite galaxies, the dependences of the characteristic 
    luminosity on halo mass of the two pseudo-colour components follow a trend 
    similar to the ones seen in the
    median central galaxy luminosity, with the difference being much larger 
    (e.g. $\sim 1.5$mag difference in the characteristic luminosity at 
    fixed halo mass). The faint-end slope of the
    CLF of the pseudo-blue satellites is generally steeper than that of the 
    pseudo-red satellites.
    
    \item[(6)] The pseudo-blue satellite component has a large spread in 
    colour at most luminosities, with a decrease of scatter at the highest
    luminosities ($\Mr \lesssim -20.5$). The pseudo-red satellite component 
    has a narrow colour distribution, with the scatter decreasing slowly 
    (from 0.06mag to 0.03mag) with luminosity.
    
    \item[(7)] At low luminosities, roughly $\Mr \gtrsim -19$, the bimodality
    of the galaxy CMD is driven largely by the contribution of blue centrals
    and red satellites. At higher luminosities, the bimodality is driven mainly
    by the blue and red central galaxy components, with satellites playing
    a steadily decreasing role. 
    
    \item[(8)] For a typical galaxy sample, colour is the strongest indicator
    of a galaxy's probability of being a satellite, with luminosity being a 
    secondary factor. The satellite fraction is higher for redder galaxies
    and higher for fainter ones, and it exceeds 50 per cent for faint
    red galaxies. The population of extremely red faint satellites likely
    originates from massive satellites whose stellar content has been 
    heavily reduced by stripping. 
    
    \item[(9)] For satellites, the median host halo mass of the 
    pseudo-blue component does not show a strong dependence on colour and 
    luminosity, while that of the pseudo-red component
    increases with satellite luminosity.
    
\end{itemize}

In \citet{Zehavi05,Zehavi11}, we model the difference in clustering of red 
and blue galaxies at fixed luminosity as a consequence of the larger
satellite fraction of red galaxies. While our analysis here (like many
others) confirms this higher satellite fraction, it also shows that the
halo masses of red central galaxies are higher than those of blue centrals
at fixed luminosity (conclusion 2 above). Galaxy-galaxy weak lensing 
measurements provide direct evidence for this mass difference 
\citep{Zu16,Zu17a}, which is derived here from clustering alone.

While the separation into pseudo-blue and pseudo-red populations in 
the model is initially phenomenological, motivated by the bimodal distribution 
of galaxies, the modelling results do show distinct differences of the 
two model populations, which indicates that the two
pseudo-colour populations have a physical origin. We speculate that 
the pseudo-blue central galaxies are dominated by galaxies with episodes of 
star formation during the past 2--3 Gyrs.
The large scatter in colour may reflect the differences in the time and 
duration of the star formation as well as effects of dust and metallicity. 
After the star formation shuts down, the stellar population evolves fast 
during the first 2--3 Gyrs, becoming fainter and redder, 
and then its colour evolves more slowly \citep[e.g.][]{Bruzual03}. 
This may lead to the pseudo-red central galaxy population, which has a 
narrow colour distribution. That is, if we
start with the pseudo-blue central CCMD, passive evolution alone 
within 2--3 Gyrs is able to `rotate' it to become close to the 
pseudo-red CCMD. It may be that the central galaxy CCMD has more 
than the two above components, but the relatively
fast transition makes the two components the dominant modes. 

For the pseudo-blue central 
galaxies, the correlation between luminosity and colour 
(with fainter galaxies appearing to be 
bluer) may be a manifestation that some effect of a transitional population 
(e.g. `pseudo-green'
galaxies) has been included. The luminosity of pseudo-blue central galaxies 
is correlated with halo mass, which may be related to the amount of cold gas 
available for star formation. For 
pseudo-red central galaxies, the colour shows a main correlation with halo 
mass and the luminosity to a less degree, which may be translated from 
the correlation between halo mass and the time of last episode of star 
formation and the mass-metallicity relation.

For satellite galaxies, the difference in the luminosity and colour 
distribution of the two pseudo-colour populations may be related to the 
dynamical evolution (tidal and ram pressure stripping and 
strangulation) of the satellites inside the parent haloes. 
The pseudo-blue satellites may be those falling into the parent halo recently, 
while pseudo-red satellites have orbited around for a longer time. 
Hydrodynamic galaxy formation simulations \citep[e.g.][]{Vogelsberger14,Schaye15}
and SAMs \citep[e.g.][]{Guo13} 
can be used to test the existence of the two populations and to better 
understand their origins, and conversely, our modelling 
results can be used to test galaxy formation models. 
We reserve such investigations for future work.

For galaxies more luminous than $\Mr = -20$, there appears to be a 
systematic trend in the projected 2PCFs for the bluest galaxy sample, 
a high clustering amplitude and steep clustering 
profile on sub-Mpc scales (see Fig.~\ref{fig:wpcomparison}). 
The CCMD model does not capture this trend well. As the signal-to-noise 
is low for each bluest sample at fixed luminosity bin, 
we performed a test with a luminous blue galaxy sample over a large range 
in luminosity, constructed using a tilted cut in the CMD. We find that the 
above small-scale clustering features persist. This could be a indication 
that the bluest satellites have a much steeper spatial distribution profile 
than assumed in our CCMD model. Further investigation with modelling
and with galaxy formation models will be useful to better understand the 
clustering of luminous blue galaxies.

Our CCMD model assumes a specific functional form for the CMD at fixed halo
mass: 2D Gaussians for the pseudo-blue and pseudo-red central and satellite
populations. Note that this is not exactly true for each satellite 
component, but the satellite CCMD can be broadly regarded as the superposition 
of a set of 2D Gaussians. 
To ensure a globally coherent description with a manageable
number of parameters, it assumes particular functional forms for the variation
of these 2D Gaussians with halo mass; these forms are motivated by previous 
observational analyses. We assume the halo mass function and halo clustering
of a specific cosmological model, and we assume that the satellite galaxies
trace the dark matter distribution within each halo while central galaxies
reside at the halo potential minimum. These assumptions could be adjusted 
without altering the basic principles of the model. The more fundamental 
physical assumption is that the statistical distributions of galaxy 
luminosity and galaxy colour depend only on halo mass, not on halo
environment or on other halo properties that are correlated with 
environment. This assumption underpins our predictions of galaxy clustering
in the two-halo regime, as we populate haloes using only their masses, and
internal profiles (for satellites).

It has been established from $N$-body simulations that the
clustering of haloes depends not only on halo mass, but also on halo assembly 
history 
\citep[e.g.][]{Sheth04,Gao05,Wechsler06,Harker06,Zhu06,Jing07,Zhu06,Li08,Salcedo17,Xu18}, 
which is termed as halo assembly bias. If galaxy properties closely tie to 
halo formation history and environment,
the inherited assembly bias from the host haloes would break the above assumption. 
If such a galaxy assembly bias effect exists and is strong, 
it would affect the clustering modelling 
and introduce systematics in cosmological constraints \citep[e.g.][]{Zentner14,Zentner16}.
Galaxy assembly bias has been searched for in galaxy 
formation simulations and models, and no
definite conclusion has been reached. SAMs 
predict significant galaxy assembly bias effects 
(e.g. on the HOD, CLF and clustering of low-mass 
galaxies; \citealt{Zhu06,Croton07,Zehavi18}). While the HODs in the hydrodynamic 
simulations studied in \citet{Berlind03} and \citet{Mehta14} show little 
dependence on halo environment, galaxy assembly bias in the EAGLE simulation 
\citep{Schaye15} leads to $\sim$25\%
effect on the galaxy 2PCFs at large scales \citep{Chaves-Montero16}. 
The different results from different models/simulations suggest that 
galaxy assembly bias in current galaxy formation models depends on the 
implementation details of star formation and feedback.
Conversely, if galaxy assembly bias can be inferred from observation, 
it would test such aspects in galaxy formation theories.

The search for assembly bias effect in galaxy survey data has not reached
a definite conclusion, either. The difficulty in detecting halo and 
galaxy assembly bias from observation lies in determining halo mass and 
finding the appropriate galaxy property as assembly proxy. 
Using an SDSS galaxy group catalog, \citet{Yang06} claim the detection 
of assembly bias (see also \citealt{Wang13,Lacerna14}), 
where halo mass is estimated in a similar way as the abundance matching 
technique \citep[e.g.][]{Conroy06} and the star formation
rate in the central galaxy is used as a proxy for halo age. 
Also with an SDSS galaxy group catalog, \citet{Berlind06} show a detection 
of assembly bias, but with a trend opposite to that in \citet{Yang06}. 
A key source of uncertainty in these studies is that groups with blue
and red central galaxies may have different halo mass at fixed total
luminosity, as suggested by our results here and by the galaxy-galaxy
weak lensing analysis of \cite{Zu16}.
\citet{Lin16} perform a more careful analysis of assembly bias in 
SDSS central galaxies,
aided by galaxy lensing measurements to control the halo mass, and no evidence
is found for assembly bias. This lack of detection suggests 
that the correlation, if any, between the chosen galaxy assembly indicator 
(e.g. star formation rate or galaxy formation 
time) and halo formation history may not be tight. 
Analyses of the luminosity dependent galaxy
clustering in the SDSS show no evidence of galaxy assembly bias for 
luminous samples and marginal
evidence for faint samples \citep[e.g.][]{Vakili16,Zentner16}. 
Recently, strong assembly bias effect is claimed to be detected with 
massive clusters, when clusters are split into subsamples
based on the average projected distance between satellite and central galaxies 
\citep{Miyatake16,More16,Baxter16}, and the unexpectedly large halo 
assembly bias was attributed to be possibly related to the nature of 
dark matter. However, the re-analysis of \citet{Zu17b} 
with a better control of projection effects shows no evidence of assembly bias. 

Our CCMD modelling results can help constrain galaxy assembly bias 
effect and we plan to work on such a study. 
We can construct a mock galaxy catalogue by populating haloes in $N$-body 
simulations according to the best-fitting CCMD model 
(and/or a series of mocks from a set of models to probe the uncertainties). 
This catalogue with galaxy luminosity and colour information would 
be an ideal control sample, free of galaxy assembly bias effect but 
reproducing the distributions of galaxy luminosity
and colour and the luminosity and colour dependent clustering of SDSS 
galaxies. With such a control sample, we plan to measure and compare 
a variety of environment dependent statistics in both the mock catalogue 
and in real data. Potential galaxy assembly signatures will be revealed 
by the difference between the two types of measurements. 
The mock catalogue will also have many other applications, 
such as helping derive accurate covariance matrices for galaxy clustering 
and testing galaxy survey designs. Another application 
is to aid the comparison between our CCMD modelling results and 
those from group catalogues, by applying group finding algorithm to 
the mock galaxy catalogue. This will in turn help understand 
systematics in various applications based on the group catalogue 
\citep[e.g.][]{Campbell15}, including those in constraining assembly 
bias and those mentioned in \S~\ref{sec:cmppre}.
The mock catalogue also makes it easy to compute cross-correlation 
functions between different samples, which will allow various tests.

In this paper, the CCMD model is constrained by galaxy number densities
and galaxy auto-correlation functions. The results can be tested with 
other clustering statistics, e.g. the cross-correlation functions
between individual colour sub-samples at a fixed magnitude bin and the full 
sample in the magnitude bin. The cross-correlation measurements can be
compared with CCMD model predictions based on the method similar to that 
presented in \S~\ref{subsec:simubased} or mock catalogues for a consistency 
check. A more effective use of the cross-correlation functions is to 
include them directly in constraining the model. The cross-correlations with 
the full sample at fixed magnitude bin not only have reduced shot noise but 
also encode information about galaxy pairs made of galaxies from colour 
sub-samples, both of which help tighten the model constraints. Similarly
galaxy lensing measurements for samples defined by fine bins of luminosity
and colour across the galaxy CMD can also serve as a consistency check or
a way to tighten model constraints, especially on the halo mass scales and
satellite fractions of different samples
\citep[e.g.][]{Leauthaud11, Leauthaud12, Tinker13}. The model can be further 
applied to and constrained by statistics like redshift-space clustering and 
higher-order clustering, which can bring in additional information, such as 
the relation between galaxy kinematics and luminosity/colour.

The CCMD modelling results can be compared to the galaxy-halo connections 
inferred from other methods. The SHAM method \citep[e.g.][]{Conroy06} maps galaxy
luminosity or stellar mass onto halo properties through matching the cumulative galaxy 
LF or stellar mass function with the cumulative abundance of haloes/sub-haloes 
(e.g. in terms of mass or circular velocity). Our result shows that the mapping
from luminosity to halo mass needs to account for the dependence on galaxy colour
(Fig.~\ref{fig:msmscen}). Using the estimate of stellar mass from galaxy luminosity 
and colour with the method in \citet{Bell03}, we find that this remains true for
connecting galaxy stellar mass to halo mass (see also \citealt{Rodriguez15, Zu16,Zu17a}). 
Whether the conclusion holds for
other halo properties (e.g. circular velocity) is yet to be tested.
The extended 
`age-matching' method \citep{Hearin13} assigns galaxy colour based on 
halo/sub-halo formation time. By construction, the method reproduces the galaxy 
CMD. Although it reasonably reproduces the colour-dependent galaxy 2PCFs 
for galaxy samples divided into red and blue sub-samples, its performance on 
modelling the clustering of galaxies in fine colour bins remains to be tested
and compared with the CCMD model. The CCMD modelling results can inform 
the range of validity of the `age-matching' method. 
Last but not least, the CCMD model that reproduces the
luminosity and colour dependent clustering can be used to test against 
predictions from galaxy formation models, including both hydrodynamic simulations
and SAMs. The differences between the empirically constrained CCMD and galaxy
formation model predictions can provide insight about physical processes in 
galaxy formation and help improve galaxy formation models.

The CCMD model presented in this paper
studies galaxies in term of the directly observed galaxy 
properties, luminosity and colour. 
In principle, the formalism can be generalised to study galaxy samples defined by 
inferred quantities, such as the stellar mass and star formation rate, 
the two quantities more physically related to galaxy formation and evolution. 
Or, as an intermediate step, one can construct and model galaxy samples defined by
galaxy stellar mass and colour. Many analyses in this paper and
proposed above can be equally done with such modelling, which will yield 
further insights into understanding galaxy formation and evolution. 
Furthermore, results of CCMD-like modelling 
of galaxy clustering at different redshifts can be used to learn about the evolution
of galaxies in the CMD and the buildup and evolution of the bimodality.

\section*{Acknowledgements}
We thank Kyle Dawson and Xiaohu Yang for useful discussions.
HX acknowledges the support by a fellowship from the Willard L. 
and Ruth P. Eccles Foundation. IZ is supported by NSF grant AST-1612085.
During the final stage, this project has been supported by a grant from Science and Technology Commission of Shanghai Municipality (Grants No.16DZ2260200) and National Natural Science Foundation of China (Grants No.11655002). 

We gratefully acknowledge the 
High Performance Computing Resource for 
Advanced Research Computing at Shanghai Astronomical Observatory. 
The support and resources from the Center for High Performance Computing 
at the University of Utah are gratefully acknowledged.

The authors gratefully acknowledge the Gauss Centre for Supercomputing 
e.V. (www.gauss-centre.eu) and the Partnership for Advanced Supercomputing 
in Europe (PRACE, www.prace-ri.eu) for funding the MultiDark simulation 
project by providing computing time on the GCS Supercomputer SuperMUC 
at Leibniz Supercomputing Centre (LRZ, www.lrz.de).
The CosmoSim database used in this paper is a service by the 
Leibniz-Institute for Astrophysics Potsdam (AIP). The MultiDark database 
was developed in cooperation with the Spanish MultiDark Consolider 
Project CSD2009-00064.

Funding for the SDSS and SDSS-II has been provided by the Alfred P. Sloan
Foundation, the Participating Institutions, the National Science Foundation,
the U.S. Department of Energy, the National Aeronautics and Space
Administration, the Japanese Monbukagakusho, the Max Planck Society, and the
Higher Education Funding Council for England. The SDSS Web Site is
\url{http://www.sdss.org/}.

The SDSS is managed by the Astrophysical Research Consortium for the
Participating Institutions. The Participating Institutions are the American
Museum of Natural History, Astrophysical Institute Potsdam, University of
Basel, University of Cambridge, Case Western Reserve University, University
of Chicago, Drexel University, Fermilab, the Institute for Advanced Study,
the Japan Participation Group, Johns Hopkins University, the Joint Institute
for Nuclear Astrophysics, the Kavli Institute for Particle Astrophysics and
Cosmology, the Korean Scientist Group, the Chinese Academy of Sciences
(LAMOST), Los Alamos National Laboratory, the Max-Planck-Institute for
Astronomy (MPIA), the Max-Planck-Institute for Astrophysics (MPA), New Mexico
State University, Ohio State University, University of Pittsburgh, University
of Portsmouth, Princeton University, the United States Naval Observatory, and
the University of Washington.

\bibliography{refdatabase}

\appendix
\section{Galaxy Samples}
\label{subsec:galsampleinfo}

Table~\ref{tab:info} lists the sample information for the 79 galaxy 
samples used in our CCMD modelling, defined 
by fine bins in colour and luminosity. The samples are constructed following 
the steps laid out in 
\S~\ref{subsec:binning}. All the samples are volume-limited,
and the samples of the same magnitude cuts (thus the same 
$z_{\rm min}$ and $z_{\rm max}$) have the same volume.  
At each magnitude bin, the division into colour sub-samples  
depends on the galaxy number within the volume,
balancing between reaching reasonable 
signal-to-noise ratios of clustering measurements and capturing
fine features along the colour direction.

\begin{table}
  \caption{The 79 volume-limited galaxy samples defined by fine bins in 
  magnitude and colour. The first and second columns 
  are the luminosity and colour ranges of the galaxy samples, respectively.
  The maximum redshift of each sample is listed
  in the third column with the same minimum redshift $z_{\rm min} = 0.02$ 
  adopted for all the samples.
  For each sample, the fourth and fifth columns are the number of galaxies 
  in the specific sample volume and 
  its number density in unit of $h^3 {\rm Mpc^{-3}}$.}
  \label{tab:info}
   \begin{tabular}{ccccc}
     \hline
     \hline
     $M_{\rm r}$ (mag) & ${\rm g-r}$ (mag) & $z_{\rm max}$ & $N_{\rm gal}$ 
     & $n_{\rm gal} (h^3 {\rm Mpc^{-3}})$ \\
     \hline
     {[-18.00, -18.25]} & [0.00,0.44]  & 0.04257   & 1667   &1.28E-03 \\
                        & [0.44,0.65]  & 0.04257   & 1556   &1.19E-03 \\
                        & [0.65,1.20]  & 0.04257   & 1651   &1.26E-03 \\
     {[-18.25, -18.50]} & [0.00,0.46]  & 0.04759   & 2145   &1.14E-03 \\
                        & [0.46,0.71]  & 0.04759   & 2202   &1.17E-03 \\
                        & [0.71,1.20]  & 0.04759   & 2188   &1.16E-03 \\
     {[-18.50, -18.75]} & [0.00,0.49]  & 0.05318   & 2826   &1.06E-03 \\
                        & [0.49,0.75]  & 0.05318   & 2873   &1.08E-03 \\
                        & [0.75,1.20]  & 0.05318   & 2847   &1.06E-03 \\
     {[-18.75, -19.00]} & [0.00,0.52]  & 0.05941   & 3635   &9.64E-04 \\
                        & [0.52,0.79]  & 0.05941   & 3456   &9.17E-04 \\
                        & [0.79,1.20]  & 0.05941   & 3558   &9.43E-04 \\
     {[-19.00, -19.25]} & [0.00,0.55]  & 0.06634   & 4800   &9.08E-04 \\
                        & [0.55,0.83]  & 0.06634   & 4714   &8.92E-04 \\
                        & [0.83,1.20]  & 0.06634   & 4820   &9.12E-04 \\
     {[-19.25, -19.50]} & [0.00,0.59]  & 0.07403   & 6485   &8.82E-04 \\
                        & [0.59,0.86]  & 0.07403   & 6461   &8.77E-04 \\
                        & [0.86,1.20]  & 0.07403   & 6308   &8.56E-04 \\
     {[-19.50, -19.75]} & [0.00,0.53]  & 0.08258   & 5036   &4.93E-04 \\
                        & [0.53,0.68]  & 0.08258   & 5068   &4.97E-04 \\
                        & [0.68,0.84]  & 0.08258   & 4997   &4.90E-04 \\
                        & [0.84,0.92]  & 0.08258   & 5172   &5.07E-04 \\
                        & [0.92,1.20]  & 0.08258   & 5011   &4.91E-04 \\
     {[-19.75, -20.00]} & [0.00,0.53]  & 0.09207   & 4920   &3.49E-04 \\
                        & [0.53,0.66]  & 0.09207   & 5211   &3.70E-04 \\
                        & [0.66,0.80]  & 0.09207   & 5157   &3.66E-04 \\
                        & [0.80,0.89]  & 0.09207   & 5013   &3.56E-04 \\
                        & [0.89,0.94]  & 0.09207   & 5357   &3.80E-04 \\
                        & [0.94,1.20]  & 0.09207   & 4946   &3.51E-04 \\
                        \hline 
   \end{tabular}
 \end{table}
 
 \begin{table}
   \contcaption{The 79 volume-limited galaxy samples defined by fine bins 
   in magnitude and colour.}
   \label{tab:infocont}
    \begin{tabular}{ccccc}
     \hline
     \hline
     $M_{\rm r}$ (mag) & ${\rm g-r}$ (mag) & $z_{\rm max}$ & $N_{\rm gal}$ 
     & $n_{\rm gal} (h^3 {\rm Mpc^{-3}})$ \\
     \hline
    {[-20.00, -20.25]}  & [0.00,0.57]  & 0.10258   & 5922   &3.05E-04 \\
                        & [0.57,0.69]  & 0.10258   & 5407   &2.79E-04 \\
                        & [0.69,0.82]  & 0.10258   & 5706   &2.94E-04 \\
                        & [0.82,0.90]  & 0.10258   & 5600   &2.89E-04 \\
                        & [0.90,0.95]  & 0.10258   & 6613   &3.41E-04 \\
                        & [0.95,1.20]  & 0.10258   & 5096   &2.63E-04 \\
    {[-20.25, -20.50]} & [0.00,0.57]  & 0.11422   & 5381   &2.02E-04 \\
                        & [0.57,0.68]  & 0.11422   & 5534   &2.08E-04 \\
                        & [0.68,0.79]  & 0.11422   & 5667   &2.13E-04 \\
                        & [0.79,0.88]  & 0.11422   & 5879   &2.21E-04 \\
                        & [0.88,0.92]  & 0.11422   & 5046   &1.90E-04 \\
                        & [0.92,0.96]  & 0.11422   & 6329   &2.38E-04 \\
                        & [0.96,1.20]  & 0.11422   & 4944   &1.86E-04 \\
     {[-20.50, -20.75]} & [0.00,0.58]  & 0.12709   & 5120   &1.41E-04 \\
                        & [0.58,0.68]  & 0.12709   & 5334   &1.47E-04 \\
                        & [0.68,0.77]  & 0.12709   & 5032   &1.38E-04 \\
                        & [0.77,0.85]  & 0.12709   & 5068   &1.39E-04 \\
                        & [0.85,0.90]  & 0.12709   & 4898   &1.35E-04 \\
                        & [0.90,0.94]  & 0.12709   & 6635   &1.82E-04 \\
                        & [0.94,0.97]  & 0.12709   & 5127   &1.41E-04 \\
                        & [0.97,1.20]  & 0.12709   & 4507   &1.24E-04 \\
     {[-20.75, -21.00]} & [0.00,0.60]  & 0.14132   & 5329   &1.08E-04 \\
                        & [0.60,0.70]  & 0.14132   & 5295   &1.07E-04 \\
                        & [0.70,0.79]  & 0.14132   & 5373   &1.08E-04 \\
                        & [0.79,0.87]  & 0.14132   & 5918   &1.19E-04 \\
                        & [0.87,0.91]  & 0.14132   & 4886   &9.86E-05 \\
                        & [0.91,0.94]  & 0.14132   & 5604   &1.13E-04 \\
                        & [0.94,0.97]  & 0.14132   & 5615   &1.13E-04 \\
                        & [0.97,1.20]  & 0.14132   & 5150   &1.04E-04 \\
     {[-21.00, -21.25]} & [0.00,0.64]  & 0.15702   & 5465   &8.13E-05 \\
                        & [0.64,0.76]  & 0.15702   & 6221   &9.25E-05 \\
                        & [0.76,0.85]  & 0.15702   & 5409   &8.04E-05 \\
                        & [0.85,0.91]  & 0.15702   & 6135   &9.12E-05 \\
                        & [0.91,0.94]  & 0.15702   & 5315   &7.90E-05 \\
                        & [0.94,0.97]  & 0.15702   & 5631   &8.37E-05 \\
                        & [0.97,1.20]  & 0.15702   & 5687   &8.46E-05 \\
     {[-21.25, -21.50]} & [0.00,0.69]  & 0.17434   & 5599   &6.16E-05 \\
                        & [0.69,0.82]  & 0.17434   & 5605   &6.17E-05 \\
                        & [0.82,0.90]  & 0.17434   & 5316   &5.85E-05 \\
                        & [0.90,0.94]  & 0.17434   & 5651   &6.22E-05 \\
                        & [0.94,0.97]  & 0.17434   & 5387   &5.93E-05 \\
                        & [0.97,1.20]  & 0.17434   & 5880   &6.47E-05 \\
     {[-21.50, -21.75]} & [0.00,0.75]  & 0.19342   & 5308   &4.34E-05 \\
                        & [0.75,0.89]  & 0.19342   & 5114   &4.18E-05 \\
                        & [0.89,0.94]  & 0.19342   & 5045   &4.12E-05 \\
                        & [0.94,0.98]  & 0.19342   & 5891   &4.82E-05 \\
                        & [0.98,1.20]  & 0.19342   & 4642   &3.79E-05 \\
     {[-21.75, -22.00]} & [0.00,0.90]  & 0.21441   & 5929   &3.61E-05 \\
                        & [0.90,0.97]  & 0.21441   & 6099   &3.71E-05 \\
                        & [0.97,1.20]  & 0.21441   & 5403   &3.29E-05 \\
   \hline
   \end{tabular}
  \end{table}
  
\section{Tests with Parameter Constraints}
\label{sec:AppendixB}

\begin{figure*}
\includegraphics[width=\textwidth]{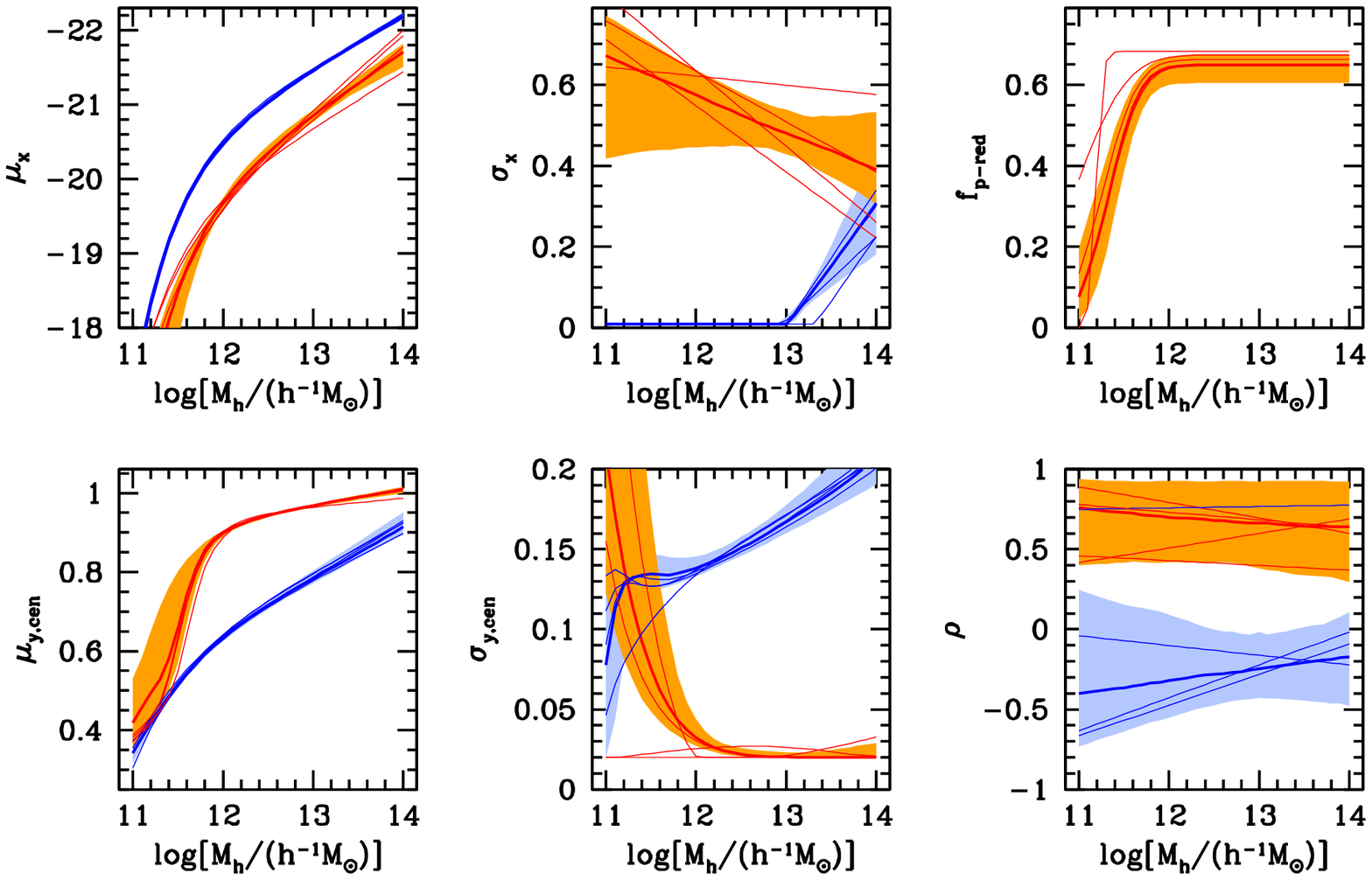}
\caption{
CCMD quantities (defined in \S~\ref{subsec:cenpara}) as a function of halo mass for central galaxies. 
Similar to Fig.\ref{fig:msmscen}, but with the shaded regions corresponding to
2$\sigma$ ranges and with thin curves added from four perturbed runs (see the text in Appendix~\ref{sec:AppendixB}).
}
\label{fig:msmscen_convergence}
\end{figure*}

\begin{figure*}
\includegraphics[width=\textwidth]{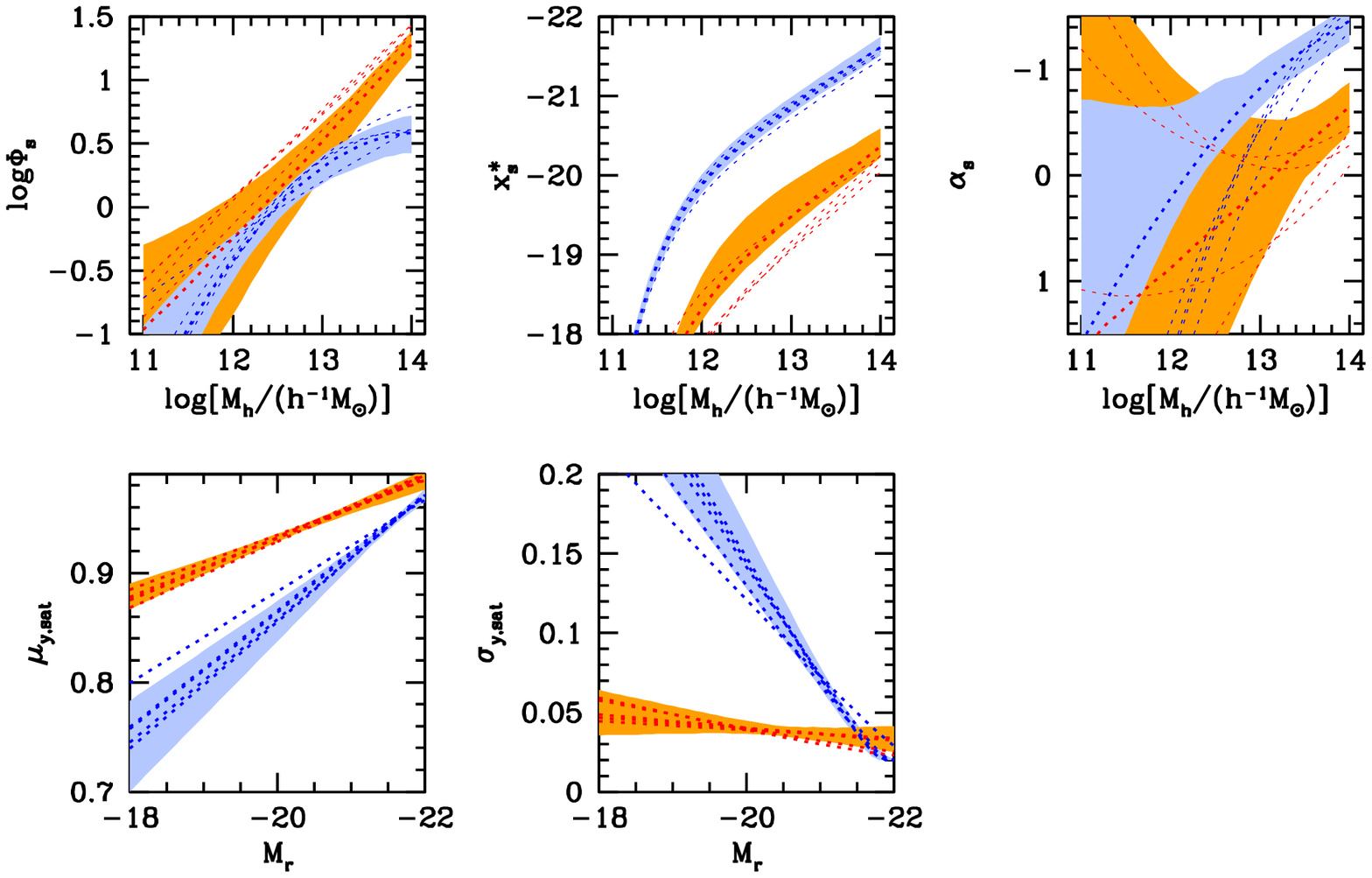}
\caption{
CCMD quantities (defined in \S~\ref{subsec:satpara}) as a function of halo mass for satellite galaxies. 
Similar to Fig.\ref{fig:msmssat}, but with the shaded regions corresponding to
2$\sigma$ ranges and with thin curves added from four perturbed runs (see the text).
}
\label{fig:msmssat_convergence}
\end{figure*}

\begin{figure*}
\includegraphics[width=\textwidth]{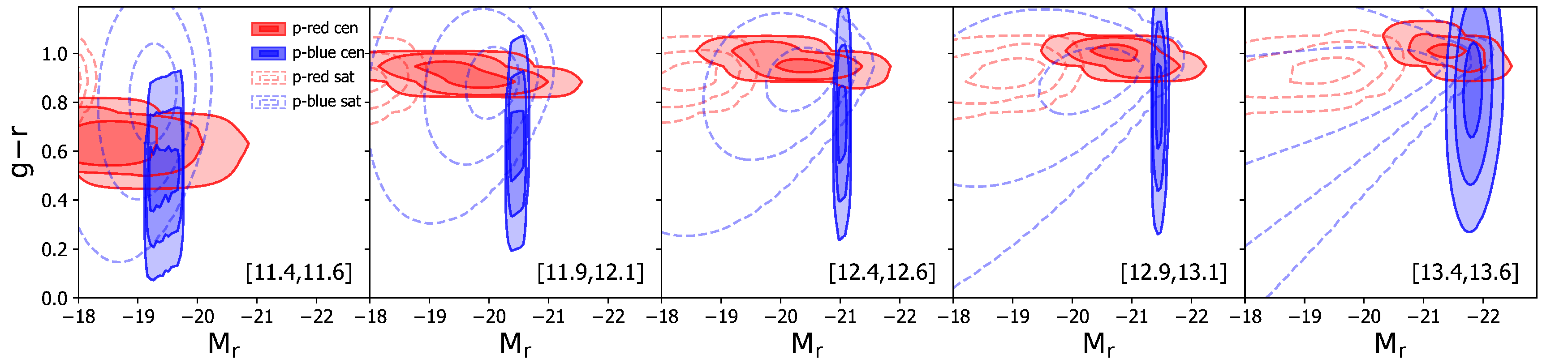}
\caption{
Similar to Fig.\ref{fig:halomassctur} but for the parameter set with a flat
$\sigma_{\rm y,cen}-\Mh$ relation shown in the bottom-middle panel in Fig.~\ref{fig:msmscen_convergence}.
}
\label{fig:halomassctur_deviationsigycen}
\end{figure*}

\begin{figure*}
\includegraphics[width=\textwidth]{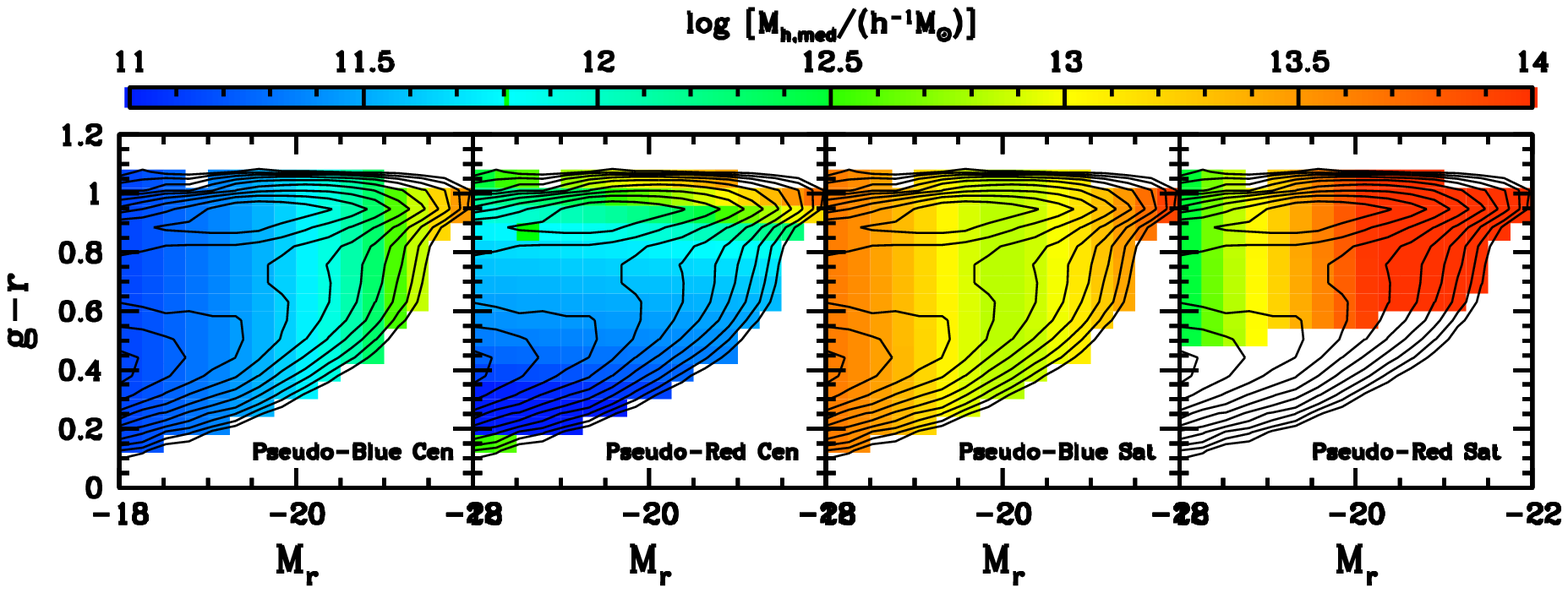}
\caption{
Similar to Fig.\ref{fig:mdmctur_4panels} but for the parameter set with flat
$\sigma_{\rm y,cen}-\Mh$ relation shown in the bottom-middle panel in Fig.~\ref{fig:msmscen_convergence}.
}
\label{fig:mdmctur_4panels_maskbinning_deviationsigycen}
\end{figure*}

\begin{figure*}
\includegraphics[width=\columnwidth]{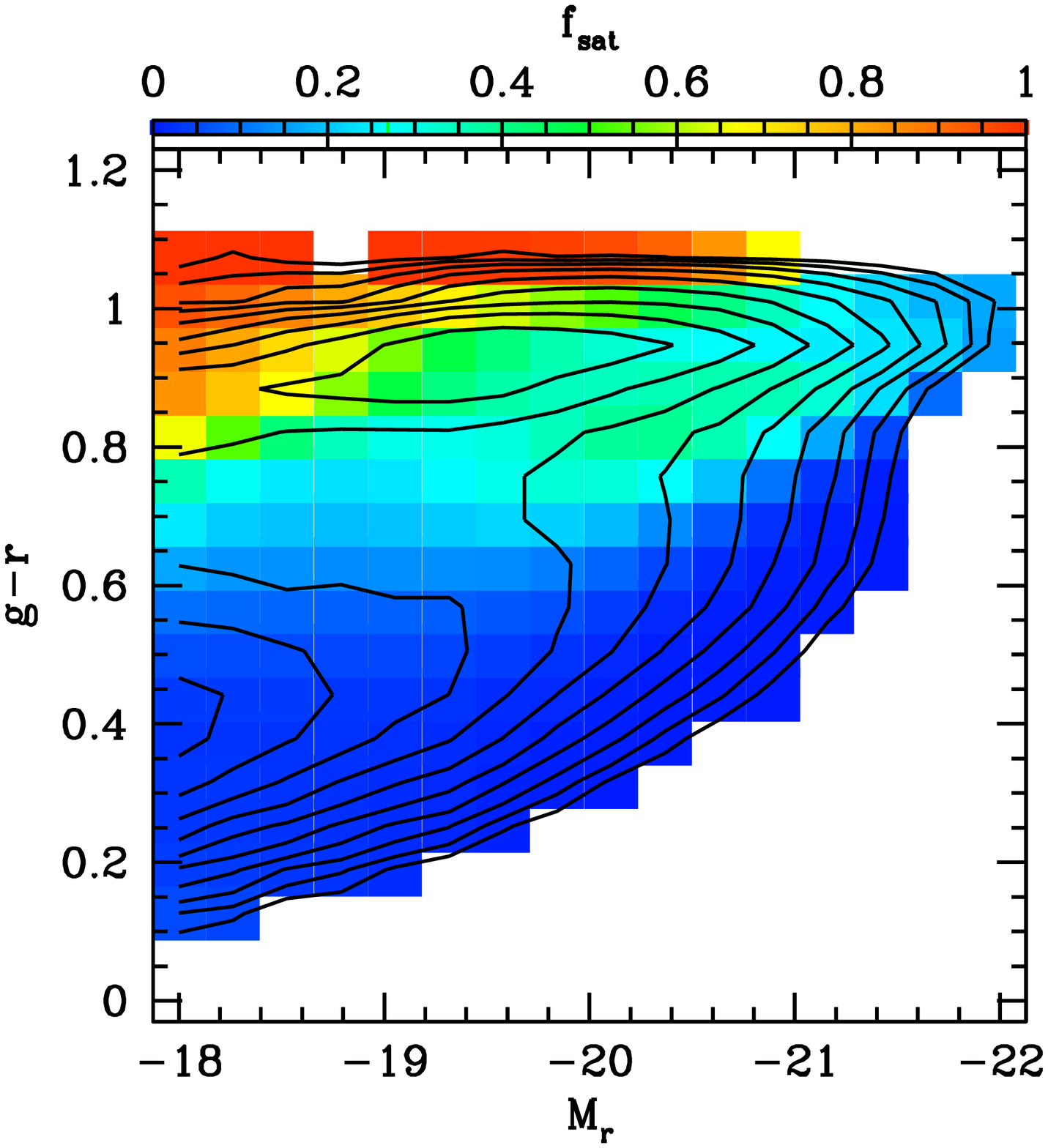}
\caption{
Similar to Fig.~\ref{fig:satcmd} but for the parameter set with flat
$\sigma_{\rm y,cen}-\Mh$ relation shown in the bottom-middle panel in Fig.~\ref{fig:msmscen_convergence}.
}
\label{fig:satcmd_deviationsigycen}
\end{figure*}

The CCMD model is characterised by 59 parameters (\S~\ref{subsec:cenpara} 
and \S~\ref{subsec:satpara}).
To help explore such a high-dimensional parameter space, we first run a set of
test MCMC chains of length around 500,000 with different starting
positions. Based on the results of the test runs, we further apply the
Gauss-Newton method to search for minima in the $\chi^2$ surface. We then use 
the position with the minimum $\chi^2$ as the starting point to the run the chain of
length $10^8$, and the results in the main text are based on such a run. 
The CCMD parameterization follows natural forms of the relations between 
galaxy properties and halo mass, which is not optimised in any way to reduce
parameter degeneracies. In terms of the results, it is the physical relations
(like those in Fig.~\ref{fig:halomassctur}, Fig.\ref{fig:msmscen}, and
Fig.~\ref{fig:msmssat}) instead of the relations among parameters themselves 
that are of interest. Also the derived quantities and trends, like those in 
Fig.~\ref{fig:mdmctur_3panels} and Fig.~\ref{fig:satcmd}, provide useful and
meaningful information for the model constraints. 

There are possibilities that the default results we adopt in the main text come 
from a local minimum or that there are locations in the parameter space with
results highly degenerate with the default ones. To see how such possibilities 
affect the derived relations and trends, we perform tests by perturbing 
the parameters to run additional MCMC chains. We start from the positions of 
the best-fitting model in the main text and randomly perturb each parameter by
-5$\sigma$, 0, and 5$\sigma$, where the value of 1-$\sigma$ range of the 
parameter is determined from the local curvature of the $\chi^2$ surface along
the parameter direction. We then randomly choose different combinations of 
the perturbed parameters to start test MCMC runs, each with length 500,000. 

In Fig.~\ref{fig:msmscen_convergence} and Fig.~\ref{fig:msmssat_convergence}, 
the thin curves show the relations between CCMD properties and halo mass from 
best-fitting models resulting from 4 such perturbed runs, all of which have
$\chi^2 \sim 950$. The shaded regions represent the 2-$\sigma$ range in the
default model. In most ranges, the curves from the perturbed runs fall into the
2-$\sigma$ range. However, there are a few noticeable deviations. For example,
in the $\sigma_{\rm y,cen}-\Mh$ panel of Fig.~\ref{fig:msmscen_convergence},
there are curves corresponding to much narrow colour distribution for the
pseudo-red central galaxies in low mass haloes, while in the 
$f_{\textrm{p-red}}-\Mh$ panel there is a case with sharp increase in
$f_{\textrm{p-red}}$ around $\log[\Mh/(\hinvMsun)]=11.2$ to the asymptotic 
value. Since at the lowest luminosity ($\Mr=-18$), the mass of the host haloes 
of the pseudo-red central galaxies is around $\log[\Mh/(\hinvMsun)]=11.4$
(top-left panel), the mass range of the above large deviations is where
the model is poorly constrained. Therefore, the deviations suggest model
degeneracies in the weakly constrained regions of the parameter space with 
the galaxy samples used in the modelling.

To see how the results are affected, we show the CCMD and derived relations 
and trends from a parameter set with $\sigma_{\rm y,cen}-\Mh$ relation being 
almost a flat curve in the bottom-middle panel of 
Fig.~\ref{fig:msmscen_convergence}. 
Its CCMD (Fig.~\ref{fig:halomassctur_deviationsigycen}) is almost the same as 
the default model (Fig.~\ref{fig:halomassctur}), in terms of the trends and 
orientations in the four pseudo-colour components. The main difference is in the
panel corresponding to the lowest mass bin, where the pseudo-red central 
component becomes slightly tighter in the colour direction and less tilted. 
The median halo mass (Fig.~\ref{fig:mdmctur_4panels_maskbinning_deviationsigycen})
for the pseudo-colour components and the luminosity/colour dependent 
satellite fraction (Fig.~\ref{fig:satcmd_deviationsigycen}) are quite 
similar to those seen in Fig.~\ref{fig:mdmctur_4panels} and 
Fig.~\ref{fig:satcmd}, with the median halo mass of pseudo-blue
central galaxies being slightly lower. We also apply the same analyses to the 
parameter set with the sharply increasing curve in the $f_{\textrm{p-red}}-\Mh$ 
panel of Fig.~\ref{fig:msmscen_convergence} and find a similar result. For this
model, the lower fraction of pseudo-blue central galaxies in the mass range
of $\log[\Mh/(\hinvMsun)]\sim$11.2-12.0 is compensated by putting more 
pseudo-blue satellites into those haloes, while approximately conserving the 
abundances and
clustering of blue galaxies. Note that the constraining power from galaxy 
clustering of the faint, blue samples are especially limited, and the abundances 
is more likely to be the driving force in the model fitting because of the 
small uncertainties. Larger survey for the faint samples would improve the 
model constraints.

Our tests show that there are regions in the parameter space that correspond to
degenerate solutions, with the CCMD relations and derived quantities/trends 
similar to those in our default models. On the one hand, this is reassuring as
the main results and conclusions of the paper hold. On the other hand, it implies
that a full exploration of the high-dimensional parameter space is difficult. 
We therefore take a practical view of our best-fitting model. It works
in reproducing the colour/luminosity dependent clustering and the abundance
distribution in the galaxy CMD, and its validity can be further tested with other
statistics (e.g. higher order clustering and galaxy lensing measurements). 
For some applications with the mock galaxy catalogues constructed based on the 
modelling results (as discussed in \S~\ref{sec:dis}), it would be necessary to
account for the systematic uncertainties from the above tests for more complete
investigations.

\end{document}